\documentclass{article}
\usepackage{jheppub} % for details on the use of the package, please see the 
\usepackage[utf8x]{inputenc}	%for german Umlaute
\usepackage[english]{babel}

\usepackage[ddmmyyyy,hhmmss]{datetime}

\usepackage{multirow}

\catcode`@=11
\allowdisplaybreaks
\usepackage{amsxtra}
\usepackage{mathtools}
\usepackage{float}
\usepackage{xcolor}
\usepackage{tikz}
\usepackage{macros}
\usepackage{hyperref}
\usepackage{amsmath}
\usepackage{graphicx}
\usepackage{rotating}  % rotating 

\newcommand{\Figref}[1]{Figure~\ref{#1}}

\newcommand{\hs}{\mathrm{HS}}
\newcommand{\hsc}{\mathrm{HS}_{\mathcal{C}}}
\newcommand{\hsh}{\mathrm{HS}_{\mathcal{H}}}
\newcommand{\pe}{\mathrm{PE}}
\usepackage{placeins}
\usetikzlibrary{calc}
\usetikzlibrary{arrows.meta, bending, patterns}
\usepackage{soul}
\usetikzlibrary{hobby,intersections} 
\usetikzlibrary{shapes.geometric}
\usetikzlibrary{shapes.misc}

\tikzset{flavour/.style={draw=none,minimum size=0.3mm,fill=white, regular polygon,regular polygon sides=4,draw}}
\tikzset{flavourr/.style={draw=none,minimum size=0.3mm,fill=red, regular polygon,regular polygon sides=4,draw}}
\tikzset{flavourb/.style={draw=none,minimum size=0.3mm,fill=blue, regular polygon,regular polygon sides=4,draw}}
\tikzset{gaugeBig/.style={inner sep=1mm,draw=none,fill=white,minimum size=2mm,circle, draw}}
\tikzset{bd/.style={circle, draw=black, inner sep=0pt, fill=black, minimum size=2mm}}
\tikzset{wd/.style={circle, draw=black, inner sep=0pt, fill=white, minimum size=2mm}}
\tikzset{Dynkin/.style={circle, draw=black, inner sep=0pt, fill=white, minimum size=2mm}}
\tikzstyle{ligne}=[draw, very thick] 
\tikzstyle{gridline}=[draw, gray] 
\tikzset{gauge/.style={circle, draw,inner sep=2.5pt}}
\tikzset{gaugeo/.style={circle, draw,inner sep=2.5pt,fill=orange}}
\tikzset{gauger/.style={circle, draw,inner sep=2.5pt,fill=red}}
\tikzset{gaugep/.style={circle, draw,inner sep=2.5pt,fill=pink}}
\tikzset{gaugeg/.style={circle, draw,inner sep=2.5pt,fill=green}}
\tikzset{gaugeb/.style={circle, draw,inner sep=2.5pt,fill=blue}}
\tikzset{gaugepu/.style={circle, draw,inner sep=2.5pt,fill=purple}}
\tikzset{gaugegoodpurple/.style={circle, draw,inner sep=2.5pt,fill=goodpurple}}
\tikzset{gaugem/.style={circle, draw,inner sep=2.5pt,fill=magenta}}
\tikzset{hasse/.style={circle, fill,inner sep=2pt}}
\tikzset{d2/.style={circle, fill,inner sep=1.3pt}}
\tikzset{shrinky/.style={circle, fill,inner sep=1pt}}
\tikzset{sized/.style={circle, draw, inner sep=1.5pt}}
\tikzset{seven/.style={circle, draw,inner sep=3pt}}
\tikzset{7brane/.style={circle, draw=black, fill=black,ultra thick,inner sep=1 pt, minimum size=1 pt,}, c/.default={4pt}}

\tikzset{gaugebl/.style={circle,draw=black,fill=black,inner sep=1.5pt}}
\tikzset{gaugeblnormal/.style={circle,draw=black,fill=black,inner sep=2.5pt}}
\tikzset{hasse/.style={circle, fill,inner sep=2pt}}
\tikzstyle{dashed_brane}=[thick, dashed]
\tikzstyle{dotted_brane}=[thick, dotted]
\tikzstyle{O3plus}=[thick, color=purple]
\tikzstyle{O3minustilde}=[thick, color=blue]
\tikzstyle{O3plustilde}=[thick, color=red]
\tikzset{D5/.style={cross out, draw=black, minimum size=7, inner sep=0pt, outer sep=0pt}, cross/.default={1pt}}
\tikzset{flavor/.style={regular polygon,regular polygon sides=4,inner sep=2.5pt, label = {}, draw}}
\tikzset{redflavor/.style={regular polygon,regular polygon sides=4,inner sep=2.5pt, color=red, label = {}, draw}}
\tikzset{redgauge/.style={inner sep=1mm,color=red,draw=none,minimum size=2mm,circle, draw}}
\tikzset{blueflavor/.style={regular polygon,regular polygon sides=4,inner sep=2.5pt, color=blue, label = {}, draw}}
\tikzset{bluegauge/.style={inner sep=1mm,color=blue,draw=none,minimum size=2mm,circle, draw}}

\makeatletter
\DeclareRobustCommand{\rvdots}{%
  \vbox{
    \baselineskip4\p@\lineskiplimit\z@
    \kern-\p@
    \hbox{.}\hbox{.}\hbox{.}
  }}
\makeatother

\newcommand{\surm}{\mathrm{SU}}

\newcommand{\urm}{\mathrm{U}}
\newcommand{\sorm}{\mathrm{SO}}

\newcommand{\orm}{\mathrm{O}}
\newcommand{\sprm}{\mathrm{Sp}}
\newcommand{\spin}{\mathrm{Spin}}
\newcommand{\hwg}{\mathrm{HWG}}

\newcommand{\hscz}{\mathrm{HS}_{\mathbb Z}^{\mathcal C}}
\newcommand{\hsczh}{\mathrm{HS}_{\frac{1}{2}\mathbb Z}^{\mathcal C}}
\newcommand{\hsz}{\mathrm{HS}_{\mathbb Z}}
\newcommand{\hszh}{\mathrm{HS}_{\mathbb{Z}_{\frac{1}{2}}}}

\tikzset{gaugebl/.style={circle,draw=black,fill=black,inner sep=1.5pt}}
\tikzset{gaugeblnormal/.style={circle,draw=black,fill=black,inner sep=2.5pt}}
\tikzset{hasse/.style={circle, fill,inner sep=2pt}}
\tikzstyle{dashed_brane}=[thick, dashed]
\tikzstyle{dotted_brane}=[thick, dotted]
\tikzstyle{O3plus}=[thick, color=purple]
\tikzstyle{O3minustilde}=[thick, color=blue]
\tikzstyle{O3plustilde}=[thick, color=red]
\tikzset{D5/.style={cross out, draw=black, minimum size=7, inner sep=0pt, outer sep=0pt}, cross/.default={1pt}}
\tikzset{flavor/.style={regular polygon,regular polygon sides=4,inner sep=2.5pt, label = {}, draw}}
\tikzset{redflavor/.style={regular polygon,regular polygon sides=4,inner sep=2.5pt, color=red, label = {}, draw}}
\tikzset{redgauge/.style={inner sep=1mm,color=red,draw=none,minimum size=2mm,circle, draw}}
\tikzset{blueflavor/.style={regular polygon,regular polygon sides=4,inner sep=2.5pt, color=blue, label = {}, draw}}
\tikzset{bluegauge/.style={inner sep=1mm,color=blue,draw=none,minimum size=2mm,circle, draw}}
\allowdisplaybreaks

\usepackage{todonotes}
\title{$\spin(N)$ Magnetic Quivers}

\preprint{Imperial/TP/26/AH/07}
\author[1]{Mohammad Akhond}
\author[2]{, Sam Bennett}
\author[2]{, Amihay Hanany}
\affiliation[1]{\it Sezione INFN Roma “Tor Vergata” \& Dipartimento di Fisica,
Universita di Roma “Tor Vergata”, Via della Ricerca Scientifica 1, 00133, Roma, Italy}
\affiliation[2]{Abdus Salam Centre for Theoretical Physics, Imperial College London, Prince Consort Road, SW7 2AZ, UK}

\emailAdd{akhond@roma2.infn.it}
\emailAdd{samuel.bennett18@imperial.ac.uk}
\emailAdd{amihay.hanany@imperial.ac.uk}
\abstract{This work introduces new magnetic quivers for $5\mathrm{d}$ $\mathcal N=1$ $\spin(N)$ gauge theories with hypermultiplets in spinor representations at both finite and infinite coupling. Among them, are new 3d $\mathcal{N}=4$ theories whose Coulomb and Higgs branches are isolated symplectic singularities, or a product of such spaces. These theories further expand the set of orthosymplectic quiver gauge theories whose Coulomb branch is a single isolated singularity, termed `minimal quivers'. Wreathings and foldings thereof realise quivers for non-simply laced nilpotent orbit closures.}
\begin{document}
\maketitle
\flushbottom
\section{Introduction}
Quantum field theory at strong coupling is notoriously opaque. However carefully a theory's dynamics, spectrum or vacuum manifold is arranged perturbatively, non-perturbative effects tend to lead to dramatic modifications. The cause is the emergence of additional massless degrees of freedom, such that perturbative physics no longer captures the system's full dynamics and new, non-perturbative techniques must be adopted in the strong-coupling regime.

Of course, the full description of strongly-coupled dynamics is unknown in general quantum field theories. For 5d $\mathcal{N}=1$ superconformal field theories admitting a description using brane webs \cite{Seiberg_1996,Aharony_1998,Zafrir_2016,Bergman_2015,Zafrir_2015,Hayashi:2018lyv,Hayashi:2019yxj,Benini:2009gi,Akhond:2020vhc}, the situation is somewhat simpler. At infinite coupling, the relevant massless degrees of freedom are gauge instantons transforming under $\urm(1)_{\mathrm{I}}$. These objects, which contribute to the theory's chiral ring, augment both the dimension and global symmetry of the theory's Higgs branch. Although these modifications are not reachable using a Lagrangian description, recent work has shown that the brane web itself offers a probe into the strong-coupling regime via the introduction of a magnetic quiver \cite{Akhond:2024nyr,Hanany:2022itc,Bourget:2020mez,Bourget:2020xdz,Bourget:2020gzi}.

Magnetic quivers are ancillaries derived from brane systems that capture the structure of the Higgs branch of a four-, five- or six-dimensional quiver gauge theory as a moduli space of dressed monopole operators. At finite-coupling, it's expected that these are often the 3d-mirror of the electric quiver in three dimensions, although such an interpretation is not thought to be universally valid \cite{Akhond:2024nyr}. Crucially, magnetic quivers offer a window to the composition of the chiral ring at infinite coupling by using brane systems to directly model the new degrees of freedom emergent in this limit. Recent work on the technicalities of the chiral ring can be found in \cite{Hanany:2025jwo,Hanany:2025ctg}.

Readers familiar with the magnetic quiver literature will no doubt be aware of previous work involving 5d $\mathcal{N}=1$ theories constructed using brane webs. Magnetic quivers for $\surm(N)$ and $\sprm(N)$ SQCD theories have been thoroughly studied in \cite{Bourget:2020gzi}, alongside a selection of theories with more than one gauge group factor. In \cite{Akhond:2022jts,Akhond:2021knl}, the construction was studied for $\spin(N)$ gauge theory with various matter -- infinite-coupling magnetic quivers were found and their chiral rings computed. Interestingly, \cite{Akhond:2022jts,Akhond:2021knl} do not study the magnetic quivers of the finite-coupling limits of these theories -- since the Higgs branches of the electric theories are straightforwardly computable using plethystic techniques and Weyl integration \cite{Feng_2007}, the magnetic quiver program does not offer any further insights into the structure of the five-dimensional theory's moduli space. However, direct construction of these theories' finite-coupling magnetic quivers uncovers a large class of previously unknown unframed 3d $\mathcal{N}=4$ orthosymplectic quivers whose Coulomb and Higgs branches are isolated symplectic singularities. Identifying these new quivers is important for several reasons. Firstly, it significantly enhances the literature's understanding of unframed orthosymplectic theories -- previously the number of such quivers with well-known moduli spaces was limited to a handful of exceptional cases \cite{Sperling_2022}. Secondly, it contributes to the classification of minimal quivers studied in \cite{Bourget:2024asp,Bourget:2025wsp}. Although the approach adopted here does not exhaust all possible minimal isolated singularities, the examples presented here must be constructed by any putative classification of minimal orthosymplectic quivers. Thirdly, these quivers appear to be useful for quiver subtraction procedures, as considered in Section \ref{sec:rank2theories}. One of the basic challenges to defining a full orthosymplectic quiver subtraction algorithm is the fact that the full set of subtraction quivers is unknown - this note expands the toolbox. Finally, at low ranks, accidental isomorphisms lead to infrared dualities between the orthosymplectic magnetic quivers studied here, and their unitary counterparts that are well studied in the literature.

Furthermore, Section \ref{sec:BC_wreathing} presents new unframed orthosymplectic Coulomb branch constructions of next-to-minimal and minimal nilpotent orbit closures of types $B$ and $C$, using wreathing and folding techniques. These quivers are believed to be completely new to the literature.

The organisation of this work is as follows. Section \ref{section:classical_bit} recapitulates on several aspects of $\spin(N)$ gauge theory at finite coupling with spinor matter, alongside various examples of unrefined Hilbert series are given for select theories. Section \ref{section:magnetic_quivers} gives all magnetic quivers at finite coupling (with some infinite-coupling examples not currently in the literature) for rank-1, rank-2 and some rank-3 theories. These examples are restricted to include only theories with a UV fixed point in five dimensions and those theories which undergo complete Higgsing. Section \ref{sec:BC_wreathing} explores wreathings and foldings of sevreral of the magnetic quivers given in Section \ref{section:magnetic_quivers}. Appendix \ref{app:spin7spin8} gives some Higgs branch Hilbert series for $\spin(7)$ and $\spin(8)$ gauge theories, while Appendix \ref{app:CB_hilbert_series} presents the Coulomb branch Hilbert series of the magnetic quivers in Sections \ref{section:magnetic_quivers} and \ref{sec:BC_wreathing}.

The following remarks make a handful of technical points regarding the techniques involved in this work and their relation to others in the wider literature.

\paragraph{Subtractions}
Magnetic quivers for rank-2 theories and higher have Coulomb branches that realise non-minimal singularities. As such, their symplectic stratification is in principle classifiable using a quiver subtraction procedure. Although details of quiver subtraction in unframed orthosymplectic theories are currently unknown, attempts are made where possible to realise the combinatorics of the Hasse diagram as a subtraction procedure on the quivers themselves. To be clear, although the quiver subtraction rules are conjectural, all Hasse diagrams for finite-coupling magnetic quivers given in this paper are exact. Furthermore, several subtraction patterns in this work realise new 3d $\mathcal{N}=4$ theories whose Coulomb and Higgs branches are nilpotent orbit closures/S\l odowy slices, giving further evidence for the validity of the given subtractions.
\paragraph{Hyper-K\"ahler Quotients and Polymerisations}
Many of the $\spin(2)$ and $\spin(3)$ magnetic quivers introduced in this work admit a straightfoward construction in terms of `quiver polymerisations' of the free theories in \Figref{tab:free_magnetic_theories}. In some cases, this is simply a direct use of the polymerisation quivers of \cite{Hanany:2024fqf} adapted via low-rank isomorphisms to be used on orthosymplectic quivers. In other cases, a straightforward extension of the unitary quiver polymerisation algorithm is necessary to complete the construction. The details of this orthosymplectic polymerisation algorithm are not given here -- the interested reader is encouraged to consult \cite{future_polymerisation} for a full specification.
\paragraph{Brane Webs}
This work uses the brane web conventions specified in \cite{Akhond:2020vhc, Akhond:2021ffo, Akhond:2021knl, Akhond:2022jts, Bourget:2020gzi, Cabrera:2018jxt,Zafrir:2015ftn}. Owing to the fact that this paper introduces no new brane web manipulations, technical information regarding these setups is omitted here. The interested reader is encouraged to refer to these references for a full summary.
\section{$\spin(N)$ Gauge Theory: Classical Higgs Branch}
\label{section:classical_bit}
The classical Higgs branch of $\spin(N)$ gauge theory with vector and (co-)spinor matter at finite coupling is readily computable without appeal to magnetic ancillaries. Many of these Higgs branches are well-known nilpotent orbit closures and have been studied extensively \cite{Bourget:2019rtl,Akhond:2022jts,Bourget:2020gzi}. In general, these theories fall into one of five families shown in \Figref{fig:quiv:generalform}, reflecting the representation structure of the $\mathfrak{so}_{2k+1}$ and $\mathfrak{so}_{2k}$ algebras. The dimensions of their various Higgs branches at finite coupling are given below. In the weakly-coupled description, the generators of the chiral ring are mesons, gaugino bilinear, and the four-spinor invariants \cite{Hanany:2025jwo,Hanany:2025ctg}.
\begin{align}
    \textrm{dim}\left[\mathcal{H}\left(\mathcal{Q}_{\ref{quiv:2k+1_1/2_mod4}}\right)\right] =&\; \textrm{dim}\left[\mathcal{H}\left(\mathcal{Q}_{\ref{quiv:2k+1_3/4_mod4}}\right)\right] =\;2^{k}N_s+(N_f-k)(2k+1)\\
     \textrm{dim}\left[\mathcal{H}\left(\mathcal{Q}_{\ref{quiv:2k_2_mod4}}\right)\right] =&\;  \textrm{dim}\left[\mathcal{H}\left(\mathcal{Q}_{\ref{quiv:2k_4_mod4}}\right)\right] =\; 2^{k-1}(N_c+N_s)+k(2N_f-2k+1)\\
     \textrm{dim}\left[\mathcal{H}\left(\mathcal{Q}_{\ref{quiv:2k_3_mod2}}\right)\right] =&\; 2^{k-1}N_s+k\left(2N_f-2k+1\right)
\label{eqn:classical_dimensions}
\end{align}
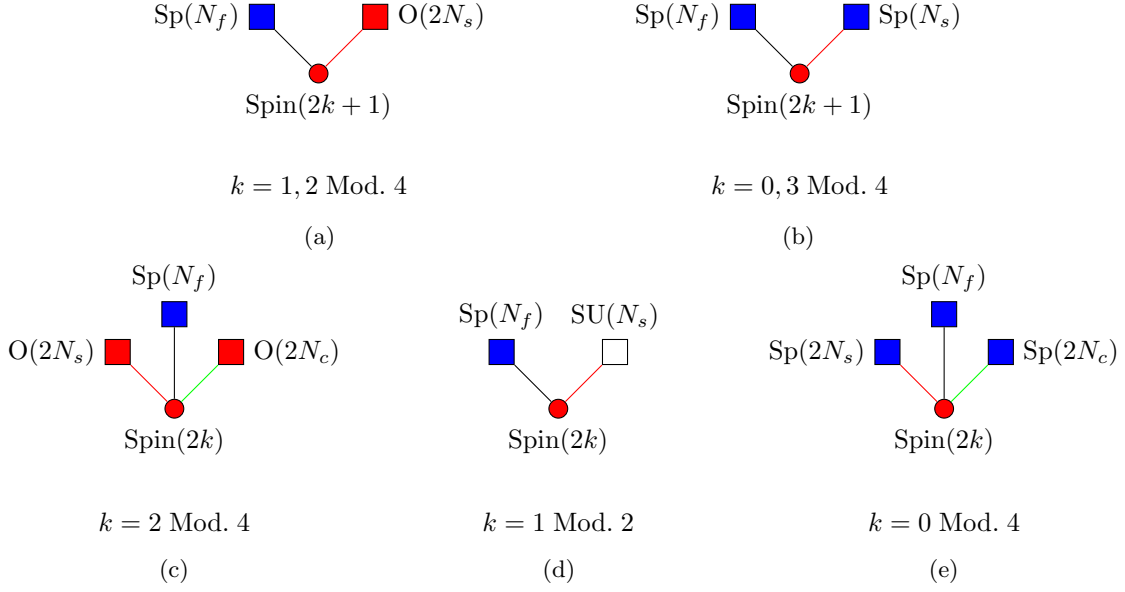
\begin{figure}[h!]
    \centering
    \begin{subfigure}{0.4\textwidth}
    \centering
    \begin{tikzpicture}
    \node[gauger, label=below:$\spin(2k+1)$] (1) at (0,0){};
    \node[flavourr, label=right:$\orm(2N_s)$] (2) at (0.75,0.75){};
    \node[flavourb, label=left:$\sprm(N_f)$] (3) at (-0.75,0.75){};
    \draw[-,red] (1)--(2);
    \draw[-] (3)--(1);
    \node at (0,-1.5) {$k=1,2\;\textrm{Mod.}\;4$};
    \end{tikzpicture}
    \caption{}
    \label{quiv:2k+1_1/2_mod4}
\end{subfigure}
\begin{subfigure}{0.4\textwidth}
    \centering
    \begin{tikzpicture}
    \node[gauger, label=below:$\spin(2k+1)$] (1) at (0,0){};
    \node[flavourb, label=right:$\sprm(N_s)$] (2) at (0.75,0.75){};
    \node[flavourb, label=left:$\sprm(N_f)$] (3) at (-0.75,0.75){};
    \draw[-,red] (1)--(2);
    \draw[-] (3)--(1);
    \node at (0,-1.5) {$k=0,3\;\textrm{Mod.}\;4$};
    \end{tikzpicture}
    \caption{}
    \label{quiv:2k+1_3/4_mod4}
\end{subfigure}
\begin{subfigure}{0.32\textwidth}
    \centering
    \begin{tikzpicture}
    \node[gauger, label=below:$\spin(2k)$] (1) at (0,0){};
    \node[flavourr, label=left:$\orm(2N_s)$] (2) at (-0.75,0.75){};
    \node[flavourr, label=right:$\orm(2N_c)$] (3) at (0.75,0.75){};
    \node[flavourb, label=above:$\sprm(N_f)$] (4) at (0,1.25){};
    \draw[-,red] (1)--(2);
    \draw[-,green] (1)--(3);
    \draw[-] (4)--(1);
    \node at (0,-1.5) {$k=2\;\textrm{Mod.}\;4$};
    \end{tikzpicture}
    \caption{}
    \label{quiv:2k_2_mod4}
\end{subfigure}
\begin{subfigure}{0.32\textwidth}
    \centering
    \begin{tikzpicture}
    \node[gauger, label=below:$\spin(2k)$] (1) at (0,0){};
    \node[flavour, label=above:$\surm(N_s)$] (2) at (0.75,0.75){};
    \node[flavourb, label=above:$\sprm(N_f)$] (3) at (-0.75,0.75){};
    \draw[-,red] (1)--(2);
    \draw[-] (3)--(1);
    \node at (0,-1.5) {$k=1\;\textrm{Mod.}\;2$};
    \end{tikzpicture}
    \caption{}
    \label{quiv:2k_3_mod2}
\end{subfigure}
\begin{subfigure}{0.32\textwidth}
    \centering
    \begin{tikzpicture}
    \node[gauger, label=below:$\spin(2k)$] (1) at (0,0){};
    \node[flavourb, label=left:$\sprm(2N_s)$] (2) at (-0.75,0.75){};
    \node[flavourb, label=right:$\sprm(2N_c)$] (3) at (0.75,0.75){};
    \node[flavourb, label=above:$\sprm(N_f)$] (4) at (0,1.25){};
    \draw[-,red] (1)--(2);
    \draw[-,green] (1)--(3);
    \draw[-] (4)--(1);
    \node at (0,-1.5) {$k=0\;\textrm{Mod.}\;4$};
    \end{tikzpicture}
    \caption{}
    \label{quiv:2k_4_mod4}
\end{subfigure}
\caption{The various $\spin(2k+1)$ SQCDs at finite coupling with $N_f$ vectors and $N_s$ spinors for different values of $k$ are given in the top row, alongside the $\spin(2k)$ gauge theories with $N_f$ vectors, $N_s$ spinors and $N_c$ cospinors in the bottom row. Red and green edges denote matter in the spinor and cospinor representations of the gauge group respectively. For $k=1\;\textrm{Mod.}\;2$ the (co-)spinor of $\spin(2k)$ is complex.}
\label{fig:quiv:generalform}
\end{figure} 
For theories of rank-1, these Higgs branches are each an isolated symplectic singularity corresponding to the moduli space of a single $\surm(M)$ (for $\spin(2)$ gauge theory) or $\sorm(2M)$ (for $\spin(3)$ gauge theory) instanton on $\mathbb{C}^{2}$, otherwise known as the closure of a minimal nilpotent orbit. Among the rank-2 theories, the Higgs branch of $\spin(5)$ SQCD at finite coupling with only spinor matter describes next-to-minimal nilpotent orbits of $\sorm(2M)$, while the $\spin(4)$ SQCDs with $N_s$ spinors and $N_c$ cospinors yield a product of $\sorm(2N_s)$ and $\sorm(2N_c)$ minimal nilpotent orbits. The full Hasse diagram structure of the rank-3 theories' Higgsings are given in \eqref{hd:spin6_hasse_diagram} and \eqref{hd:spin7_hasse_diagram}.
As \eqref{hd:spin6_hasse_diagram} shows, there are six different SQCDs hidden in this figure, all with unitary or special unitary gauge groups and with fundamental matter (of course, this is simply a result of the fact that $\spin(6)\simeq\surm(4)$ and the spinor of the former is the fundamental of the latter). Note that for $N=6$ spinors the theory is a union of two cones owing to the $d_2$ transition at the top of the Hasse diagram. For larger $N$ the theory's Higgs branch is a single cone, while for smaller $N$ there are both mesonic and baryonic cones \cite{Bourget:2019rtl}.
The same analysis can be applied to the Higgs branch of $\spin(7)$ gauge theory. In this case there are ten different SQCDs hidden in this Hasse diagram \eqref{hd:spin7_hasse_diagram}, including that for the exceptional $G_2$ gauge theory with fundamental matter, and $\mathbb{Z}_{2}$ gauge theory with matter rotated under $\sprm(N)$. This complementarity between the Higgs branches of these theories goes some way to explain the remarkable overlap between many of their magnetic quivers.

The Higgs branches of rank-3 theories are of height \footnote{Note that `height' here refers to the number of symplectic leaves less one for a linear Hasse diagram.} three and grow in dimension faster than their lower-rank cousins. Current techniques can only compute the Weyl integration of the Higgs branch of $\spin(7)$ theories exactly up to $N_s=2$, $N_c=2$ before computation time becomes unreasonable, as shown in Table \ref{tab:spin7_bits}. For $\spin(8)$ this becomes even more unfeasible -- the results in Table \ref{tab:spin8_bits} are exact only up to order $t^{6}$. All of these results are unrefined.
\begin{equation}
\raisebox{-0.5\height}{\begin{tikzpicture}
\node (1) [hasse] at (0,0) {};
\node (1a) at (1.8,0) {};
\node (2) [hasse] at (0,1) {};
\node (2a) at (1.8,1) {};
\node (3) [hasse] at (0,2) {};
\node (3a) at (1.8,2) {};
\node (4) [hasse] at (0,3) {};
\draw[-] (1)--(2) node[pos=0.5,midway, left]{$a_{N-1}$} --(3) node[pos=0.5,midway, left]{$a_{N-3}$}--(4) node[pos=0.5,midway, left]{$d_{N-4}$};
\draw[-] (0.5,0)--(0.75,0)--(0.75,3);
\draw[-] (0.5,1)--(0.7,1) (0.8,1)--(0.9,1)--(0.9,3);

\draw[-] (0.5,2)--(0.7,2) (0.95,2)--(1.05,2)--(1.05,3);
\draw[-] (0.5,3)--(0.75,3);

\draw[-] (-1,0)--(-1.25,0)--(-1.25,0.95)--(-1,0.95);
\draw[-] (-1,1.05)--(-1.25,1.05)--(-1.25,1.95)--(-1,1.95);

\draw[-] (-3,0)--(-3.25,0)--(-3.25,1.95)--(-3,1.95);
\draw[-] (0.78,2)--(0.84,2) (0.78,3)--(0.9,3) (0.94,3)--(1.05,3);
\node (a) at (2,0.5) {\scalebox{0.75}{\begin{tikzpicture}
    \node[gauger, label=below:$\spin(6)$] (1a) at (0,0){};
    \node[flavourb, label=below:$\surm(N)$] (2a) at (1.75,0){};
    \draw[-,red] (1a)--(2a);
\end{tikzpicture}}};
\node (b) at (2.15,1.5) {\scalebox{0.75}{\begin{tikzpicture}
    \node[gaugeBig, label=below:$\surm(3)$] (1a) at (0,0){};
    \node[flavour, label=below:$\surm(N-2)$] (2a) at (1.75,0){};
    \draw[-] (1a)--(2a);
\end{tikzpicture}}};
\node (c) at (2.4,2.5) {\scalebox{0.75}{\begin{tikzpicture}
    \node[gaugeBig, label=below:$\surm(2)$] (1a) at (0,0){};
    \node[flavour, label=below:$\sorm(2N-8)$] (2a) at (1.75,0){};
    \draw[-] (1a)--(2a);
\end{tikzpicture}}};
\node (e) at (-2.25,0.5) {\scalebox{0.75}{\begin{tikzpicture}
    \node[gaugeBig, label=below:$\urm(1)$] (1a) at (0,0){};
    \node[flavourb, label=below:$\surm(N)$] (2a) at (-1.25,0){};
    \draw[-] (1a)--(2a);
\end{tikzpicture}}};
\node (f) at (-2.25,1.5) {\scalebox{0.75}{\begin{tikzpicture}
    \node[gaugeBig, label=below:$\urm(1)$] (1a) at (0,0){};
    \node[flavourb, label=below:$\surm(N-2)$] (2a) at (-1.25,0){};
    \draw[-] (1a)--(2a);
\end{tikzpicture}}};
\node (h) at (-3.75,1) {\scalebox{0.75}{\begin{tikzpicture}
    \node[gaugeBig, label=below:$\urm(2)$] (1a) at (0,0){};
    \node[flavour, label=above:$\surm(N)$] (2a) at (0,1){};
    \draw[-] (1a)--(2a);
\end{tikzpicture}}};
\end{tikzpicture}}
\label{hd:spin6_hasse_diagram}
\end{equation}
\begin{equation}
\raisebox{-0.5\height}{\begin{tikzpicture}
\node (1) [hasse] at (0,0) {};
\node (1a) at (1.8,0) {};
\node (2) [hasse] at (0,1) {};
\node (2a) at (1.8,1) {};
\node (3) [hasse] at (0,2) {};
\node (3a) at (1.8,2) {};
\node (4) [hasse] at (0,3) {};
\node (3a) at (1.8,3) {};
\node (5) [hasse] at (0,4) {};
\draw[-] (1)--(2) node[pos=0.5,midway, left]{$c_N$} --(3) node[pos=0.5,midway, left]{$c_{N-1}$}--(4) node[pos=0.5,midway, left]{$a_{2N-5}$} --(5) node[pos=0.5,midway, left]{$d_{2N-6}$};
\draw[-] (0.5,0)--(0.75,0)--(0.75,4);
\draw[-] (0.5,1)--(0.7,1) (0.8,1)--(0.9,1)--(0.9,4);
\draw[-] (0.5,2)--(0.7,2) (0.95,2)--(1.05,2)--(1.05,4);
\draw[-] (0.5,3)--(0.7,3) (1.1,3)--(1.25,3)--(1.25,4);
\draw[-] (0.5,0)--(0.75,0)--(0.75,4);
\draw[-] (-1,0)--(-1.25,0)--(-1.25,0.95)--(-1,0.95);
\draw[-] (-1,1.05)--(-1.25,1.05)--(-1.25,1.95)--(-1,1.95);
\draw[-] (-1,2.05)--(-1.25,2.05)--(-1.25,2.95)--(-1,2.95);
\draw[-] (-3,0)--(-3.25,0)--(-3.25,1.95)--(-3,1.95);
\draw[-] (-4.5,1)--(-4.75,1)--(-4.75,2.95)--(-4.5,2.95);
\draw[-] (-6,0)--(-6.25,0)--(-6.25,2.95)--(-6,2.95);
\draw[-] (1.25,4)--(0.5,4);
\draw[-] (0.78,2)--(0.84,2) (0.78,3)--(0.84,3) (0.94,3)--(1,3);
\node (a) at (2,0.5) {\scalebox{0.75}{\begin{tikzpicture}
    \node[gauger, label=below:$\spin(7)$] (1a) at (0,0){};
    \node[flavourb, label=below:$\sprm(N)$] (2a) at (1.75,0){};
    \draw[-,red] (1a)--(2a);
\end{tikzpicture}}};
\node (b) at (2.15,1.5) {\scalebox{0.75}{\begin{tikzpicture}
    \node[gaugeg, label=below:$G_2$] (1a) at (0,0){};
    \node[flavourb, label=below:$\sprm(N-1)$] (2a) at (1.75,0){};
    \draw[-] (1a)--(2a);
\end{tikzpicture}}};
\node (c) at (2.4,2.5) {\scalebox{0.75}{\begin{tikzpicture}
    \node[gauge, label=below:$\surm(3)$] (1a) at (0,0){};
    \node[flavour, label=below:$\surm(2N-6)$] (2a) at (1.75,0){};
    \draw[-] (1a)--(2a);
\end{tikzpicture}}};
\node (d) at (2.75,3.5) {\scalebox{0.75}{\begin{tikzpicture}
    \node[gauge, label=below:$\surm(2)$] (1a) at (0,0){};
    \node[flavourr, label=below:$\sorm(4N-12)$] (2a) at (1.75,0){};
    \draw[-] (1a)--(2a);
\end{tikzpicture}}};
\node (e) at (-2.25,0.5) {\scalebox{0.75}{\begin{tikzpicture}
    \node[gauge, label=below:$\mathbb{Z}_{2}$] (1a) at (0,0){};
    \node[flavourb, label=below:$\sprm(N)$] (2a) at (-1.25,0){};
    \draw[-] (1a)--(2a);
\end{tikzpicture}}};
\node (f) at (-2.25,1.5) {\scalebox{0.75}{\begin{tikzpicture}
    \node[gauge, label=below:$\mathbb{Z}_{2}$] (1a) at (0,0){};
    \node[flavourb, label=below:$\sprm(N-1)$] (2a) at (-1.25,0){};
    \draw[-] (1a)--(2a);
\end{tikzpicture}}};
\node (g) at (-2.5,2.5) {\scalebox{0.75}{\begin{tikzpicture}
    \node[gauge, label=below:$\urm(1)$] (1a) at (0,0){};
    \node[flavour, label=below:$\surm(2N-4)$] (2a) at (-1.75,0){};
    \draw[-] (1a)--(2a);
\end{tikzpicture}}};
\node (h) at (-3.75,1) {\scalebox{0.75}{\begin{tikzpicture}
    \node[gauger, label=below:$\orm(2)$] (1a) at (0,0){};
    \node[flavourb, label=above:$\sprm(N)$] (2a) at (0,1){};
    \draw[-] (1a)--(2a);
\end{tikzpicture}}};
\node (h) at (-5.5,2) {\scalebox{0.75}{\begin{tikzpicture}
    \node[gauger, label=below:$\sorm(3)$] (1a) at (0,0){};
    \node[flavourb, label=above:$\sprm(N-1)$] (2a) at (0,1){};
    \draw[-] (1a)--(2a);
\end{tikzpicture}}};
\node (h) at (-7,1.5) {\scalebox{0.75}{\begin{tikzpicture}
    \node[gauger, label=below:$\sorm(4)$] (1a) at (0,0){};
    \node[flavourb, label=above:$\sprm(N)$] (2a) at (0,1){};
    \draw[-] (1a)--(2a);
\end{tikzpicture}}};
\end{tikzpicture}}
\label{hd:spin7_hasse_diagram}
\end{equation}
\section{Magnetic Quivers for $\spin(N)$ SQCD with Spinor Matter}
\label{section:magnetic_quivers}
This section recovers various magnetic quivers for $\spin(N)$ gauge theory with spinor matter and uses Hilbert series techniques to explicitly check agreement across the moduli spaces of the two theories. The majority of these magnetic quivers are valid at finite-coupling in the electric theory -- their infinite-coupling counterparts were previously found in \cite{Akhond:2021knl,Akhond:2022jts}. Exceptions include the $\spin(5)$ and $\spin(7)$ theories, whose infinite-coupling magnetic quivers are given for the first time.
\subsection{Free Theories -- $\spin(0)$}
The theories in Table \ref{tab:free_magnetic_theories}, which are magnetic quivers for $\spin(0)$ gauge theory with spinor matter, recur throughout this section. Their Coulomb branches are smooth and evaluate to $\mathbb{H}^{n}$, making them useful tools for constructing theories with singular Coulomb branches using an orthosymplectic form of quiver polymerisation \cite{Hanany:2024fqf}.
\begin{table}[h!]
\centering
\begin{tabular}{|c|c|c|c|c|}
\hline
$N_\textrm{S}$ & Magnetic Quiver & CB & HB & $G_{\mathcal{\mathrm{Outer}}}$\\ \hline
$2$ & $\raisebox{-0.5\height}{\begin{tikzpicture}
    \node[gauger, label=below:$D_1$] (1) at (0,0){};
    \node[gauger, label=below:$D_1$] (2) at (1,0){};
    \draw[-] (1)--(2);
\end{tikzpicture}}$ & $\mathbb{H}^{2}$ & --- & $S_2$ \\ \hline
$4$ & $\raisebox{-0.5\height}{\begin{tikzpicture}
    \node[gauger, label=below:$D_1$] (1) at (0,0){};
    \node[gaugeb, label=below:$C_1$] (2) at (1,0){};
    \node[gauger, label=below:$D_1$] (3) at (2,0){};
    \node[gauger, label=left:$D_1$] (4) at (1,1){};
    \draw[-] (1)--(2)--(3) (2)--(4);
\end{tikzpicture}}$ & $\mathbb{H}^{4}$ & --- & $S_3$\\ \hline
$8$ & $\raisebox{-0.5\height}{\begin{tikzpicture}
    \node[gauger, label=below:$D_1$] (1) at (0,0){};
    \node[gaugeb, label=below:$C_1$] (2) at (1,0){};
    \node[gauger, label=below:$D_2$] (3) at (2,0){};
    \node[gaugeb, label=below:$C_1$] (4) at (3,0){};
    \node[gauger, label=below:$D_1$] (5) at (4,0){};
    \node[gaugeb, label=right:$C_1$] (6) at (2,1){};
    \node[gauger, label=right:$D_1$] (7) at (2,2){};
    \draw[-] (1)--(2)--(3)--(4)--(5) (3)--(6)--(7);
\end{tikzpicture}}$ & $\mathbb{H}^{8}$ & --- & $S_3$\\ \hline
$16$ & $\raisebox{-0.5\height}{\begin{tikzpicture}
    \node[gauger, label=below:$D_1$] (1) at (0,0){};
    \node[gaugeb, label=below:$C_1$] (2) at (1,0){};
    \node[gauger, label=below:$D_2$] (3) at (2,0){};
    \node[gaugeb, label=below:$C_2$] (4) at (3,0){};
    \node[gauger, label=below:$D_3$] (5) at (4,0){};
    \node[gaugeb, label=below:$C_2$] (6) at (5,0){};
    \node[gauger, label=below:$D_2$] (7) at (6,0){};
    \node[gaugeb, label=below:$C_1$] (8) at (7,0){};
    \node[gauger, label=below:$D_1$] (9) at (8,0){};
    \node[gaugeb, label=left:$C_1$] (10) at (4,1){};
    \draw(1)--(2)--(3)--(4)--(5)--(6)--(7)--(8)--(9) (5)--(10);
\end{tikzpicture}}$ & $\mathbb{H}^{16}$ & --- & $S_2$\\ \hline
\end{tabular}
\caption{Various magnetic quivers for $\spin(0)$ gauge theory at finite coupling with $N_{\textrm{S}}$ spinors. These theories construct Coulomb branches of quaternionic dimension 2, 4, 8 and 16 respectively; their Higgs branches are trivial. These theories, which are IR dual to free hypermultiplets, are used throughout the following sections as building blocks for magnetic quivers under quiver polymerisation \cite{Hanany:2024fqf}. The precise details of this procedure will be explained in future work \cite{future_polymerisation}.}
\label{tab:free_magnetic_theories}
\end{table}

\subsection{Rank-1 Theories}
This section considers the magnetic quivers for $\spin(2)$ and $\spin(3)$ gauge theory with spinor matter at finite coupling. The magnetic quivers for the infinite-coupling theories were enumerated in \cite{Akhond:2022jts}. At finite coupling, the moduli spaces of interest are minimal nilpotent orbit closures of the $\mathfrak{sl}_n$ and  $\mathfrak{so}_{2n}$ algebras. Interestingly, the finite-coupling magnetic quivers characteristically undergo symmetry enhancement from the half-integer lattice, which embeds several Lie-group factors from the integer lattice into a single factor of Type-$A$ or Type-$D$ acting on the entire Coulomb branch. Many of these theories are later used in to populate the symplectic stratification of the of the rank-2 theories' Higgs branches. Several of them can be wreathed to produce new quivers whose Coulomb branch is a nilpotent orbit closure of the $\mathfrak{sp}_{n}$ or $\mathfrak{so}_{2n+1}$ algebras, as explained in Section \ref{sec:BC_wreathing}.
\subsubsection{$\spin(2)$}
\label{sec:spin2}
Brane webs for $\spin(2)$ gauge theory support spinor contributions  of the form $2^{n}+2^{m}$, $0\leq m,n\leq5$, where the two factors arise on the left and right independently \cite{Akhond:2022jts}. The upper bound ensures that these theories have a UV fixed point in five dimensions instead of a decompactification limit. Interestingly, this means that the number of new orthosymplectic magnetic quivers in this paper is finite.

The set of magnetic quivers for $\spin(2)$ theories with $N_{\textrm{S}}$ spinors are given in Tables \ref{tab:spin2_magquivs_i} and \ref{tab:spin2_magquivs_ii}. The Coulomb branches of these theories evaluate as closures of minimal nilpotent orbits of $\mathfrak{sl}_{n}$. Interestingly, the Higgs branches of these theories turn out to be Klein singularities of type-$A$, leading to the possibility that these theories are weak-weak dual to the mirrors of $\spin(N)$ gauge theory in three dimensions. Many of these theories have a systematic constrcution in terms of $\spin(2)$ quiver polymerisations of the theories in Table \ref{tab:free_magnetic_theories}. \footnote{The details of this procedure will be given in upcoming work.} Interestingly, the magnetic quiver balance algorithm \cite{Gaiotto:2008ak, Cabrera:2019izd} is not sufficient to read the global symmetries of the theories given here -- although balanced nodes lead to contributions from the integer lattice, the half-integer lattice also generically contributes to the global symmetry. In the Coulomb branch Hilbert series, this looks like a $2^{k}t^{2}$ term, where $k$ is the number of gauge nodes in the quiver with $\mathfrak{so}_{2n}$ gauge algebra.

The magnetic quivers for $5$, $9$ and $17$ spinors appear to have a slightly different construction. These theories arise from adding a single charge-2 hypermultiplet to a $\urm(1)$ gauge node in the theories in Table \ref{tab:free_magnetic_theories}. From the perspective of the Coulomb branch, adding this hypermultiplet transforms the space from a smooth multiplet of $\mathbb{H}^{n}$ to a symplectic singularity.

Tables \ref{tab:spin2_magquivs_i} and \ref{tab:spin2_magquivs_ii} also contain information about the outer automorphisms of these theories, inherited from those of Table \ref{tab:free_magnetic_theories} via the polymerisation process. This is one source of finite-group gaugings that orbifold the Coulomb branch. These procedures will be discussed further in future work.

The integer and half-integer contributions to the Coulomb branch Hilbert series are given in Table \ref{tab:spin2_hs} in Appendix \ref{app:CB_hilbert_series}.
\begin{table}[h]
\centering
\begin{tabular}{|c|c|c|c|c|c|}
\hline
$N_\textrm{S}$ & Magnetic Quiver & CB & HB & $(n,m)$ & $G_{\mathcal{\mathrm{Outer}}}$\\ \hline
4 & $\raisebox{-0.5\height}{\begin{tikzpicture}
    \node[gauge, label=below:$1$] (1) at (0,0){};
    \node[gauger, label=below:$D_1$] (2) at (1,0){};
    \node[gauge, label=below:$1$] (3) at (2,0){};
    \draw(1)--(2)--(3);
\end{tikzpicture}}$ & $a_3$ & $A_3$ & $(2,2)$ & $S_2$ \\ \hline
5 & $\raisebox{-0.5\height}{\begin{tikzpicture}
    \node[gaugeBig, label=below:$1$] (1) at (0,0){};
    \node[gaugeb, label=below:$C_1$] (2) at (1,0){};
    \node[gauger, label=below:$D_1$] (3) at (2,-0.5){};
    \node[gauger, label=below:$D_1$] (4) at (2,0.5){};
    \node[flavour, label=below:$\urm(1)$] (5) at (-1,0){};
    \node at (-0.5,0.25) {$\mathrm{S}^2$};
    \draw(5)--(1)--(2)--(3) (2)--(4);
    
\end{tikzpicture}}$& $a_4$ & $A_4$ & -- & $S_2$ \\ \hline
6 & $\raisebox{-0.5\height}{\begin{tikzpicture}
    \node[gaugeBig, label=below:$1$] (1) at (0,0){};
    \node[gauger, label=below:$D_1$] (2) at (1,0){};
    \node[gaugeb, label=below:$C_1$] (3) at (2,0){};
    \node[gauger, label=below:$D_1$] (4) at (3,-0.5){};
    \node[gauger, label=below:$D_1$] (5) at (3,0.5){};
    \draw(1)--(2)--(3)--(4) (3)--(5);
\end{tikzpicture}}$ & $a_5$ & $A_5$ & $(2,4)$ & $S_2$\\ \hline
8 & $\raisebox{-0.5\height}{\begin{tikzpicture}
    \node[gauger, label=below:$D_1$] (1) at (0,-0.5){};
    \node[gaugeb, label=below:$C_1$] (2) at (1,0){};
    \node[gauger, label=below:$D_1$] (3) at (2,0){};
    \node[gaugeb, label=below:$C_1$] (4) at (3,0){};
    \node[gauger, label=below:$D_1$] (5) at (4,-0.5){};
    \node[gauger, label=below:$D_1$] (6) at (0,0.5){};
    \node[gauger, label=below:$D_1$] (7) at (4,0.5){};
    \draw(1)--(2)--(3)--(4)--(5) (2)--(6) (4)--(7);
\end{tikzpicture}}$ & $a_7$ & $A_7$ & $(4,4)$ & $S_2\times S_2$\\ \hline
9 & $\raisebox{-0.5\height}{\begin{tikzpicture}
    \node[gauge, label=below:$1$] (1) at (0,0){};
    \node[gaugeb, label=below:$C_1$] (2) at (1,0){};
    \node[gauger, label=below:$D_2$] (3) at (2,0){};
    \node[gaugeb, label=below:$C_1$] (4) at (3,-0.5){};
    \node[gauger, label=below:$D_1$] (5) at (4,-0.5){};
    \node[flavour, label=below:$\urm(1)$] (6) at (-1,0){};
    \node[gaugeb, label=above:$C_1$] (7) at (3,0.5){};
    \node[gauger, label=above:$D_1$] (8) at (4,0.5){};
    \draw(6)--(1)--(2)--(3)--(4)--(5) (3)--(7)--(8);
    \node at (-0.5,0.25) {$\mathrm{S}^2$};
\end{tikzpicture}}$ & $a_8$ & $A_8$ & -- & $S_2$\\ \hline
10 & $\raisebox{-0.5\height}{\begin{tikzpicture}
    \node[gauger, label=below:$D_1$] (1) at (0,0){};
    \node[gaugeb, label=below:$C_1$] (2) at (1,0){};
    \node[gauger, label=below:$D_2$] (3) at (2,0){};
    \node[gaugeb, label=below:$C_1$] (4) at (3,-0.5){};
    \node[gauger, label=below:$D_1$] (5) at (4,-0.5){};
    \node[gaugeBig, label=below:$1$] (6) at (-1,0){};
    \node[gaugeb, label=below:$C_1$] (7) at (3,0.5){};
    \node[gauger, label=below:$D_1$] (8) at (4,0.5){};
    \draw(1)--(2)--(3)--(4)--(5) (1)--(6) (3)--(7)--(8);
\end{tikzpicture}}$& $a_{9}$ & $A_{9}$ & $(2,8)$ & $S_2$\\ \hline
12 & $\raisebox{-0.5\height}{\begin{tikzpicture}
    \node[gauger, label=below:$D_1$] (1) at (0,-0.5){};
    \node[gaugeb, label=below:$C_1$] (2) at (1,0){};
    \node[gauger, label=below:$D_1$] (3) at (2,0){};
    \node[gaugeb, label=below:$C_1$] (4) at (3,0){};
    \node[gauger, label=below:$D_2$] (5) at (4,0){};
    \node[gaugeb, label=below:$C_1$] (6) at (5,-0.5){};
    \node[gauger, label=below:$D_1$] (7) at (6,-0.5){};
    \node[gauger, label=below:$D_1$] (8) at (0,0.5){};
    \node[gaugeb, label=below:$C_1$] (9) at (5,0.5){};
    \node[gauger, label=below:$D_1$] (10) at (6,0.5){};
    \draw(1)--(2)--(3)--(4)--(5)--(6)--(7) (2)--(8) (5)--(9)--(10);
\end{tikzpicture}}$& $a_{11}$ & $A_{11}$ & $(4,8)$ & $S_2\times S_2$\\ \hline
16 & $\raisebox{-0.5\height}{\begin{tikzpicture}
    \node[gauger, label=below:$D_1$] (1) at (0,-0.5){};
    \node[gaugeb, label=below:$C_1$] (2) at (1,-0.5){};
    \node[gauger, label=below:$D_2$] (3) at (2,0){};
    \node[gaugeb, label=below:$C_1$] (4) at (3,0){};
    \node[gauger, label=below:$D_1$] (5) at (4,0){};
    \node[gaugeb, label=below:$C_1$] (6) at (5,0){};
    \node[gauger, label=below:$D_2$] (7) at (6,0){};
    \node[gaugeb, label=below:$C_1$] (8) at (7,-0.5){};
    \node[gauger, label=below:$D_1$] (9) at (8,-0.5){};
    \node[gaugeb, label=below:$C_1$] (10) at (1,0.5){};
    \node[gauger, label=below:$D_1$] (11) at (0,0.5){};
    \node[gaugeb, label=below:$C_1$] (12) at (7,0.5){};
    \node[gauger, label=below:$D_1$] (13) at (8,0.5){};
    \draw(1)--(2)--(3)--(4)--(5)--(6)--(7)--(8)--(9)(3)--(10)--(11) (7)--(12)--(13);
\end{tikzpicture}}$& $a_{15}$ & $A_{15}$ & $(8,8)$ & $S_2\times S_2$\\ \hline
\end{tabular}
\caption{Magnetic quivers for $\spin(2)$ gauge theory with $N_{\mathrm{S}}$ spinors, $N_{\mathrm{S}} =4,\cdots,12$. The Coulomb branches of these theories are minimal nilpotent orbit closure of $\mathfrak{sl}_{n}$, in agreement with the Higgs branch of the electric theory. Direct computation also confirms that the Higgs branches of these theories are Klein singularities of type-$A$. The fifth colunmn of this table gives the polymerisation construction of these theories in terms of the theories of Table \ref{tab:free_magnetic_theories}. $G_{\mathrm{Outer}}$ records the outer automorphism symmetry of the magnetic quiver. Note that the $\mathrm{S}^2$ edge denotes a charge 2 hypermultiplet.}
\label{tab:spin2_magquivs_i}
\end{table}
\begin{table}[h!]
\centering
\begin{tabular}{|c|c|c|c|c|c|}
\hline
$N_\textrm{S}$ & Magnetic Quiver & CB & HB & $(n,m)$ & $G_{\mathcal{\mathrm{Outer}}}$\\ \hline
17 & $\raisebox{-0.5\height}{\begin{tikzpicture}
    \node[gaugeBig, label=below:$1$] (1) at (4,0.5){};
    \node[gaugeb, label=below:$C_1$] (2) at (3,0.5){};
    \node[gauger, label=below:$D_2$] (3) at (2,0.5){};
    \node[gaugeb, label=below:$C_2$] (4) at (1,0.5){};
    \node[gauger, label=below:$D_3$] (5) at (0,0){};
    \node[gaugeb, label=below:$C_2$] (6) at (1,-0.5){};
    \node[gauger, label=below:$D_2$] (7) at (2,-0.5){};
    \node[gaugeb, label=below:$C_1$] (8) at (3,-0.5){};
    \node[gauger, label=below:$D_1$] (9) at (4,-0.5){};
    \node[gaugeb, label=below:$C_1$] (10) at (-1,0){};
    \node[flavour, label=right:$\urm(1)$] (11) at (5,0.5){};
    \draw(11)--(1)--(2)--(3)--(4)--(5) (5)--(6)--(7)--(8)--(9) (5)--(10);
    \node at (4.5,0.25) {$\mathrm{S}^2$};
\end{tikzpicture}}$ & $a_{16}$ & $A_{16}$ & -- & $1$\\ \hline
18 & $\raisebox{-0.5\height}{\begin{tikzpicture}
    \node[gaugeBig, label=below:$1$] (11) at (5,0.5){};
    \node[gauger, label=below:$D_1$] (1) at (4,0.5){};
    \node[gaugeb, label=below:$C_1$] (2) at (3,0.5){};
    \node[gauger, label=below:$D_2$] (3) at (2,0.5){};
    \node[gaugeb, label=below:$C_2$] (4) at (1,0.5){};
    \node[gauger, label=below:$D_3$] (5) at (0,0){};
    \node[gaugeb, label=below:$C_2$] (6) at (1,-0.5){};
    \node[gauger, label=below:$D_2$] (7) at (2,-0.5){};
    \node[gaugeb, label=below:$C_1$] (8) at (3,-0.5){};
    \node[gauger, label=below:$D_1$] (9) at (4,-0.5){};
    \node[gaugeb, label=below:$C_1$] (10) at (-1,0){};
    \draw (11)--(1)--(2)--(3)--(4)--(5)--(6)--(7)--(8)--(9) (5)--(10);
\end{tikzpicture}}$ & $a_{17}$ & $A_{17}$ & $(2,16)$ & $1$\\ \hline
20 & $\raisebox{-0.5\height}{\begin{tikzpicture}
    \node[gauger, label=below:$D_1$] (13) at (6,1){};
    \node[gauger, label=below:$D_1$] (12) at (6,0){};
    \node[gaugeb, label=below:$C_1$] (11) at (5,0.5){};
    \node[gauger, label=below:$D_1$] (1) at (4,0.5){};
    \node[gaugeb, label=below:$C_1$] (2) at (3,0.5){};
    \node[gauger, label=below:$D_2$] (3) at (2,0.5){};
    \node[gaugeb, label=below:$C_2$] (4) at (1,0.5){};
    \node[gauger, label=below:$D_3$] (5) at (0,0){};
    \node[gaugeb, label=below:$C_2$] (6) at (1,-0.5){};
    \node[gauger, label=below:$D_2$] (7) at (2,-0.5){};
    \node[gaugeb, label=below:$C_1$] (8) at (3,-0.5){};
    \node[gauger, label=below:$D_1$] (9) at (4,-0.5){};
    \node[gaugeb, label=below:$C_1$] (10) at (-1,0){};
    \draw (11)--(1)--(2)--(3)--(4)--(5)--(6)--(7)--(8)--(9) (5)--(10) (13)--(11)--(12);
\end{tikzpicture}}$ & $a_{19}$ & $A_{19}$ & $(4,16)$ & $S_2$\\ \hline
24 & $\raisebox{-0.5\height}{\begin{tikzpicture}
    \node[gauger, label=below:$D_1$] (1) at (8,1){};
    \node[gaugeb, label=below:$C_1$] (2) at (7,1){};
    \node[gauger, label=below:$D_2$] (3) at (6,0.5){};
    \node[gaugeb, label=below:$C_1$] (4) at (5,0.5){};
    \node[gauger, label=below:$D_1$] (5) at (4,0.5){};
    \node[gaugeb, label=below:$C_1$] (6) at (3,0.5){};
    \node[gauger, label=below:$D_2$] (7) at (2,0.5){};
    \node[gaugeb, label=below:$C_2$] (8) at (1,0.5){};
    \node[gauger, label=below:$D_3$] (9) at (0,0){};
    \node[gaugeb, label=below:$C_2$] (10) at (1,-0.5){};
    \node[gauger, label=below:$D_2$] (11) at (2,-0.5){};
    \node[gaugeb, label=below:$C_1$] (12) at (3,-0.5){};
    \node[gauger, label=below:$D_1$] (13) at (4,-0.5){};
    \node[gaugeb, label=below:$C_1$] (14) at (7,0){};
    \node[gauger, label=below:$D_1$] (15) at (8,0){};
    \node[gaugeb, label=below:$C_1$] (16) at (-1,0){};
    \draw (1)--(2)--(3)--(4)--(5)--(6)--(7)--(8)--(9)--(10)--(11)--(12)--(13) (3)--(14)--(15) (9)--(16);
    \end{tikzpicture}}$& $a_{23}$ & $A_{23}$ & $(8,16)$ & $S_2$\\ \hline
32 & $\raisebox{-0.5\height}{\begin{tikzpicture}
    \node[gauger, label=below:$D_1$] (1) at (0,0){};
    \node[gaugeb, label=below:$C_1$] (2) at (1,-0.75){};
    \node[gauger, label=below:$D_2$] (3) at (2,-0.75){};
    \node[gaugeb, label=below:$C_2$] (4) at (3,-0.75){};
    \node[gauger, label=above:$D_3$] (5) at (4,-0.75){};
    \node[gaugeb, label=below:$C_2$] (6) at (5,-0.75){};
    \node[gauger, label=below:$D_2$] (7) at (6,-0.75){};
    \node[gaugeb, label=below:$C_1$] (8) at (7,-0.75){};
    \node[gauger, label=below:$D_1$] (9) at (8,-0.75){};
    \node[gaugeb, label=left:$C_1$] (10) at (4,-1.5){};
    \node[gaugeb, label=below:$C_1$] (11) at (1,0.75){};
    \node[gauger, label=below:$D_2$] (12) at (2,0.75){};
    \node[gaugeb, label=below:$C_2$] (13) at (3,0.75){};
    \node[gauger, label=below:$D_3$] (14) at (4,0.75){};
    \node[gaugeb, label=below:$C_2$] (15) at (5,0.75){};
    \node[gauger, label=below:$D_2$] (16) at (6,0.75){};
    \node[gaugeb, label=below:$C_1$] (17) at (7,0.75){};
    \node[gauger, label=below:$D_1$] (18) at (8,0.75){};
    \node[gaugeb, label=left:$C_1$] (19) at (4,1.5){};
    \draw[-] (1)--(2)--(3)--(4)--(5)--(6)--(7)--(8)--(9) (5)--(10);
    \draw[-] (1)--(11)--(12)--(13)--(14)--(15)--(16)--(17)--(18) (14)--(19);
    \end{tikzpicture}}$& $a_{31}$ & $A_{31}$ & $(16,16)$ & $S_2$\\ \hline
\end{tabular}
\caption{Magnetic quivers for $\spin(2)$ gauge theory with $N_{\mathrm{S}}$ spinors, $N_{\mathrm{S}}$ as specified in the first column. The Coulomb branches of these theories are minimal nilpotent orbit closure of $\mathfrak{sl}_{n}$, in agreement with the Higgs branch of the electric theory. Direct computation also confirms that the Higgs branches of these theories are Klein singularities of type-$A$. The fifth colunmn of this table gives the polymerisation construction of these theories in terms of the theories of Table \ref{tab:free_magnetic_theories}. $G_{\mathrm{Outer}}$ records the outer automorphism symmetry of the magnetic quiver. Note that the $\mathrm{S}^{2}$ denotes a charge 2 hypermultiplet.}
\label{tab:spin2_magquivs_ii}
\end{table}
\subsubsection{$\spin(3)$}
\label{sec:spin3}
The magnetic quivers for $\spin(3)$ gauge theory with $N_{\mathrm{S}}$ spinors are given in Table \ref{tab:spin3_magquivs}. Like the Higgs branch of the electric theory, their Coulomb branches are closures of minimal nilpotent orbits of $\mathfrak{so}_{2n}$. Interestingly, their Higgs branches are Klein singularities of type $D_{n}$, as expected from Barbasch-Vogan duality \cite{barbasch_vogan}. As in the $\spin(2)$ case, many of these theories admit a construction in terms of the polymerisation of the theories in Table \ref{tab:free_magnetic_theories}; the specification for the polymerisation quivers is given in the fifth column of Table \ref{tab:spin3_magquivs}. As before, many of these theories inherit outer automorphisms from those of Table \ref{tab:free_magnetic_theories}. Note that the $N_{\mathrm{S}}=16$ case leads to a magnetic quiver for the moduli space of a single $\sorm(32)$ instanton -- this differs from the construction in Equation (50) of \cite{Bennett:2026xpm}. The magnetic quivers for $N_{\mathrm{S}}=6$ and $N_{\mathrm{S}}=10$ admit a wreathing that recovers a new quiver whose Coulomb branch constructs the closure of the next-to-minimal nilpotent orbits of $\sorm(11)$ and $\sorm(19)$ respectively. The integer and half-integer contributions to the Coulomb branch Hilbert series are given in Table \ref{tab:spin3_hs} in Appendix \ref{app:CB_hilbert_series}.
\begin{table}[h!]
\centering
\begin{tabular}{|c|c|c|c|c|c|}
\hline
$N_\textrm{S}$ & Magnetic Quiver & CB & HB & $(n,m)$ & $G_{\mathcal{\mathrm{Outer}}}$\\ \hline
5 & $\raisebox{-0.5\height}{\begin{tikzpicture}
    \node[gauger, label=below:$D_1$] (1) at (2,0.5){};
    \node[gaugeb, label=below:$C_1$] (2) at (1,0.5){};
    \node[gauger, label=below:$D_2$] (3) at (0,0){};
    \node[gaugeb, label=below:$C_1$] (4) at (1,-0.5){};
    \node[gauger, label=below:$D_1$] (5) at (2,-0.5){};
    \node[gauge, label=below:$1$] (6) at (-1,0){};
    \draw[-] (1)--(2)--(3)--(4)--(5) (3)--(6);
\end{tikzpicture}}$ & $d_5$ & $D_5$ & $(2,8)$ & $S_2$\\ \hline
6 & $\raisebox{-0.5\height}{\begin{tikzpicture}
    \node[gauger, label=below:$D_1$] (1) at (2,-0.5){};
    \node[gaugeb, label=below:$C_1$] (2) at (1,-0.5){};
    \node[gauger, label=below:$D_2$] (3) at (0,0){};
    \node[gaugeb, label=below:$C_1$] (4) at (1,0.5){};
    \node[gauger, label=below:$D_1$] (5) at (2,0.5){};
    \node[gaugeb, label=below:$C_1$] (6) at (-1,0){};
    \node[gauger, label=below:$D_1$] (7) at (-2,0.5){};
    \node[gauger, label=below:$D_1$] (8) at (-2,-0.5){};
    \draw[-] (1)--(2)--(3)--(4)--(5) (3)--(6)--(7) (6)--(8);
\end{tikzpicture}}$& $d_6$ & $D_6$ & $(4,8)$ & $S_2\times S_2$\\ \hline
8 & $\raisebox{-0.5\height}{\begin{tikzpicture}
    \node[gauger, label=below:$D_1$] (1) at (0,-0.5){};
    \node[gaugeb, label=below:$C_1$] (2) at (1,-0.5){};
    \node[gauger, label=below:$D_2$] (3) at (2,0){};
    \node[gaugeb, label=below:$C_1$] (4) at (3,0){};
    \node[gauger, label=below:$D_2$] (5) at (4,0){};
    \node[gaugeb, label=below:$C_1$] (6) at (5,-0.5){};
    \node[gauger, label=below:$D_1$] (7) at (6,-0.5){};
    \node[gaugeb, label=below:$C_1$] (8) at (1,0.5){};
    \node[gauger, label=below:$D_1$] (9) at (0,0.5){};
    \node[gaugeb, label=below:$C_1$] (10) at (5,0.5){};
    \node[gauger, label=below:$D_1$] (11) at (6,0.5){};
    \draw[-] (1)--(2)--(3)--(4)--(5)--(6)--(7) (3)--(8)--(9) (5)--(10)--(11);
\end{tikzpicture}}$ & $d_8$ & $D_8$ & $(8,8)$ & $S_2\times S_2$\\ \hline
9 & $\raisebox{-0.5\height}{\begin{tikzpicture}
    \node[gauger, label=below:$D_1$] (1) at (5,-0.5){};
    \node[gaugeb, label=below:$C_1$] (2) at (4,-0.5){};
    \node[gauger, label=below:$D_2$] (3) at (3,-0.5){};
    \node[gaugeb, label=below:$C_2$] (4) at (2,-0.5){};
    \node[gauger, label=below:$D_3$] (5) at (1,0){};
    \node[gaugeb, label=below:$C_2$] (6) at (2,0.5){};
    \node[gauger, label=below:$D_2$] (7) at (3,0.5){};
    \node[gauge, label=below:$1$] (8) at (4,0.5){};
    \node[gaugeb, label=below:$C_1$] (10) at (0,0){};
    \draw[-] (1)--(2)--(3)--(4)--(5)--(6)--(7)--(8) (5)--(10);
\end{tikzpicture}}$ & $d_9$ & $D_9$ & $(2,16)$ & $1$\\ \hline
10 & $\raisebox{-0.5\height}{\begin{tikzpicture}
    \node[gauger, label=below:$D_1$] (1) at (4,1.25){};
    \node[gauger, label=below:$D_1$] (1a) at (4,0.25){};
    \node[gaugeb, label=below:$C_1$] (2) at (3,0.75){};
    \node[gauger, label=below:$D_2$] (3) at (2,0.75){};
    \node[gaugeb, label=below:$C_2$] (4) at (1,0.75){};
    \node[gauger, label=below:$D_3$] (5) at (0,0){};
    \node[gaugeb, label=below:$C_2$] (6) at (1,-0.75){};
    \node[gauger, label=below:$D_2$] (7) at (2,-0.75){};
    \node[gaugeb, label=below:$C_1$] (8) at (3,-0.75){};
    \node[gauger, label=below:$D_1$] (9) at (4,-0.75){};
    \node[gaugeb, label=below:$C_1$] (10) at (-1,0){};
    \draw(1a)--(2)--(3)--(4)--(5)--(6)--(7)--(8)--(9) (5)--(10) (2)--(1);
\end{tikzpicture}}$& $d_{10}$ & $D_{10}$ & $(4,16)$ & $S_2$\\ \hline
12 & $\raisebox{-0.5\height}{\begin{tikzpicture}
    \node[gauger, label=below:$D_1$] (0b) at (6,-0.25){};
    \node[gaugeb, label=below:$C_1$] (0a) at (5,-0.25){};
    \node[gauger, label=below:$D_1$] (0) at (6,-1.25){};
    \node[gaugeb, label=below:$C_1$] (1) at (5,-1.25){};
    \node[gauger, label=below:$D_2$] (1a) at (4,-0.75){};
    \node[gaugeb, label=below:$C_1$] (2) at (3,-0.75){};
    \node[gauger, label=below:$D_2$] (3) at (2,-0.75){};
    \node[gaugeb, label=below:$C_2$] (4) at (1,-0.75){};
    \node[gauger, label=below:$D_3$] (5) at (0,0){};
    \node[gaugeb, label=below:$C_2$] (6) at (1,0.75){};
    \node[gauger, label=below:$D_2$] (7) at (2,0.75){};
    \node[gaugeb, label=below:$C_1$] (8) at (3,0.75){};
    \node[gauger, label=below:$D_1$] (9) at (4,0.75){};
    \node[gaugeb, label=below:$C_1$] (10) at (-1,0){};
    \draw (1a)--(2)--(3)--(4)--(5)--(6)--(7)--(8)--(9) (5)--(10) (0b)--(0a)--(1a)--(1)--(0);
\end{tikzpicture}}$& $d_{12}$ & $D_{12}$ & $(8,16)$ & $S_2$\\ \hline
16 & $\raisebox{-0.5\height}{\begin{tikzpicture}
    \node[gaugeb, label=below:$C_1$] (2) at (1,0){};
    \node[gauger, label=below:$D_2$] (3) at (2,-0.75){};
    \node[gaugeb, label=below:$C_2$] (4) at (3,-0.75){};
    \node[gauger, label=above:$D_3$] (5) at (4,-0.75){};
    \node[gaugeb, label=below:$C_2$] (6) at (5,-0.75){};
    \node[gauger, label=below:$D_2$] (7) at (6,-0.75){};
    \node[gaugeb, label=below:$C_1$] (8) at (7,-0.75){};
    \node[gauger, label=below:$D_1$] (9) at (8,-0.75){};
    \node[gaugeb, label=left:$C_1$] (10) at (4,-1.5){};
    \node[gauger, label=below:$D_2$] (12) at (2,0.75){};
    \node[gaugeb, label=below:$C_2$] (13) at (3,0.75){};
    \node[gauger, label=below:$D_3$] (14) at (4,0.75){};
    \node[gaugeb, label=below:$C_2$] (15) at (5,0.75){};
    \node[gauger, label=below:$D_2$] (16) at (6,0.75){};
    \node[gaugeb, label=below:$C_1$] (17) at (7,0.75){};
    \node[gauger, label=below:$D_1$] (18) at (8,0.75){};
    \node[gaugeb, label=left:$C_1$] (19) at (4,1.5){};
    \draw[-] (2)--(3)--(4)--(5)--(6)--(7)--(8)--(9) (5)--(10);
    \draw[-] (2)--(12)--(13)--(14)--(15)--(16)--(17)--(18) (14)--(19);
\end{tikzpicture}}$& $d_{16}$ & $D_{16}$ & $(16,16)$ & $S_2$\\ \hline
\end{tabular}
\caption{Magnetic quivers for $\spin(3)$ gauge theory with $N_{\mathrm{S}}$ spinors at finite coupling. The Coulomb branches of these theories are the closure of the minimal nilpotent orbits of $\mathfrak{so}_{2n}$ -- the Higgs branches are Klein singularities of Type-$D$. The label $(n,m)$ refers to the theories in Table \ref{tab:free_magnetic_theories} that construct these magnetic quivers via a $\spin(3)$ polymerisation. Note that $G_{\mathrm{Outer}}$ records the outer automorphism symmetries inherited by these theories from those of Table \ref{tab:free_magnetic_theories}. Only two of these theories are strictly star-shaped, those for five and nine spinors, and all undergo a characteristic symmetry enhancement from the half-integer lattice.}
\label{tab:spin3_magquivs}
\end{table}
\subsection{Rank-2 Theories}
\label{sec:rank2theories}
\subsubsection{$\spin(4)$}
\label{sec:spin4}
Magnetic quivers for $\spin(4)$ gauge theory with $(N_{\mathrm{S}}, N_{\mathrm{C}})$ spinors and cospinors are given in Table \ref{tab:spin4_magquivs}. These theories differ from those in Sections \ref{sec:spin2} and \ref{sec:spin3} in that now their Coulomb branches construct a product of two minimal nilpotent orbit closures, each of $D$-type (modulo $A_3\simeq D_3$). Interestingly, their Higgs branches also factorise into a product of two $D$-type Klein singularities, motivating the interpretation of these theories as 3d-mirrors of the electric quiver in three dimensions. This is despite the fact that the quivers themselves are connected -- for such theories, the conclusion that their moduli space is a product of two smaller spaces is surprising. That these theories construct product spaces is an immediate consequence of the fact that the electric theory with $\spin(4)\simeq\surm(2)\times\surm(2)$ gauge group factorsies into two independent SQCDs. For this the spinor matter is crucial -- recall that the spinor and cospinor representations of $\spin(4)$ are simply the fundamental representations of each factor of $\surm(2)$. In the absence of vector $\spin(4)$ matter (which becomes a bifundamental under $\surm(2)\times\surm(2)$) these two gauge groups are decoupled from each other. Hence, the moduli space reduces to a simple product. The fifth column of Table \ref{tab:spin4_magquivs} records the theories in Table \ref{tab:free_magnetic_theories} necessary to construct these theories as $\spin(4)$ polymerisations.

The integer and half-integer contributions to the Coulomb branch Hilbert series are given in Table \ref{tab:spin4_hs} in Appendix \ref{app:CB_hilbert_series}. The $(2,3)$ example consists of two identical cones on its Higgs branch as it Higgses to $\sprm(1)$ with two flavours. These two cones are both given by the same magnetic quiver.
\begin{table}[h!]
\centering
\begin{tabular}{|c|c|c|c|c|c|}
\hline
$(N_\textrm{S},\;N_\textrm{C})$ & Magnetic Quiver & CB & HB & $(n,m)$ & $G_{\mathcal{\mathrm{Outer}}}$\\ \hline
$(2,3)$ & $\raisebox{-0.5\height}{\begin{tikzpicture}
    \node[gauger, label=below:$D_1$] (1) at (0,0){};
    \node[gaugeb, label=below:$C_1$] (2) at (1,0){};
    \node[gaugeBig, label=below:$1$] (3) at (2,0){};
    \node[gauger, label=below:$D_1$] (4) at (3,0){};
    \draw[transform canvas={yshift=1.3pt}](2)--(3);
    \draw[transform canvas={yshift=-1.3pt}](2)--(3);
    \draw[-] (1)--(2) (3)--(4);
\end{tikzpicture}}$ & $a_1 \times a_3$ & $A_1 \times A_3$ & & $1$ \\ \hline
$(3,3)$ & $\raisebox{-0.5\height}{\begin{tikzpicture}
    \node[gauger, label=below:$D_1$] (1) at (0,0){};
    \node[gaugeb, label=below:$C_1$] (2) at (1,0){};
    \node[gauger, label=below:$D_1$] (3) at (2,0){};
    \node[gaugeb, label=below:$C_1$] (4) at (3,0){};
    \node[gauger, label=below:$D_1$] (5) at (4,0){};
    \node[gauger, label=above:$D_1$] (6) at (2,1){};
    \draw[-] (1)--(2)--(3)--(4)--(5) (2)--(6)--(4);
\end{tikzpicture}}$ & $a_3 \times a_3$ & $A_3 \times A_3$ & $(4,8)$ & $S_2$\\ \hline
$(4,4)$ & $\raisebox{-0.5\height}{\begin{tikzpicture}
    \node[gauger, label=below:$D_1$] (1) at (0,-0.5){};
    \node[gaugeb, label=below:$C_1$] (2) at (1,-0.5){};
    \node[gauger, label=below:$D_2$] (3) at (2,0){};
    \node[gaugeb, label=above:$C_1$] (4) at (3,0.5){};
    \node[gauger, label=above:$D_1$] (5) at (4,0.5){};
    \node[gaugeb, label=below:$C_1$] (6) at (3,-0.5){};
    \node[gauger, label=below:$D_1$] (7) at (4,-0.5){};
    \node[gaugeb, label=above:$C_1$] (8) at (1,0.5){};
    \node[gauger, label=above:$D_1$] (9) at (0,0.5){};
    \draw[-] (1)--(2)--(3)--(4)--(5) (7)--(6)--(3)--(8)--(9);
\end{tikzpicture}}$ & $d_4 \times d_4$ & $D_4 \times D_4$ & $(8,8)$ & $S_2 \times S_2$ \\ \hline
$(5,4)$ & $\raisebox{-0.5\height}{\begin{tikzpicture}
    \node[gauger, label=below:$D_1$] (1) at (0,0){};
    \node[gaugeb, label=below:$C_1$] (2) at (1,0){};
    \node[gauger, label=below:$D_2$] (3) at (2,0){};
    \node[gaugeb, label=below:$C_2$] (4) at (3,0){};
    \node[gauger, label=below:$D_3$] (5) at (4,0){};
    \node[gaugeb, label=right:$C_1$] (6) at (5,0.75){};
    \node[gaugeb, label=right:$C_1$] (7) at (5,-0.75){};
    \node[gauge, label=above:$1$] (8) at (4,1){};
    \draw[-] (1)--(2)--(3)--(4)--(5)--(6) (5)--(7) (5)--(8);
\end{tikzpicture}}$ & $d_4 \times d_5$ & $D_4 \times D_5$ & $(2,16)$ & $S_2$ \\ \hline
$(5,5)$ & $\raisebox{-0.5\height}{\begin{tikzpicture}
    \node[gauger, label=below:$D_1$] (1) at (0,0){};
    \node[gaugeb, label=below:$C_2$] (2) at (1,0){};
    \node[gauger, label=below:$D_3$] (3) at (2,0){};
    \node[gaugeb, label=below:$C_2$] (4) at (3,0){};
    \node[gauger, label=below:$D_2$] (5) at (4,0){};
    \node[gaugeb, label=below:$C_1$] (6) at (5,0){};
    \node[gauger, label=below:$D_1$] (7) at (6,0){};
    \node[gauger, label=above:$D_1$] (8) at (1,1){};
    \node[gaugeb, label=above:$C_1$] (9) at (2,1){};
    \draw[-] (1)--(2)--(3)--(4)--(5)--(6)--(7) (2)--(8) (3)--(9);
\end{tikzpicture}}$ & $d_5 \times d_5$ & $D_5 \times D_5$ & $(4,16)$ & $1$ \\ \hline
$(6,6)$ & $\raisebox{-0.5\height}{\begin{tikzpicture}
    \node[gauger, label=below:$D_1$] (1) at (4,-0.75){};
    \node[gaugeb, label=below:$C_1$] (2) at (3,-0.75){};
    \node[gauger, label=below:$D_2$] (3) at (2,-0.75){};
    \node[gaugeb, label=below:$C_2$] (4) at (1,-0.75){};
    \node[gauger, label=below:$D_3$] (5) at (0,0){};
    \node[gaugeb, label=below:$C_1$] (6) at (3,1.25){};
    \node[gaugeb, label=below:$C_1$] (7) at (-1,0){};
    \node[gaugeb, label=below:$C_2$] (8) at (1,0.75){};
    \node[gauger, label=below:$D_1$] (11) at (4,0.25){};
    \node[gaugeb, label=below:$C_1$] (10) at (3,0.25){};
    \node[gauger, label=below:$D_2$] (9) at (2,0.75){};
    \node[gauger,label=below:{$D_1$}] (12) at (4,1.25){};
    \draw[-] (1)--(2)--(3)--(4)--(5)--(8)--(9)--(10)--(11) (7)--(5) (6)--(12) (9)--(6);
\end{tikzpicture}}$ & $d_6 \times d_6$ & $D_6 \times D_6$ & $(8,16)$ & $S_2$ \\ \hline
$(8,8)$ & $\raisebox{-0.5\height}{\begin{tikzpicture}
    \node[gauger, label=below:$D_2$] (3) at (2,0){};
    \node[gaugeb, label=below:$C_2$] (4) at (3,-0.75){};
    \node[gauger, label=above:$D_3$] (5) at (4,-0.75){};
    \node[gaugeb, label=below:$C_2$] (6) at (5,-0.75){};
    \node[gauger, label=below:$D_2$] (7) at (6,-0.75){};
    \node[gaugeb, label=below:$C_1$] (8) at (7,-0.75){};
    \node[gauger, label=below:$D_1$] (9) at (8,-0.75){};
    \node[gaugeb, label=below:$C_1$] (10) at (4,-1.5){};
    \node[gaugeb, label=below:$C_2$] (13) at (3,0.75){};
    \node[gauger, label=below:$D_3$] (14) at (4,0.75){};
    \node[gaugeb, label=below:$C_2$] (15) at (5,0.75){};
    \node[gauger, label=below:$D_2$] (16) at (6,0.75){};
    \node[gaugeb, label=below:$C_1$] (17) at (7,0.75){};
    \node[gauger, label=below:$D_1$] (18) at (8,0.75){};
    \node[gaugeb, label=above:$C_1$] (19) at (4,1.5){};
    \draw[-] (3)--(4)--(5)--(6)--(7)--(8)--(9) (5)--(10);
    \draw[-] (3)--(13)--(14)--(15)--(16)--(17)--(18) (14)--(19);
\end{tikzpicture}}$ & $d_8 \times d_8$ & $D_8 \times D_8$ & $(16,16)$ & $S_2$ \\ \hline
\end{tabular}
\caption{Magnetic quivers for $\spin(4)$ gauge theory with $(N_{\mathrm{S}},N_{\mathrm{C}})$ spinors and cospinors. As expected from the electric theory, the Coulomb branches of these quivers are products of $\mathfrak{so}_{2n}$ nilpotent orbit closures. Their Higgs branches are also products of $D$-type Klein singularities. These theories admit a construction in terms of $\spin(4)$ polymerisations of quivers in Table \ref{tab:free_magnetic_theories}, and inherit outer automorphisms given by $G_{\mathrm{Outer}}$.}
\label{tab:spin4_magquivs}
\end{table}
\subsubsection{$\spin(5)$}
Increasing the rank of the gauge group reduces the number of matter configurations with UV fixed-points in five dimensions. This section will consider $\spin(5)$ SQCD with two, four, five, six and eight spinors. Note that under the isomorphism $\spin(5)\simeq\sprm(2)$ the infinite-coupling limits of these theories were studied in \cite{Bourget:2020gzi}. Hence the calculations given here are immediately verifiable against known results. As stated in the Introduction, the purpose of recording these quivers is to use existing knowledge relating to theories in five dimensions to better understand the moduli space structure of the magnetic quivers themselves.
\paragraph{Four Spinors}
At finite coupling, $\spin(5)$ with four spinors is a union of two identical cones, as expected from the height-two orbits inside the $\mathfrak{so}_8$ nilcone. Both these cones are readily seen in the brane system, and are given below in \eqref{quiv:spin5_4S_FC_both_cones} alongside their associated Coulomb branch, $\overline{n.min.\sorm(8)}$. The magnetic quiver giving rise to one cone is analysed in more detail in \eqref{quiv:B2_4S_Mag_FC}, in which a conjectural Coulomb branch quiver subtraction pattern is given, matching the known Hasse diagram. From the top-down perspective, the first subtraction slice can be taken to be a unitary quiver (SQED with two flavours). Subsequent rebalancing using a $C_1$ gauge node yields an unframed orthosymplectic quiver with Coulomb branch $d_4$. Note that this theory has two independent $S_2$ outer autmorphism symmetries, one of which acts on the (left or right) bouquet of $D_1$ gauge nodes and the other as a reflection symmetry through the middle of the quiver. The integer and half-integer lattice contributions to the Coulomb branch Hilbert series, as well as the Higgs branch Hilbert series, are given below in \eqref{hs:b2_4s_fc_i} to \eqref{hs:b2_4s_fc_iv}. As expected from inversion, the Higgs branch of this quiver is the S\l odowy slice $\mathcal{S}^{D_4}_{\left[4^{2}\right]}$.
\begin{equation}
\raisebox{-0.5\height}{\begin{tikzpicture}
\node (a) at (0,0) {\scalebox{0.75}{\begin{tikzpicture}
   \node[gauger, label=left:$D_1$] (1) at (0,-1){};
    \node[gauger, label=left:$D_1$] (2) at (0,1){};
    \node[gaugeb, label=below:$C_1$] (3) at (1,0){};
    \node[gaugeb, label=below:$C_1$] (4) at (2.5,0){};
    \node[gauger, label=right:$D_1$] (5) at (3.5,-1){};
    \node[gauger, label=right:$D_1$] (6) at (3.5,1){};
    \draw[-] (1)--(3) (2)--(3)--(4)--(5) (4)--(6);
\end{tikzpicture}}};
\node (b) at (3,0) {\scalebox{0.75}{\begin{tikzpicture}
\node at (0,0){\Large $\cup$};
\end{tikzpicture}}};
\node (c) at (6,0) {\scalebox{0.75}{\begin{tikzpicture}
    \node[gauger, label=left:$D_1$] (1) at (0,-1){};
    \node[gauger, label=left:$D_1$] (2) at (0,1){};
    \node[gaugeb, label=below:$C_1$] (3) at (1,0){};
    \node[gaugeb, label=below:$C_1$] (4) at (2.5,0){};
    \node[gauger, label=right:$D_1$] (5) at (3.5,-1){};
    \node[gauger, label=right:$D_1$] (6) at (3.5,1){};
    \draw[-] (1)--(3) (2)--(3)--(4)--(5) (4)--(6);
\end{tikzpicture}}};
\node (c) at (10,0) {\begin{tikzpicture}
\node (aa1) at (0,-0.5) {$\mathcal{N}^{D_4}$};
\node (ab1) at (0.75,1) {\tiny $\left[2^2,1^4\right]$};
\node (ab2) at (1.5,2) {\tiny$\left[2^4_{\textrm{II}}\right]$};
\node (ab1) at (-1.5,2) {\tiny$\left[2^4_{\textrm{I}}\right]$};
\node (0) at (0,2.5) {$\mathcal{C}$};
    \node[hasse] (1a) at (0,0){};
    \node[hasse] (2a) at (0,1){};
    \node[hasse] (3a) at (-1,2){};
    \node[hasse] (4a) at (1,2){};
\draw[-] (1a)--(2a)node[midway, right]{$d_4$}--(3a) node[midway, left]{$A_1$} (2a)--(4a) node[midway, right]{$A_1$};
\end{tikzpicture}};
\end{tikzpicture}}
\label{quiv:spin5_4S_FC_both_cones}
\end{equation}
Note that there is a subtlety regarding the particular orbits in $\mathfrak{so}_{8}$ under consideration. The $2^{4}$ orbit is `very even', and as such is better thought of as a union of two orbits $2^{4}_{\mathrm{I}}$ and $2^{4}_{\mathrm{II}}$ \cite{Distler:2022yse}. This is entirely analogous to the spinor/cospinor symmetry of the Dynkin diagram, and so for $\mathfrak{so}_{8}$ it becomes instead a triality including the vector representation also. Refined Hilbert series distinguish the Higgs branch of $\spin(5)$ with four spinors as the union of orbits given in \eqref{quiv:spin5_4S_FC_both_cones}. The integer lattice contributes $\surm(2)^4$ to the global symmetry, enhanced via the half-integer lattice to $\sorm(8)$.
\begin{equation}
\raisebox{-0.5\height}{\begin{tikzpicture}
\node (a) at (0,0) {\begin{tikzpicture}
    \node[gauger, label=left:$D_1$] (1) at (0,-1){};
    \node[gauger, label=left:$D_1$] (2) at (0,1){};
    \node[gaugeb, label=below:$C_1$] (3) at (1,0){};
    \node[gaugeb, label=below:$C_1$] (4) at (2.5,0){};
    \node[gauger, label=right:$D_1$] (5) at (3.5,-1){};
    \node[gauger, label=right:$D_1$] (6) at (3.5,1){};
    \draw[-] (1)--(3) (2)--(3)--(4)--(5) (4)--(6);
\end{tikzpicture}};
\node (b) at (5.5,0) {\begin{tikzpicture}
\node (0) at (0,3.5) {$\mathcal{C}$};
\node (aa1) at (0,-0.5) {$\bar{\mathcal{O}}^{D_4}_{\left[2^{4}_{\mathrm{I}/\mathrm{II}}\right]}$};
\node[hasse] (1a) at (0,0){};
\node[hasse] (2a) at (0,1.5){};
\node[hasse] (3a) at (0,3){};
\draw[-] (1a)--(2a)node[midway, right]{$d_4$}--(3a) node[midway, right]{$a_1$};

\node (c) at (-1,2.25) {\scalebox{0.5}{\begin{tikzpicture}
    \node[gaugeBig, label=below:$1$] (1b) at (1,0){};
    \node[gaugeBig, label=below:$1$] (2b) at (2,0){};
    \draw[transform canvas={yshift=1.3pt}](1b)--(2b);
    \draw[transform canvas={yshift=-1.3pt}](1b)--(2b);
    \end{tikzpicture}}};
\node (d) at (-1,0.75) {\scalebox{0.5}{\begin{tikzpicture}
    \node[gaugeb, label=below:$C_1$] (1d) at (0,0){};
    \node[gauger, label=above:$D_1$] (2d) at (0.5,0.5){};
    \node[gauger, label=left:$D_1$] (3d) at (-0.75,0){};
    \node[gauger, label=above:$D_1$] (4d) at (-0.5,0.5){};
    \node[gauger, label=right:$D_1$] (5d) at (0.75,0){};
    \draw[-] (1d)--(2d) (1d)--(3d) (1d)--(4d) (1d)--(5d);
    \end{tikzpicture}}};
\end{tikzpicture}};
\node (3) at (9,0) {\begin{tikzpicture}
\node (0) at (0,3.5) {$\mathcal{H}$};
\node (aa1) at (0,-0.5) {$\mathcal{S}^{D_4}_{\left[4^{2}\right]}$};
\node[hasse] (1a) at (0,0){};
\node[hasse] (2a) at (0,1.5){};
\node[hasse] (3a) at (0,3){};
\draw[-] (1a)--(2a)node[midway, right]{$A_1$}--(3a) node[midway, right]{$D_4$};
\end{tikzpicture}};
\end{tikzpicture}}
\label{quiv:B2_4S_Mag_FC}
\end{equation}
\begin{align}
    \hsz\left[\ref{quiv:B2_4S_Mag_FC}\right]=&\;\frac{1+6t^{2}+112t^{4}+302t^{6}+1143t^{8}+1484t^{10}+2352t^{12}+\cdots+t^{24}}{(1-t^{2})^{6}(1-t^{4})^{6}},\label{hs:b2_4s_fc_i}\\
    \hszh\left[\ref{quiv:B2_4S_Mag_FC}\right]=&\;\frac{16t^{2}\left(1+4t^{2}+28t^{4}+52t^{6}+123t^{8}+112t^{10}+\cdots+t^{20}\right)}{(1-t^{2})^{6}(1-t^{4})^{6}},\label{hs:b2_4s_fc_ii}\\
    \hs\left[\ref{quiv:B2_4S_Mag_FC}\right]=&\;\frac{(1+t^{2})^{2}(1+14t^{2}+36t^{4}+14t^{6}+t^{8})}{(1-t^{2})^{12}}\rightarrow\overline{n.min.\sorm(8)},\label{hs:b2_4s_fc_iii}\\
    \hsh\left[\ref{quiv:B2_4S_Mag_FC}\right]=&\frac{(1-t^{8})(1-t^{12})}{(1-t^{2})^{3}(1-t^{6})^{3}}.\;\label{hs:b2_4s_fc_iv}
\end{align}
At infinite coupling one of the cones is unchanged, while the other is described by the magnetic theory in \eqref{quiv:B2_4S_Mag_IC}. The integer and half-integer contributions to the Coulomb branch Hilbert series are as below in \eqref{eqn:B2_4S_Mag_IC_i} and \eqref{eqn:B2_4S_Mag_IC_ii}. The theory's Higgs branch Hilbert series is given in \eqref{eqn:B2_4S_Mag_IC_iv}. Note that now there are two independent outer automorphisms, acting on each bouquet of $\urm(1)$ gauge nodes. The global symmetry of the infinite coupling magnetic quiver's Coulomb branch is $\sorm(8)\times\urm(1)_{\mathrm I}$.
\begin{equation}
\raisebox{-0.5\height}{\begin{tikzpicture}
\node (a) at (0,0) {\begin{tikzpicture}
    \node[gauger, label=left:$D_1$] (1) at (0,-1){};
    \node[gauger, label=left:$D_1$] (2) at (0,1){};
    \node[gaugeb, label=below:$C_1$] (3) at (1,0){};
    \node[gaugeBig, label=right:$2$] (4) at (2.5,0){};
    \node[gaugeBig, label=right:$1$] (5) at (3.5,-1){};
    \node[gaugeBig, label=right:$1$] (6) at (3.5,1){};
    \node[flavour, label=above:$\urm(1)$] (7) at (2.5,1){};
    \node at (2.25,0.5) {$\wedge^2$};
    \draw[-] (1)--(3) (2)--(3)--(4)--(5) (4)--(6) (4)--(7);
\end{tikzpicture}}; 
\node (b) at (6,0) {\begin{tikzpicture}
    \node (0) at (0,3.5) {$\mathcal{C}$};
    \node[hasse] (1) at (0,0){};
    \node[hasse] (2) at (0,1.5){};
    \node[hasse] (3) at (0,3){};
    \draw[-] (1)--(2)node[midway, right]{$d_{4}$}--(3) node[midway, right]{$a_2$};
\end{tikzpicture}};
\node (3) at (10.5,0) {\begin{tikzpicture}
    \node (0) at (0,3.5) {$\mathcal{H}$};
    \node[hasse] (1a) at (0,0){};
    \node[hasse] (2a) at (0,1.5){};
    \node[hasse] (3a) at (0,3){};
    \draw[-] (1a)--(2a)node[midway, right]{$A_2$}--(3a) node[midway, right]{$D_4$};
    \end{tikzpicture}};
\end{tikzpicture}}
\label{quiv:B2_4S_Mag_IC}
\end{equation}
\begin{align}
    \hsz\left[\ref{quiv:B2_4S_Mag_IC}\right]=&1+13t^{2}+196t^{4}+1584t^{6}+9938t^{8}+49056t^{10}+\cdots\;\label{eqn:B2_4S_Mag_IC_i}\\
    \hszh\left[\ref{quiv:B2_4S_Mag_IC}\right]=&16t^{2}+184t^{4}+1608t^{6}+9872t^{8}+49168t^{10}+\cdots\;\label{eqn:B2_4S_Mag_IC_ii}\\
    \hs\left[\ref{quiv:B2_4S_Mag_IC}\right]=&1+29t^{2}+380t^{4}+3192t^{6}+19810t^{8}+98224t^{10}+\cdots\;\label{eqn:B2_4S_Mag_IC_iii}\\
    \hsh\left[\ref{quiv:B2_4S_Mag_IC}\right]=&\frac{1-t+t^{2}+t^{3}-t^{4}+t^{5}+t^{6}+t^{8}+t^{9}-t^{10}+t^{11}+t^{12}-t^{13}+t^{14}}{(1-t)(1-t^{4})(1-t^{6})(1-t^{7})}\;\label{eqn:B2_4S_Mag_IC_iv}
\end{align}
This is weak-weak dual to the theory in Table 8 of \cite{Hanany:2025ctg} given below alongside its Coulomb branch HWG.
\begin{equation}
\raisebox{-0.5\height}{\begin{tikzpicture}
    \node[gauge, label=below:$1$] (0) at (0,0){};
    \node[gauge, label=below:$2$] (1) at (1,0){};
    \node[gauge, label=below:$2$] (2) at (2,0){};
    \node[gauge, label=below:$1$] (3) at (3,0){};
    \node[gauge, label=left:$1$] (4) at (1,1){};
    \node[gauge, label=left:$1$] (5) at (2,1){};
    \draw[-] (0)--(1)--(2)--(3) (1)--(4) (2)--(5)--(3);
    \node at (8,0.35) {$\pe\left[\left(1+\mu_2\right)t^{2}+\left(q+q^{-1}\right)\mu_4 t^3+\mu_4^2t^4-\mu_4^2 t^6\right]$};
\end{tikzpicture}}
\end{equation}
\noindent\textbf{Five Spinors  } At finite coupling, the magnetic quiver for $\spin(5)$ with five spinors is given below alongside its associated Coulomb branch, $\overline{n.min.\sorm(10)}$. 
\begin{equation}
\raisebox{-0.5\height}{\begin{tikzpicture}
\node (a) at (0,0) {\begin{tikzpicture}
    \node[gauger, label=below:$D_1$] (1) at (0,0){};
    \node[gaugeb, label=below:$C_1$] (2) at (1,0){};
    \node[gauger, label=below:$D_2$] (3) at (2,0){};
    \node[gaugeb, label=below:$C_2$] (4) at (3,0){};
    \node[gauger, label=below:$D_2$] (5) at (4,0){};
    \node[gaugeb, label=above:$C_1$] (6) at (4,1){};
    \node[gauger, label=above:$D_1$] (7) at (3,1){};
    \draw[-] (1)--(2)--(3)--(4)--(5)--(6)--(7)--(4);
\end{tikzpicture}};
\node (b) at (5.5,0) {\begin{tikzpicture}
\node (0) at (0,3.5) {$\mathcal{C}$};
\node (aa1) at (0,-0.5) {$\bar{\mathcal{O}}^{D_5}_{\left[2^{4},1^{2}\right]}$};
    \node[hasse] (1) at (0,0){};
    \node[hasse] (2) at (0,1.5){};
    \node[hasse] (3) at (0,3){};
    \node (ab3) at (0.6,1.5) {\tiny$\left[2^2,1^6\right]$};
    \node (ab3) at (0.6,3) {\tiny$\left[2^4,1^2\right]$};
    \draw[-] (1)--(2)node[midway, right]{$d_{5}$}--(3) node[midway, right]{$a_3$};
\node (c) at (-1,2.25) {\scalebox{0.5}{\begin{tikzpicture}
    \node[gauge, label=below:$1$] (1b) at (1,0){};
    \node[gauge, label=below:$1$] (2b) at (2,0){};
    \node[gauge, label=above:$1$] (3b) at (2,1){};
    \node[gauge, label=above:$1$] (4b) at (1,1){};
    \draw[-] (1b)--(2b)--(3b)--(4b)--(1b);
    \end{tikzpicture}}};
\node (d) at (-1.5,0.75) {\scalebox{0.5}{\begin{tikzpicture}
    \node[gauger, label=below:$D_1$] (1d) at (0,0){};
    \node[gaugeb, label=below:$C_1$] (2d) at (1,0){};
    \node[gauger, label=below:$D_2$] (3d) at (2,0){};
    \node[gaugeb, label=below:$C_1$] (4d) at (3,0){};
    \node[gauger, label=below:$D_1$] (5d) at (4,0){};
    \node[gauge, label=above:$1$] (6d) at (2,1){};
    \draw[-] (1d)--(2d)--(3d)--(4d)--(5d) (3d)--(6d);
    \end{tikzpicture}}};
\end{tikzpicture}};
\node (3) at (9,0) {\begin{tikzpicture}
\node (0) at (0,3.5) {$\mathcal{H}$};
\node[hasse] (1a) at (0,0){};
\node[hasse] (2a) at (0,1.5){};
\node (ab3) at (0.6,1.5) {\tiny$\left[7,3\right]$};
\node (ab3) at (0.6,3) {\tiny$\left[9,1\right]$};
\node[hasse] (3a) at (0,3){};
\draw[-] (1a)--(2a)node[midway, right]{$A_3$}--(3a) node[midway, right]{$D_5$};
\node (aa1) at (0,-0.5) {$\mathcal{S}^{D_5}_{\left[5^{2}\right]}$};
\end{tikzpicture}};
\end{tikzpicture}}
\label{quiv:B2_5S_Mag_FC}
\end{equation}
\begin{align}
    \hscz\left[\ref{quiv:B2_5S_Mag_FC}\right]=&\;1+29t^{2}+532t^{4}+6992t^{6}+68532t^{8}+527209t^{10}+\cdots,\\
    \hsczh\left[\ref{quiv:B2_5S_Mag_FC}\right]=\;&16t^{2}+448t^{4}+6592t^{6}+67008t^{8}+522224t^{10}+\cdots,\\
    \hsc\left[\ref{quiv:B2_5S_Mag_FC}\right]=&\;1+45t^{2}+980t^{4}+13584t^{6}+135540t^{8}+1049433t^{10}+\cdots\rightarrow\overline{n.min.\sorm(10)},\\
    \hsh\left[\ref{quiv:B2_5S_Mag_FC}\right]&=\frac{(1-t^{12})(1-t^{16})}{(1-t^{2})(1-t^{4})^{2}(1-t^{6})(1-t^{8})^{2}}.
\end{align}
A conjectural Coulomb branch quiver subtraction pattern begins with a subtraction of the \emph{unitary} affine $A_3$ quiver, after which relabancing with a $D_1$ gauge node produces the magnetic quiver for $\spin(3)$ with five spinors given in Table \ref{tab:spin3_magquivs}. Note that this magnetic quiver has no obvious outer automorphism. A perturbative calculation of its Coulomb branch Hilbert series is given below to $\mathcal{O}(t^{10})$, which agrees with that for $\overline{n.min.\sorm(10)}$. Direct calculation of its Higgs branch confirms the expectation from Lusztig-Spaltenstein duality.
At infinite coupling the magnetic quiver is modified to be that given below in \eqref{quiv:B2_5S_Mag_IC}. The integer and half-integer contributions to the Coulomb branch Hilbert series are as in \eqref{eqn:quiv:B2_5S_Mag_IC_i} and \eqref{eqn:quiv:B2_5S_Mag_IC_ii}. Clearly, outer automorphism symmetries do not appear at infinite coupling. Note that the Higgs branch, while not known to be a nilpotent orbit, it conjectured to have the Hasse diagram below.
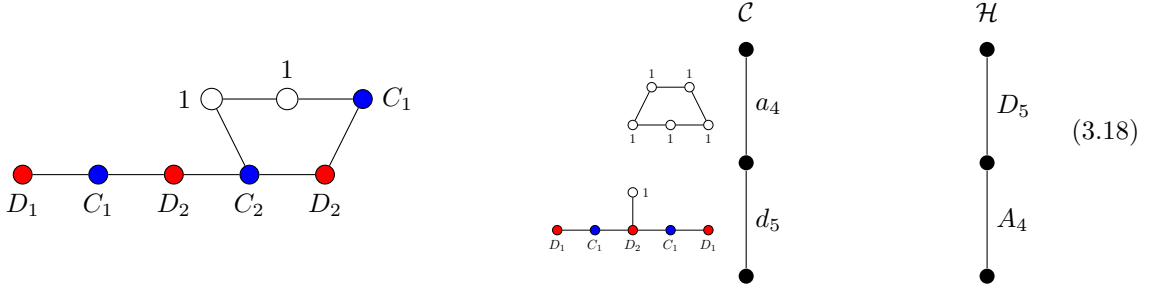
\begin{equation}
\raisebox{-0.5\height}{\begin{tikzpicture}
\node (a) at (0,0) {\begin{tikzpicture}
    \node[gauger, label=below:$D_1$] (1) at (0,0){};
    \node[gaugeb, label=below:$C_1$] (2) at (1,0){};
    \node[gauger, label=below:$D_2$] (3) at (2,0){};
    \node[gaugeb, label=below:$C_2$] (4) at (3,0){};
    \node[gauger, label=below:$D_2$] (5) at (4,0){};
    \node[gaugeb, label=right:$C_1$] (6) at (4.5,1){};
    \node[gaugeBig, label=left:$1$] (7) at (2.5,1){};
    \node[gaugeBig, label=above:$1$] (8) at (3.5,1){};
    \draw[-] (1)--(2)--(3)--(4)--(5)--(6) (7)--(4) (6)--(8) (7)--(8);
\end{tikzpicture}}; 
\node (b) at (6,0) {\begin{tikzpicture}
    \node (0) at (0,3.5) {$\mathcal{C}$};
    \node[hasse] (1) at (0,0){};
    \node[hasse] (2) at (0,1.5){};
    \node[hasse] (3) at (0,3){};
    \draw[-] (1)--(2)node[midway, right]{$d_{5}$}--(3) node[midway, right]{$a_4$};
    \node (c) at (-1,2.25) {\scalebox{0.5}{\begin{tikzpicture}
    \node[gauge, label=below:$1$] (1b) at (0,0){};
    \node[gauge, label=below:$1$] (1bb) at (1,0){};
    \node[gauge, label=below:$1$] (2b) at (2,0){};
    \node[gauge, label=above:$1$] (3b) at (1.5,1){};
    \node[gauge, label=above:$1$] (4b) at (0.5,1){};
    \draw[-] (1b)--(1bb)--(2b)--(3b)--(4b)--(1b);
    \end{tikzpicture}}};
\node (d) at (-1.5,0.75) {\scalebox{0.5}{\begin{tikzpicture}
    \node[gauger, label=below:$D_1$] (1d) at (0,0){};
    \node[gaugeb, label=below:$C_1$] (2d) at (1,0){};
    \node[gauger, label=below:$D_2$] (3d) at (2,0){};
    \node[gaugeb, label=below:$C_1$] (4d) at (3,0){};
    \node[gauger, label=below:$D_1$] (5d) at (4,0){};
    \node[gauge, label=right:$1$] (6d) at (2,1){};
    \draw[-] (1d)--(2d)--(3d)--(4d)--(5d) (3d)--(6d);
    \end{tikzpicture}}};
\end{tikzpicture}};
\node (3) at (10.5,0) {\begin{tikzpicture}
    \node (0) at (0,3.5) {$\mathcal{H}$};
    \node[hasse] (1a) at (0,0){};
    \node[hasse] (2a) at (0,1.5){};
    \node[hasse] (3a) at (0,3){};
    \draw[-] (1a)--(2a)node[midway, right]{$A_4$}--(3a) node[midway, right]{$D_5$};
    \end{tikzpicture}};
\end{tikzpicture}}
\label{quiv:B2_5S_Mag_IC}
\end{equation}
\begin{align}
    \hscz\left[\ref{quiv:B2_5S_Mag_IC}\right]&=\begin{aligned}1&+30t^{2}+16t^{3}+562t^{4}+576t^{6}+7694t^{6}+10080t^{7}+80034t^{8}\\&+116304t^{9}+661567t^{10}+\cdots\end{aligned}\label{eqn:quiv:B2_5S_Mag_IC_i}\\
    \hsczh\left[\ref{quiv:B2_5S_Mag_IC}\right]&=\begin{aligned}16t^{2}&+16t^{3}+464t^{4}+576t^{5}+7168t^{6}+10080t^{7}+77760t^{8}\\&+116304t^{9}+653120t^{10}+\cdots\end{aligned}\;\label{eqn:quiv:B2_5S_Mag_IC_ii}\\
    \hsc\left[\ref{quiv:B2_5S_Mag_IC}\right]&=\begin{aligned}1&+46t^{2}+32t^{3}+1026t^{4}+1152t^{5}+14862t^{6}+20160t^{7}+157794t^{8}\\&+232608t^{9}+1314687t^{10}+\cdots\end{aligned}\;\label{eqn:quiv:B2_5S_Mag_IC_iii}\\
    \hsh\left[\ref{quiv:B2_5S_Mag_IC}\right]&=\frac{(1-t^{10})(1-t^{14})^{2}(1-t^{17})^{2}(1-t^{18})}{(1-t^{2})^{2}(1-t^{5})^{2}(1-t^{6})(1-t^{8})(1-t^{9})^{2}(1-t^{12})(1-t^{19})^{2}}.\label{eqn:quiv:B2_5S_Mag_IC_iv}
\end{align}
This is weak-weak dual to the theory in Table 8 of \cite{Hanany:2025ctg} given below alongside its Coulomb branch HWG.
\begin{equation}
    \raisebox{-0.5\height}{\begin{tikzpicture}
    \node[gauge, label=below:$1$] (0) at (0,0){};
    \node[gauge, label=below:$2$] (1) at (1,0){};
    \node[gauge, label=below:$3$] (2) at (2,0){};
    \node[gauge, label=below:$2$] (3) at (3,0){};
    \node[gauge, label=below:$1$] (4) at (4,0){};
    \node[gauge, label=left:$2$] (5) at (2,1){};
    \node[gauge, label=right:$1$] (6) at (3,1){};
    \draw[-] (0)--(1)--(2)--(3)--(4) (2)--(5)--(6)--(4);
    \node at (9,0.3) {$\pe\left[\left(1+\mu_2\right)t^{2}+\left(q\mu_4+q^{-1}\mu_5\right)t^3+\mu_4\mu_5 t^4-\mu_4\mu_5 t^6\right]$};
\end{tikzpicture}}
\end{equation}
\noindent\textbf{Six Spinors  } At finite coupling, the magnetic quiver for $\spin(5)$ with six spinors is given below alongside its associated Coulomb branch, $\overline{n.min.\sorm(12)}$. A conjectural Coulomb branch quiver subtraction pattern begins with the subtraction of the affine $D_4$ unitary quiver, followed by rebalancing using a $C_1$ gauge node to produce the magnetic quiver for $\spin(3)$ with six spinors given in Table \ref{tab:spin3_magquivs}. This quiver has an $S_2$ outer automorphism that permutes the two legs of $D_1-C_1$ connected to the central $D_3$ gauge node. A perturbative calculation of its Coulomb branch Hilbert series is given below to $\mathcal{O}(t^{10})$, which agrees with that for $\overline{n.min.\sorm(12)}$. Direct calculation of the Higgs branch confirms the expectation from Lusztig-Spaltenstein duality.
\begin{equation}
\raisebox{-0.5\height}{\begin{tikzpicture}
\node (a) at (0,0) {\begin{tikzpicture}
    \node[gauger, label=below:$D_1$] (1) at (0,0){};
    \node[gaugeb, label=below:$C_1$] (2) at (1,0){};
    \node[gauger, label=below:$D_2$] (3) at (2,0){};
    \node[gaugeb, label=below:$C_2$] (4) at (3,0){};
    \node[gauger, label=below:$D_3$] (5) at (4,0){};
    \node[gaugeb, label=above:$C_1$] (6) at (5,1){};
    \node[gauger, label=above:$D_1$] (7) at (6,1){};
    \node[gaugeb, label=below:$C_1$] (8) at (5,-1){};
    \node[gauger, label=below:$D_1$] (9) at (6,-1){};
    \node[gaugeb, label=right:$C_1$] (10) at (5,0){};
    \draw[-] (1)--(2)--(3)--(4)--(5)--(6)--(7) (5)--(8)--(9) (5)--(10);
\end{tikzpicture}};
\node (b) at (5.5,0) {\begin{tikzpicture}
\node (0) at (0,3.5) {$\mathcal{C}$};
\node (aa1) at (0,-0.5) {$\bar{\mathcal{O}}^{D_6}_{\left[2^{4},1^{4}\right]}$};
    \node[hasse] (1) at (0,0){};
    \node[hasse] (2) at (0,1.5){};
    \node[hasse] (3) at (0,3){};
    \node (ab3) at (0.6,1.5) {\tiny$\left[2^2,1^8\right]$};
    \node (ab4) at (0.6,3) {\tiny$\left[2^4,1^4\right]$};
    \draw[-] (1)--(2)node[midway, right]{$d_6$}--(3) node[midway, right]{$d_4$};
\node (c) at (-1,2.25) {\scalebox{0.5}{\begin{tikzpicture}
    \node[gauge, label=below:$2$] (1b) at (0,0){};
    \node[gauge, label=below:$1$] (2b) at (0.75,0){};
    \node[gauge, label=below:$1$] (3b) at (-0.75,0){};
    \node[gauge, label=above:$1$] (4b) at (0.5,0.5){};
    \node[gauge, label=above:$1$] (5b) at (-0.5,0.5){};
    \draw[-] (1b)--(2b) (1b)--(3b) (1b)--(4b) (5b)--(1b);
    \end{tikzpicture}}};
\node (d) at (-1.5,0.75) {\scalebox{0.5}{\begin{tikzpicture}
    \node[gauger, label=below:$D_1$] (1) at (2,-0.5){};
    \node[gaugeb, label=below:$C_1$] (2) at (1,-0.5){};
    \node[gauger, label=below:$D_2$] (3) at (0,0){};
    \node[gaugeb, label=below:$C_1$] (4) at (1,0.5){};
    \node[gauger, label=below:$D_1$] (5) at (2,0.5){};
    \node[gaugeb, label=below:$C_1$] (6) at (-1,0){};
    \node[gauger, label=below:$D_1$] (7) at (-2,0.5){};
    \node[gauger, label=below:$D_1$] (8) at (-2,-0.5){};
    \draw[-] (1)--(2)--(3)--(4)--(5) (3)--(6)--(7) (6)--(8);
    \end{tikzpicture}}};
\end{tikzpicture}};
\node (3) at (9,0) {\begin{tikzpicture}
\node (0) at (0,3.5) {$\mathcal{H}$};
\node (aa1) at (0,-0.5) {$\mathcal{S}^{D_6}_{\left[7,5\right]}$};
\node[hasse] (1a) at (0,0){};
\node[hasse] (2a) at (0,1.5){};
\node[hasse] (3a) at (0,3){};
\node (ab3) at (0.6,1.5) {\tiny$\left[9,3\right]$};
\node (ab3) at (0.6,3) {\tiny$\left[11,1\right]$};
\draw[-] (1a)--(2a)node[midway, right]{$D_4$}--(3a) node[midway, right]{$D_6$};
\end{tikzpicture}};
\end{tikzpicture}}
\label{quiv:B2_6S_Mag_FC}
\end{equation}
\begin{align}
\hscz\left(\ref{quiv:B2_6S_Mag_FC}\right)&=1+34t^{2}+1077t^{4}+22073t^{6}+328443t^{8}+3772266t^{10}+\cdots\\
    \hsczh\left(\ref{quiv:B2_6S_Mag_FC}\right)&=32t^{2}+1056t^{4}+22048t^{6}+328224t^{8}+3772128t^{10}+\cdots\\
    \hsc\left(\ref{quiv:B2_6S_Mag_FC}\right)&=1+66t^{2}+2133t^{4}+44121t^{6}+656667t^{8}+7544394t^{10}+\cdots\rightarrow\overline{n.min.\sorm(12)},\\
    \hsh\left[\ref{quiv:B2_6S_Mag_FC}\right]&=\frac{(1-t^{16})(1-t^{20})}{(1-t^{4})^{2}(1-t^{6})(1-t^{8})^{2}(1-t^{10})}.\;
\end{align}
At infinite coupling the magnetic quiver is modified to become that given below in \eqref{quiv:spin5_6S_magquiv_IC}. Note that it no longer has an outer automorphism symmetry. The integer-lattice and half-integer lattice contributions to the Coulomb branch Hilbert series are given in \eqref{eqn:spin5_6S_magquiv_IC_i} and \eqref{eqn:spin5_6S_magquiv_IC_ii}. Its Higgs branch Hilbert series is also straightforward, given in \eqref{eqn:spin5_6S_magquiv_IC_iv}, and is conjectured to have the Hasse diagram below.
\begin{equation}
\raisebox{-0.5\height}{\begin{tikzpicture}
\node (a) at (0,0) {\begin{tikzpicture}
    \node[gauger, label=below:$D_1$] (1) at (0,0){};
    \node[gaugeb, label=below:$C_1$] (2) at (1,0){};
    \node[gauger, label=below:$D_2$] (3) at (2,0){};
    \node[gaugeb, label=below:$C_2$] (4) at (3,0){};
    \node[gauger, label=below:$D_3$] (5) at (4,0){};
    \node[gaugeb, label=below:$C_1$] (6) at (5,0){};
    \node[gauger, label=below:$D_1$] (7) at (6,0){};
    \node[gaugeb, label=left:$C_2$] (8) at (4,1){};
    \node[gaugeBig, label=left:$2$] (9) at (4,2){};
    \node[gaugeBig, label=left:$1$] (10) at (4,3){};
    \draw[-] (1)--(2)--(3)--(4)--(5)--(6)--(7) (5)--(8)--(9)--(10);
\end{tikzpicture}}; 
\node (b) at (6,0) {\begin{tikzpicture}
\node (0) at (0,3.5) {$\mathcal{C}$};
     \node (0) at (0,3.5) {$\mathcal{C}$};
    \node[hasse] (1) at (0,0){};
    \node[hasse] (2) at (0,1.5){};
    \node[hasse] (3) at (0,3){};
    \draw[-] (1)--(2)node[midway, right]{$d_{6}$}--(3) node[midway, right]{$d_5$};
\end{tikzpicture}};
\node (3) at (8.5,0) {\begin{tikzpicture}
    \node (0) at (0,3.5) {$\mathcal{H}$};
    \node[hasse] (1a) at (0,0){};
    \node[hasse] (2a) at (0,1.5){};
    \node[hasse] (3a) at (0,3){};
    \draw[-] (1a)--(2a)node[midway, right]{$D_5$}--(3a) node[midway, right]{$D_6$};
    \end{tikzpicture}};
\end{tikzpicture}}
\label{quiv:spin5_6S_magquiv_IC}
\end{equation}
\begin{align}
\hscz\left(\ref{quiv:spin5_6S_magquiv_IC}\right)&=\begin{aligned}1&+35t^{2}+32t^{3}+1112t^{4}+1760t^{5}+23899t^{6}+47520t^{7}+384532t^{8}\\&+845856t^{9}+4897165t^{10}+\cdots\end{aligned}\label{eqn:spin5_6S_magquiv_IC_i}\\
    \hsczh\left(\ref{quiv:spin5_6S_magquiv_IC}\right)&=\begin{aligned}32t^{2}&+32t^{3}+1088t^{4}+1760t^{5}+23808t^{6}+47520t^{7}+384192t^{8}\\&+845856t^{9}+4895904t^{10}+\cdots
    \end{aligned}\label{eqn:spin5_6S_magquiv_IC_ii}\\
    \hsc\left(\ref{quiv:spin5_6S_magquiv_IC}\right)&=\begin{aligned}1&+67t^{2}+64t^{3}+2200t^{4}+3520t^{5}+47707t^{6}+95040t^{7}+768724t^{8}\\&+1691712t^{9}+9793069t^{10}+\cdots
    \end{aligned}\label{eqn:spin5_6S_magquiv_IC_iii}\\
    \hsh\left[\ref{quiv:spin5_6S_magquiv_IC}\right]&=\frac{(1-t^{20})^{2}}{(1-t^{4})(1-t^{6})(1-t^{8})^{2}(1-t^{10})(1-t^{12})^{2}}.\label{eqn:spin5_6S_magquiv_IC_iv}\;
\end{align}
This is weak-weak dual to the theory in Table 8 of \cite{Hanany:2025ctg} given below alongside its Coulomb branch HWG.
\begin{equation}
    \raisebox{-0.5\height}{\begin{tikzpicture}
    \node[gauge, label=below:$1$] (0) at (0,0){};
    \node[gauge, label=below:$2$] (1) at (1,0){};
    \node[gauge, label=below:$3$] (2) at (2,0){};
    \node[gauge, label=below:$4$] (3) at (3,0){};
    \node[gauge, label=below:$3$] (4) at (4,0){};
    \node[gauge, label=below:$1$] (5) at (5,0){};
    \node[gauge, label=left:$2$] (6) at (3,1){};
    \node[gauge, label=left:$1$] (7) at (4,1){};
    \draw[-] (0)--(1)--(2)--(3)--(4)--(5) (3)--(6) (4)--(7);
    \node at (9,0.3) {$\pe\left[\left(1+\mu_2\right)t^{2}+\left(q+q^{-1}\right)\mu_6t^3+\mu_4 t^4\right]$};
\end{tikzpicture}}
\end{equation}
\textbf{Eight Spinors  } At finite coupling, the magnetic quiver for $\spin(5)$ with eight spinors is given below alongside its associated Coulomb branch, $\overline{n.min.\sorm(16)}$. This theory arises from the brane web \eqref{web:spin5_8spinors}. A conjectural Coulomb branch quiver subtraction pattern begins with the subtraction of the affine $D_6$ unitary quiver, followed by rebalancing using a $C_1$ gauge node to produce the magnetic quiver for $\spin(3)$ with eight spinors given in Table \ref{tab:spin3_magquivs}. This quiver has an $S_2$ outer automorphism realised by reflection about the central $\sprm(2)$ gauge node. A perturbative calculation of its Coulomb branch Hilbert series is given below to $\mathcal{O}(t^{10})$, which agrees with that for $\overline{n.min.\sorm(16)}$. Direct calculation of its Higgs branch Hilbert series agrees with expectation from Lusztig-Spaltenstein duality.
\begin{equation}
    \begin{array}{c}
         \scalebox{0.85}{\begin{tikzpicture}
            \draw[dashed](0,0)--(1,0);
            \draw(1,0)--(12,0);
            \draw(15,0)--(13,0);
            \draw[dashed](16,0)--(15,0);
            \node[7brane]at(1,0){};
            \node[7brane]at(2,0){};
            \node[7brane]at(3,0){};
            \node[7brane]at(4,0){};
            \node[7brane]at(5,0){};
            \node[7brane]at(6,0){};
            \node[7brane]at(9,0){};
            \node[7brane]at(10,0){};
            \node[7brane]at(11,0){};
            \node[7brane]at(12,0){};
            \node[7brane]at(13,0){};
            \node[7brane]at(14,0){};
            \node[7brane]at(15,0){};
            \draw(7,0)--(7,1);
            \draw(8,0)--(7.5,.5);
            \node[7brane]at(7,1){};
            \node[7brane]at(7.5,0.5){};
            \node[label=below:{$\frac{1}{2}$}]at(1.5,0){};
            \node[label=below:{$1$}]at(2.5,0){};
            \node[label=below:{$\frac{3}{2}$}]at(3.5,0){};
            \node[label=below:{$2$}]at(4.5,0){};
            \node[label=below:{$\frac{5}{2}$}]at(5.5,0){};
            \node[label=below:{$3$}]at(6.5,0){};
            \node[label=left:{$2$}]at(7,.5){};
            \node[label=below:{$3$}]at(7.5,0){};
            \node[label=right:{$2$}]at(7.5,0.5){};
            \node[label=below:{$5$}]at(8.5,0){};
            \node[label=below:{$\frac{7}{2}$}]at(9.5,0){};
            \node[label=below:{$3$}]at(10.5,0){};
            \node[label=below:{$\frac{5}{2}$}]at(11.5,0){};
            \node[label=below:{$1$}]at(13.5,0){};
            \node[label=below:{$\frac{1}{2}$}]at(14.5,0){};
            \node at (12.5,0){$\cdots$};
         \end{tikzpicture}}
    \end{array}
\label{web:spin5_8spinors}
\end{equation}
\begin{equation}
\raisebox{-0.5\height}{\begin{tikzpicture}
\node (a) at (0,0) {\begin{tikzpicture}
    \node[gauger, label=below:$D_1$] (1) at (0,0){};
    \node[gaugeb, label=below:$C_1$] (2) at (1,0){};
    \node[gauger, label=below:$D_2$] (3) at (2,0){};
    \node[gaugeb, label=below:$C_2$] (4) at (3,0){};
    \node[gauger, label=below:$D_3$] (5) at (4,0){};
    \node[gaugeb, label=above:$C_1$] (6) at (5,0.5){};
    \node[gaugeb, label=right:$C_2$] (7) at (5,-0.5){};
    \node[gauger, label=below:$D_3$] (8) at (4,-1){};
    \node[gaugeb, label=below:$C_2$] (9) at (3,-1){};
    \node[gauger, label=below:$D_2$] (10) at (2,-1){};
    \node[gaugeb, label=below:$C_1$] (11) at (1,-1){};
    \node[gauger, label=below:$D_1$] (12) at (0,-1){};
    \node[gaugeb, label=below:$C_1$] (13) at (5,-1.5){};
    \draw[-] (1)--(2)--(3)--(4)--(5)--(7)--(8)--(9)--(10)--(11)--(12) (5)--(6) (13)--(8);
\end{tikzpicture}};
\node (b) at (6.5,0) {\begin{tikzpicture}
    \node (aa1) at (0,-0.5) {$\bar{\mathcal{O}}^{D_8}_{\left[2^{4},1^{8}\right]}$};
    \node (0) at (0,3.5) {$\mathcal{C}$};
    \node[hasse] (1) at (0,0){};
    \node[hasse] (2) at (0,1.5){};
    \node[hasse] (3) at (0,3){};
    \node (ab3) at (0.6,1.5) {\tiny$\left[2^2,1^{12}\right]$};
    \node (ab3) at (0.6,3) {\tiny$\left[2^4,1^{8}\right]$};
    \draw[-] (1)--(2)node[midway, right]{$d_8$}--(3) node[midway, right]{$d_6$};
\node (c) at (-1.25,2.25) {\scalebox{0.5}{\begin{tikzpicture}
    \node[gauge, label=below:$2$] (1b) at (0,0){};
    \node[gauge, label=below:$2$] (2b) at (0.75,0){};
    \node[gauge, label=below:$2$] (3b) at (1.5,0){};
    \node[gauge, label=below:$1$] (4b) at (-0.75,0){};
    \node[gauge, label=above:$1$] (5b) at (0,0.5){};
    \node[gauge, label=below:$1$] (6b) at (2.25,0){};
    \node[gauge, label=above:$1$] (7b) at (1.5,0.5){};
    \draw[-] (4b)--(1b)--(2b)--(3b)--(6b) (1b)--(5b) (3b)--(7b);
    \end{tikzpicture}}};
\node (d) at (-1.75,0.75) {\scalebox{0.5}{\begin{tikzpicture}
    \node[gauger, label=below:$D_1$] (1) at (0,-0.5){};
    \node[gaugeb, label=below:$C_1$] (2) at (1,-0.5){};
    \node[gauger, label=below:$D_2$] (3) at (2,0){};
    \node[gaugeb, label=below:$C_1$] (4) at (3,0){};
    \node[gauger, label=below:$D_2$] (5) at (4,0){};
    \node[gaugeb, label=below:$C_1$] (6) at (5,-0.5){};
    \node[gauger, label=below:$D_1$] (7) at (6,-0.5){};
    \node[gaugeb, label=below:$C_1$] (8) at (1,0.5){};
    \node[gauger, label=below:$D_1$] (9) at (0,0.5){};
    \node[gaugeb, label=below:$C_1$] (10) at (5,0.5){};
    \node[gauger, label=below:$D_1$] (11) at (6,0.5){};
    \draw[-] (1)--(2)--(3)--(4)--(5)--(6)--(7) (3)--(8)--(9) (5)--(10)--(11);
    \end{tikzpicture}}};
\end{tikzpicture}};
\node (3) at (9.5,0) {\begin{tikzpicture}
    \node (aa1) at (0,-0.5) {$\mathcal{S}^{D_8}_{\left[11,5\right]}$};
    \node (0) at (0,3.5) {$\mathcal{H}$};
    \node[hasse] (1a) at (0,0){};
    \node[hasse] (2a) at (0,1.5){};
    \node[hasse] (3a) at (0,3){};
    \node (ab3) at (0.6,1.5) {\tiny$\left[13,3\right]$};
    \node (ab3) at (0.6,3) {\tiny$\left[15,1\right]$};
    \draw[-] (1a)--(2a)node[midway, right]{$D_6$}--(3a) node[midway, right]{$D_8$};
    \end{tikzpicture}};
\end{tikzpicture}}
\label{quiv:B2_8S_Mag_FC}
\end{equation}

\begin{align}
    \hscz\left(\ref{quiv:B2_8S_Mag_FC}\right)&=1+56t^{2}+3604t^{4}+135303t^{6}+3695140t^{8}+77072320t^{10}+\cdots\\
    \hsczh\left(\ref{quiv:B2_8S_Mag_FC}\right)&=64t^{2}+3520t^{4}+135744t^{6}+3692480t^{8}+77083136t^{10}+\cdots\\
    \hsc\left(\ref{quiv:B2_8S_Mag_FC}\right)&=\begin{aligned}1+&120t^{2}+7124t^{4}+271047t^{6}+7387620t^{8}\\+&154155456t^{10}+\cdots\rightarrow\overline{n.min.\sorm(16)}\end{aligned},\\
    \hsh\left[\ref{quiv:B2_8S_Mag_FC}\right]&=\frac{(1-t^{24})(1-t^{28})}{(1-t^{4})(1-t^{8})^{2}(1-t^{10})(1-t^{12})(1-t^{14})}.\;
\end{align}
At infinite coupling the magnetic quiver becomes that given below in \eqref{quiv:B2_8S_Mag_IC}. Note that in line with the results of \cite{Bourget:2020gzi}, the Coulomb branch is expected to have the closure of the minimal nilpotent orbit of $\mathfrak{e}_{7}$ as its top slice. However, the $\sprm(4)$ gauge node in this theory has negative balance and at present canot be reliably calculated. Hence, the true identity of this theory's Coulomb branch remains unclear. So too is the Higgs branch unclear -- the $\sorm(10)$ gauge node is technically too computationally intensive for present techniques. Note that the $S_2$ outer automorphism present at finite coupling has now vanished.
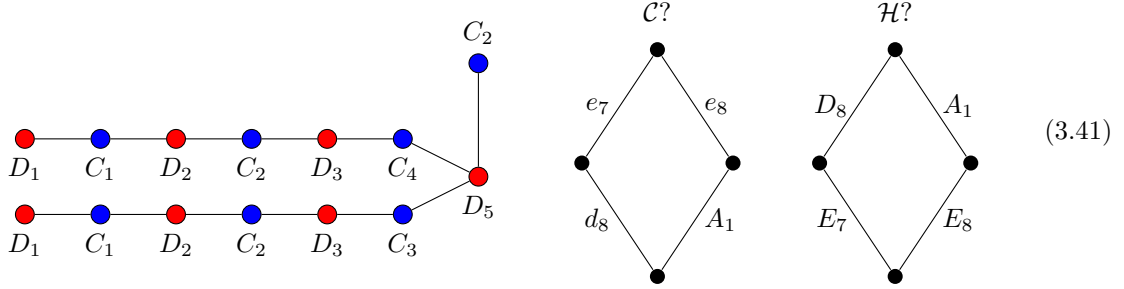
\begin{equation}
\raisebox{-0.5\height}{\begin{tikzpicture}
\node (a) at (0,0) {\begin{tikzpicture}
    \node[gauger, label=below:$D_1$] (1) at (0,0){};
    \node[gaugeb, label=below:$C_1$] (2) at (1,0){};
    \node[gauger, label=below:$D_2$] (3) at (2,0){};
    \node[gaugeb, label=below:$C_2$] (4) at (3,0){};
    \node[gauger, label=below:$D_3$] (5) at (4,0){};
    \node[gaugeb, label=below:$C_4$] (6) at (5,0){};
    \node[gauger, label=below:$D_5$] (7) at (6,-0.5){};
    \node[gaugeb, label=below:$C_3$] (8) at (5,-1){};
    \node[gauger, label=below:$D_3$] (9) at (4,-1){};
    \node[gaugeb, label=below:$C_2$] (10) at (3,-1){};
    \node[gauger, label=below:$D_2$] (11) at (2,-1){};
    \node[gaugeb, label=below:$C_1$] (12) at (1,-1){};
    \node[gauger, label=below:$D_1$] (13) at (0,-1){};
    \node[gaugeb, label=above:$C_2$] (14) at (6,1){};
    \draw[-] (1)--(2)--(3)--(4)--(5)--(6)--(7)--(8)--(9)--(10)--(11)--(12)--(13) (7)--(14);
\end{tikzpicture}};
\node (b) at (5.4,0) {\begin{tikzpicture}
    \node (0) at (0,3.5) {$\mathcal{C}?$};
    \node[hasse] (1) at (0,0){};
    \node[hasse] (2a) at (-1,1.5){};
    \node[hasse] (2b) at (1,1.5){};
    \node[hasse] (3) at (0,3){};
    \draw[-] (1)--(2a)node[midway, left]{$d_8$}--(3) node[midway, left]{$e_7$};
     \draw[-] (1)--(2b)node[midway, right]{$A_1$}--(3) node[midway, right]{$e_8$};
\end{tikzpicture}};
\node (3) at (8.5,0) {\begin{tikzpicture}
    \node (0) at (0,3.5) {$\mathcal{H}?$};
    \node[hasse] (1) at (0,0){};
    \node[hasse] (2a) at (-1,1.5){};
    \node[hasse] (2b) at (1,1.5){};
    \node[hasse] (3) at (0,3){};
    \draw[-] (1)--(2a)node[midway, left]{$E_7$}--(3) node[midway, left]{$D_8$};
     \draw[-] (1)--(2b)node[midway, right]{$E_8$}--(3) node[midway, right]{$A_1$};
    \end{tikzpicture}};
\end{tikzpicture}}
\label{quiv:B2_8S_Mag_IC}
\end{equation}
This is weak-weak dual to the theory in Table 8 of \cite{Hanany:2025ctg} given below alongside its Coulomb branch HWG.
\begin{equation}
    \raisebox{-0.5\height}{\begin{tikzpicture}
    \node[gauge, label=below:$1$] (0) at (0,0){};
    \node[gauge, label=below:$2$] (1) at (1,0){};
    \node[gauge, label=below:$3$] (2) at (2,0){};
    \node[gauge, label=below:$4$] (3) at (3,0){};
    \node[gauge, label=below:$5$] (4) at (4,0){};
    \node[gauge, label=below:$6$] (5) at (5,0){};
    \node[gauge, label=below:$4$] (6) at (6,0){};
    \node[gauge, label=below:$2$] (7) at (7,0){};
    \node[gauge, label=below:$1$] (8) at (8,0){};
    \node[gauge, label=left:$3$] (9) at (5,1){};
    \draw[-] (0)--(1)--(2)--(3)--(4)--(5)--(6)--(7)--(8) (5)--(9);
    \node at (11,0.3) {$\pe\left[\begin{aligned}&\left(\mu_2+\nu^{2}\right)t^{2}+\nu\mu_8 t^3\\&+\left(\mu_4+1\right) t^4+\nu\mu_8 t^5\\& +\mu_6t^6+\mu_8t^8-\nu^2\mu_8^2t^{10}\end{aligned}\right]$};
\end{tikzpicture}}
\end{equation}
\subsection{Rank-3 Theories}
\label{sec:rank3theories}
The set of $\spin(N)$ gauge theories admitting a five-dimensional UV fixed-point at rank-3 is limited to a handful of examples. This section considers $\spin(6)$ gauge theories with spinor matter at finite coupling (infinite-coupling quivers were studied in \cite{Akhond:2022jts}) and $\spin(7)$ theories with spinor matter at both finite and infinite coupling.
\subsubsection{$\spin(6)$} Previous work on $\spin(6)$ gauge theory at infinite coupling \cite{Akhond:2022jts}, as well as the natural restrcitions on matter content that arise from five-dimensional field theories in the UV, lead to only three different examples of $\spin(6)$ magnetic quivers being considered in this section. Specifically, these are five, six and eight spinors. Examples with less matter lead to incomplete Higgsing (and familiar magnetic quivers), while adding more matter leads to the UV fixed point entering a decompactification limit. Note that seven spinors is not possible since the brane system does not have a massless limit. Despite these restrictions, the magnetic quivers in this section uncover new examples of unframed orthosymplectic 3d $\mathcal{N}=4$ theories whose Coulomb/Higgs branches realise orbit closures and S\l odowy intersections in the nilpotent cones of classical Lie algebras. These claims are checked directly against explicit monopole formula and Weyl integration calculations, and their associated Hasse diagrams provide a clear structure for conjectures about the Coulomb branch quiver subtraction procedure for these theories.

\noindent\textbf{Five Spinors  } As shown in \cite{Bourget:2019rtl}, the Higgs branch of this electric theory has two cones, the baryonic cone and the mesonic cone. The baryonic cone has the stratification $a_4-A_2$ and is accordingly of dimension 5. This is given by the magnetic quiver on the left-hand side of \eqref{quiv:spin6_5spinors}. The mesonic cone, which has stratification $a_4-a_2$ and is of dimension 6, is given by the magnetic quiver on the right-hand side of \eqref{quiv:spin6_5spinors}. As is described in more detail below, this constructs a union of S\l odowy intersections in the nilpotent cone of $\mathfrak{sl}_8$.
\begin{equation}
\raisebox{-0.5\height}{\begin{tikzpicture}
\node (a) at (0,0) {\begin{tikzpicture}
    \node (0) at (1,-1) {$\mathrm{Baryonic}\;\mathrm{Cone}$};
    \node[gauger, label=below:$D_1$] (1) at (0,0){};
    \node[gaugeb, label=below:$C_1$] (2) at (1,0){};
    \node[gauger, label=below:$D_1$] (3) at (2,0){};
    \node[gauger, label=left:$D_1$] (4) at (1,1){};
    \node[gaugeb, label=left:$C_1$] (5) at (1,2){};
    \draw[-] (1)--(2)--(3) (2)--(4) (4)--(5);
    \draw[transform canvas={xshift=1.3pt}] (4)--(5);
    \draw[transform canvas={xshift=-1.3pt}] (4)--(5);
\end{tikzpicture}};
\node (b) at (2.25,-0.5) {\begin{tikzpicture}
\node at (0,0){\Large $\cup$};
\end{tikzpicture}};
\node (c) at (5,0) {\begin{tikzpicture}
    \node (0) at (1.75,-1) {$\mathrm{Mesonic}\;\mathrm{Cone}$};
    \node[gauger, label=below:$D_1$] (1) at (0,0){};
    \node[gaugeb, label=below:$C_1$] (2) at (1,0){};
    \node[gauger, label=below:$D_2$] (3) at (2,0){};
    \node[gaugeb, label=below:$C_1$] (4) at (3,0){};
    \node[gaugeBig,label=above:$1$] (5) at (2.5,1){};
    \draw[-] (1)--(2)--(3)--(4)--(5)--(3);
\end{tikzpicture}};
\node (c) at (10,0) {\begin{tikzpicture}
    \node (0) at (0,3.5) {$\mathcal{C}$};
    \node (0) at (-1,3.4) {$\mathrm{B}$};
    \node (0) at (1,3.4) {$\mathrm{M}$};
    \node[hasse] (1a) at (0,0){};
    \node[hasse] (2a) at (0,1.5){};
    \node[hasse] (3a) at (-1,3){};
    \node[hasse] (4a) at (1,3){};
\draw[-] (1a)--(2a)node[midway, left]{$a_4$}--(3a) node[midway, left]{$A_2$};
\draw[-] (2a)--(4a) node[midway, right]{$a_2$};
\node (aa1) at (0,-0.5) {$\mathcal{N}^{A_7}$};
\node (aa3) at (0.5,0) {\tiny $\left[3,1^{5}\right]$};
\node (aa3) at (0.6,1.5) {\tiny $\left[3,2,1^{3}\right]$};
\node (aa3) at (1.6,3) {\tiny $\left[3,2^2,1\right]$};
\node (aa3) at (-1.5,3) {\tiny $\left[4,1^{4}\right]$};
\end{tikzpicture}};
\end{tikzpicture}}
\label{quiv:spin6_5spinors}
\end{equation}
Consider the magnetic quiver on the left-hand side of \eqref{quiv:spin6_5spinors}, given in \eqref{quiv:D3_5S_Mag_FC_i}. The Coulomb branch of this theory realises the S\l odowy intersection in $\mathfrak{sl}_8$ consisting of the orbits $\left[3,1^5\right]$, $\left[3,2,1^3\right]$ and $\left[4,1^4\right]$. As expected by Lusztig-Spaltenstein duality, the Higgs branch constructs the S\l owody intersection consisting of the orbits $\left[5,1^3\right]$, $\left[5,2,1\right]$ and $\left[6,1^2\right]$. A conjectural quiver subtraction pattern is given for the Coulomb branch in \eqref{quiv:D3_5S_Mag_FC_i} that appears to realise this stratification, although the presence of the charge 2 hypermultiplet in the lower slice is broadly unmotivated outside of the brane context. Note that this magnetic quiver has an $S_2$ outer automorphism symmetry permuting the two $\sorm(2)$ gauge nodes connected to the central $\sprm(1)$. This quiver's unrefined Coulomb and Higgs branch Hilbert series are given explicitly in \eqref{hs:D3_5S_Mag_FC_i}, \eqref{hs:D3_5S_Mag_FC_ii}, \eqref{hs:D3_5S_Mag_FC_iii}, and \eqref{hs:D3_5S_Mag_FC_iv}.
\begin{equation}
\raisebox{-0.5\height}{\begin{tikzpicture}
\node (a) at (0,0) {\begin{tikzpicture}
    \node[gauger, label=below:$D_1$] (1) at (0,0){};
    \node[gaugeb, label=below:$C_1$] (2) at (1,0){};
    \node[gauger, label=below:$D_1$] (3) at (2,0){};
    \node[gauger, label=left:$D_1$] (4) at (1,1){};
    \node[gaugeb, label=left:$C_1$] (5) at (1,2){};
    \draw[-] (1)--(2)--(3) (2)--(4) (4)--(5);
    \draw[transform canvas={xshift=1.3pt}] (4)--(5);
    \draw[transform canvas={xshift=-1.3pt}] (4)--(5);
\end{tikzpicture}};
\node (b) at (5.5,0) {\begin{tikzpicture}
\node (0) at (0,3.5) {$\mathcal{C}$};
\node (aa1) at (0,-0.5) {$\mathcal{N}^{A_7}$};
\node (aa2) at (0.5,3) {\tiny $\left[4,1^{4}\right]$};
\node (aa3) at (0.6,1.5) {\tiny $\left[3,2,1^{3}\right]$};
\node (aa3) at (0.5,0) {\tiny $\left[3,1^{5}\right]$};
\node[hasse] (1a) at (0,0){};
\node[hasse] (2a) at (0,1.5){};
\node[hasse] (3a) at (0,3){};
\draw[-] (1a)--(2a)node[midway, right]{$a_4$}--(3a) node[midway, right]{$A_2$};

\node (c) at (-1,2.25) {\scalebox{0.5}{\begin{tikzpicture}
    \node[gaugeBig, label=below:$1$] (1b) at (1,0){};
    \node[gaugeBig, label=below:$1$] (2b) at (2,0){};
    \draw[transform canvas={yshift=1.3pt}](1b)--(2b);
    \draw[transform canvas={yshift=-1.3pt}](1b)--(2b);
    \draw[-] (1b)--(2b);
    \end{tikzpicture}}};

\node (c) at (-1,0.75) {\scalebox{0.5}{\begin{tikzpicture}
\node[gaugeBig, label=below:$1$] (1) at (0,0){};
    \node[gaugeb, label=below:$C_1$] (2) at (1,0){};
    \node[gauger, label=below:$D_1$] (3) at (2,-0.5){};
    \node[gauger, label=below:$D_1$] (4) at (2,0.5){};
    \node[flavour, label=below:$\urm(1)$] (5) at (-1,0){};
    \draw(1)--(2)--(3) (2)--(4);
    \draw[decoration = {zigzag,segment length = 3mm, amplitude = 1mm},decorate] (1)--(5);
\end{tikzpicture}}};

\end{tikzpicture}};
\node (3) at (9,0) {\begin{tikzpicture}
\node (0) at (0,3.5) {$\mathcal{H}$};
\node (aa1) at (0,-0.5) {$\mathcal{N}^{A_7}$};
\node (aa2) at (0.5,3) {\tiny $\left[6,1^{2}\right]$};
\node (aa3) at (0.6,1.5) {\tiny $\left[5,2,1\right]$};
\node (aa3) at (0.5,0) {\tiny $\left[5,1^{3}\right]$};
\node[hasse] (1a) at (0,0){};
\node[hasse] (2a) at (0,1.5){};
\node[hasse] (3a) at (0,3){};
\draw[-] (1a)--(2a)node[midway, right]{$a_2$}--(3a) node[midway, right]{$A_4$};
\end{tikzpicture}};
\end{tikzpicture}}
\label{quiv:D3_5S_Mag_FC_i}
\end{equation}
\begin{align}
\hsz\left[\ref{quiv:D3_5S_Mag_FC_i}\right]&=\frac{\left(\begin{aligned}1&+12t^{2}+63t^{4}+204t^{6}+550t^{8}+1094t^{10}+1906t^{12}\\+&2707t^{14}+3432t^{16}+3596t^{28}+\cdots+t^{36}\end{aligned}\right)}{(1-t^{2})^{5}(1-t^{4})(1-t^{8})^{4}}\label{hs:D3_5S_Mag_FC_i}\\
\hszh\left[\ref{quiv:D3_5S_Mag_FC_i}\right]&=\frac{8t^{2}\left(\begin{aligned}1&+7t^{2}+27t^{4}+67t^{6}+142t^{8}+236t^{10}\\+&344t^{12}+418t^{14}+458t^{16}+\cdots+t^{32}\end{aligned}\right)}{(1-t^{2})^{5}(1-t^{4})(1-t^{8})^{4}}\label{hs:D3_5S_Mag_FC_ii}\\
\hs\left[\ref{quiv:D3_5S_Mag_FC_i}\right]&=\frac{1+20t^{2}+115t^{4}+340t^{6}+620t^{8}+750t^{10}+\cdots+t^{20}}{(1-t^{2})^{5}(1-t^{4})^{5}}\label{hs:D3_5S_Mag_FC_iii}\\
\hsh\left[\ref{quiv:D3_5S_Mag_FC_i}\right]&=\frac{(1+t^{2})(1+5t^{2}+7t^{4}+14t^{6}+10t^{8}+14t^{10}+7t^{12}+5t^{14}+t^{16})}{(1-t^{2})^{3}(1-t^{6})^{3}}\label{hs:D3_5S_Mag_FC_iv}
\end{align}
The rightmost cone of \eqref{quiv:spin6_5spinors} is analysed in more detail below in \eqref{quiv:D3_5S_Mag_FC_ii}. The Coulomb branch of this theory is the S\l odowy intersection in the $\mathfrak{sl}_8$ nilcone consisting of the orbits $\left[3,1^5\right]$, $\left[3,2,1^3\right]$, and $\left[3,2^2,1\right]$. A conjectural Coulomb branch quiver subtraction pattern realising the stratification is given below in \eqref{quiv:D3_5S_Mag_FC_ii} -- as in \eqref{quiv:D3_5S_Mag_FC_i}, the presence of the charge 2 hypermultiplet is unclear without recourse to the brane system. The Higgs branch of this theory, which is the S\l odowy intersection consisting of the orbits $\left[5,1^3\right]$, $\left[5,2,1\right]$, and $\left[6,1^2\right]$, agrees with the expectation from Lusztig-Spaltenstein duality. Note that this theory does not possess an outer autmorphism symmetry. The Coulomb and Higgs branches are given explicitly in \eqref{hs:D3_5S_Mag_FC_ii_I}, \eqref{hs:D3_5S_Mag_FC_ii_II}, \eqref{hs:D3_5S_Mag_FC_ii_III}, and \eqref{hs:D3_5S_Mag_FC_ii_IV}. Note that \cite{Akhond:2022jts} gives unitary IR dual theories for those considered above. As expected, the Hilbert series of the moduli spaces agree.
\begin{equation}
\raisebox{-0.5\height}{\begin{tikzpicture}
\node (a) at (0,0) {\begin{tikzpicture}
    \node[gauger, label=below:$D_1$] (1) at (0,0){};
    \node[gaugeb, label=below:$C_1$] (2) at (1,0){};
    \node[gauger, label=below:$D_2$] (3) at (2,0){};
    \node[gaugeb, label=below:$C_1$] (4) at (3,0){};
    \node[gaugeBig,label=above:$1$] (5) at (2.5,1){};
    \draw[-] (1)--(2)--(3)--(4)--(5)--(3);
\end{tikzpicture}};
\node (b) at (5.5,0) {\begin{tikzpicture}
\node (0) at (0,3.5) {$\mathcal{C}$};
\node (aa1) at (0,-0.5) {$\mathcal{N}^{A_7}$};
\node (aa2) at (0.6,3) {\tiny $\left[3,2^{2},1\right]$};
\node (aa3) at (0.6,1.5) {\tiny $\left[3,2,1^3\right]$};
\node (aa3) at (0.5,0) {\tiny $\left[3,1^{5}\right]$};
\node[hasse] (1a) at (0,0){};
\node[hasse] (2a) at (0,1.5){};
\node[hasse] (3a) at (0,3){};
\draw[-] (1a)--(2a)node[midway, right]{$a_4$}--(3a) node[midway, right]{$a_2$};

\node (c) at (-1,2.25) {\scalebox{0.5}{\begin{tikzpicture}
    \node[gaugeBig, label=below:$1$] (1a) at (0,0){};
    \node[gaugeBig, label=below:$1$] (1b) at (1,0){};
    \node[gaugeBig, label=above:$1$] (1c) at (0.5,1){};
    \draw[-] (1a)--(1b)--(1c)--(1a);
    \end{tikzpicture}}};

\node (c) at (-1,0.75) {\scalebox{0.5}{\begin{tikzpicture}
\node[gaugeBig, label=below:$1$] (1) at (0,0){};
    \node[gaugeb, label=below:$C_1$] (2) at (1,0){};
    \node[gauger, label=below:$D_1$] (3) at (2,-0.5){};
    \node[gauger, label=below:$D_1$] (4) at (2,0.5){};
    \node[flavour, label=below:$\urm(1)$] (5) at (-1,0){};
    \draw(1)--(2)--(3) (2)--(4);
    \draw[decoration = {zigzag,segment length = 3mm, amplitude = 1mm},decorate] (1)--(5);
\end{tikzpicture}}};

\end{tikzpicture}};
\node (3) at (9,0) {\begin{tikzpicture}
\node (0) at (0,3.5) {$\mathcal{H}$};
\node (aa1) at (0,-0.5) {$\mathcal{N}^{A_7}$};
\node (aa2) at (0.5,3) {\tiny $\left[6,1^2\right]$};
\node (aa3) at (0.6,1.5) {\tiny $\left[5,2,1\right]$};
\node (aa3) at (0.5,0) {\tiny $\left[5,1^{3}\right]$};
\node[hasse] (1a) at (0,0){};
\node[hasse] (2a) at (0,1.5){};
\node[hasse] (3a) at (0,3){};
\draw[-] (1a)--(2a)node[midway, right]{$A_2$}--(3a) node[midway, right]{$A_4$};
\end{tikzpicture}};
\end{tikzpicture}}
\label{quiv:D3_5S_Mag_FC_ii}
\end{equation}
\begin{align}
\hsz\left[\ref{quiv:D3_5S_Mag_FC_ii}\right]&=\frac{1 + 8 t^2 + 51 t^4 + 192 t^6 + 380 t^8 + 496 t^{10} + 380 t^{12} + 
 192 t^{14} + 51 t^{16} + 8 t^{18} + t^{20}}{(1-t^{2})^{8}(1-t^{4})^{4}}\label{hs:D3_5S_Mag_FC_ii_I}\\
\hszh\left[\ref{quiv:D3_5S_Mag_FC_ii}\right]&=\frac{8t^{2}\left(1 + 7 t^2 + 23 t^4 + 49 t^6 + 60 t^8 + 49 t^{10} + 23 t^{12} + 
 7 t^{14} + t^{16}\right)}{(1-t^{2})^{8}(1-t^{4})^{4}}\label{hs:D3_5S_Mag_FC_ii_II}\\
\hs\left[\ref{quiv:D3_5S_Mag_FC_ii}\right]&=\frac{1 + 12 t^2 + 53 t^4 + 88 t^6 + 53 t^8 + 12 t^{10} + t^{12}}{(1-t^{2})^{12}}\label{hs:D3_5S_Mag_FC_ii_III}\\
\hsh\left[\ref{quiv:D3_5S_Mag_FC_ii}\right]&=\frac{(1-t^{8})(1-t^{10})}{(1-t^{2})(1-t^{3})^{2}(1-t^{4})(1-t^{5})^{2}}\label{hs:D3_5S_Mag_FC_ii_IV}
\end{align}
\textbf{Six Spinors  } The theory with six spinors is also a union of two cones, although in this case it morally follows from the fact that the theory Higgses to $\sprm(1)$ with two flavours, well-known to have a union of two $A_1$ cones on its Higgs branch. As mentioned in \cite{Bourget:2019rtl}, these two cones can be termed baryonic and mesonic interchangeably. Analysis of the brane web finds two different magnetic quivers, given below in \eqref{quiv:spin6_6spinors}. These construct the union of S\l odowy intersections in the nilpotent cone of $\mathfrak{sl}_6$.
\begin{equation}
\raisebox{-0.5\height}{\begin{tikzpicture}
\node (a) at (0,0) {\scalebox{0.75}{\begin{tikzpicture}
    \node (0) at (2,-1) {$\mathrm{Baryonic}\;\mathrm{Cone}$};
    \node[gauger, label=below:$D_1$] (1) at (0,0){};
    \node[gaugeb, label=below:$C_1$] (2) at (1,0){};
    \node[gauger, label=below:$D_2$] (3) at (2,0){};
    \node[gaugeb, label=below:$C_1$] (4) at (3,0){};
    \node[gauger, label=below:$D_1$] (5) at (4,0){};
    \node[gaugeb, label=above:$C_1$] (6) at (1.5,1){};
    \node[gauger, label=above:$D_1$] (7) at (2.5,1){};
    \node[gaugeb, label=above:$C_1$] (8) at (3.5,1){};
    \draw[-] (1)--(2)--(3)--(4)--(5) (3)--(6)--(7)--(8)--(4);
\end{tikzpicture}}};
\node (b) at (2.5,-0.25) {\scalebox{0.75}{\begin{tikzpicture}
\node at (0,0){\Large $\cup$};
\end{tikzpicture}}};
\node (c) at (5,0.25) {\scalebox{0.75}{\begin{tikzpicture}
    \node (0) at (2,-1) {$\mathrm{Mesonic}\;\mathrm{Cone}$};
    \node[gauger, label=below:$D_1$] (1) at (0,0){};
    \node[gaugeb, label=below:$C_1$] (2) at (1,0){};
    \node[gauger, label=below:$D_2$] (3) at (2,0){};
    \node[gaugeb, label=below:$C_2$] (4) at (3,0){};
    \node[gauger, label=below:$D_1$] (5) at (4,0){};
    \node[gaugeb, label=left:$C_1$] (6) at (3,1){};
    \node[gauger, label=left:$D_1$] (7) at (3,2){};
    \draw[-] (1)--(2)--(3)--(4)--(5) (4)--(6)--(7);
\end{tikzpicture}}};
\node (c) at (9.5,0) {\begin{tikzpicture}
    \node (0) at (0,5) {$\mathcal{C}$};
    \node (0) at (-1.3,4.9) {$\mathrm{B}$};
    \node (0) at (1.3,4.9) {$\mathrm{M}$};
    \node[hasse] (1a) at (0,0){};
    \node[hasse] (2a) at (0,1.5){};
    \node[hasse] (3a) at (0,3){};
    \node[hasse] (4a) at (-1.3,4.5){};
    \node[hasse] (5a) at (1.3,4.5){};
\draw[-] (1a)--(2a)node[midway, left]{$a_5$}--(3a) node[midway, left]{$a_3$}--(4a) node[midway, left]{$a_1$};
\draw[-] (3a)--(5a)node[midway, right]{$a_1$};
\node (ab1) at (1.8,4.5) {\tiny$\left[2^{3}\right]$};
\node (ab2) at (0.6,3) {\tiny$\left[2^2,1^{2}\right]$};
\node (ab3) at (0.5,1.5) {\tiny$\left[2,1^{4}\right]$};
\node (aa1) at (0,-0.5) {$\mathcal{N}^{A_5}$};
\end{tikzpicture}};
\end{tikzpicture}}
\label{quiv:spin6_6spinors}
\end{equation}
Considering the magnetic quiver on the left of \eqref{quiv:spin6_6spinors} in more detail, \eqref{quiv:D3_6S_Mag_FC_I} shows that its Coulomb branch appears to support the subtraction pattern identified in the Hasse diagram. Similar to the example with five spinors, the choice made to rebalance the quiver with a $\urm(1)$ gauge node for the lowest $a_5$ slice is unclear without a supporting brane system. The Higgs branch of this theory agrees with expectation from inversion. Following Tables 3 and 4 of \cite{Akhond:2022jts}, the Coulomb and Higgs branch Hilbert series is given in \eqref{hs:D3_6S_Mag_FC_I_i}, \eqref{hs:D3_6S_Mag_FC_I_ii}, and \eqref{hs:D3_6S_Mag_FC_I_iii} and \eqref{hs:D3_6S_Mag_FC_I_iv}.
\begin{equation}
\raisebox{-0.5\height}{\begin{tikzpicture}
\node (a) at (0,0) {\begin{tikzpicture}
    \node[gauger, label=below:$D_1$] (1) at (0,0){};
    \node[gaugeb, label=below:$C_1$] (2) at (1,0){};
    \node[gauger, label=below:$D_2$] (3) at (2,0){};
    \node[gaugeb, label=below:$C_1$] (4) at (3,0){};
    \node[gauger, label=below:$D_1$] (5) at (4,0){};
    \node[gaugeb, label=above:$C_1$] (6) at (1.5,1){};
    \node[gauger, label=above:$D_1$] (7) at (2.5,1){};
    \node[gaugeb, label=above:$C_1$] (8) at (3.5,1){};
    \draw[-] (1)--(2)--(3)--(4)--(5) (3)--(6)--(7)--(8)--(4);
\end{tikzpicture}};
\node (b) at (5.5,0) {\begin{tikzpicture}
\node (0) at (0,5) {$\mathcal{C}$};
\node[hasse] (1a) at (0,0){};
\node[hasse] (2a) at (0,1.5){};
\node[hasse] (3a) at (0,3){};
\node[hasse] (4a) at (0,4.5){};
\draw[-] (1a)--(2a)node[midway, right]{$a_5$}--(3a) node[midway, right]{$a_3$}--(4a) node[midway, right]{$a_1$};

\node (c) at (-1,3.75) {\scalebox{0.5}{\begin{tikzpicture}
    \node[gaugeBig, label=below:$1$] (1b) at (1,0){};
    \node[gaugeBig, label=below:$1$] (2b) at (2,0){};
    \draw[transform canvas={yshift=1.3pt}](1b)--(2b);
    \draw[transform canvas={yshift=-1.3pt}](1b)--(2b);
    \end{tikzpicture}}};
\node (d) at (-1,2.25) {\scalebox{0.5}{\begin{tikzpicture}
    \node[gauge, label=below:$1$] (1b) at (1,0){};
    \node[gauge, label=below:$1$] (2b) at (2,0){};
    \node[gauge, label=above:$1$] (3b) at (2,1){};
    \node[gauge, label=above:$1$] (4b) at (1,1){};
    \draw[-] (1b)--(2b)--(3b)--(4b)--(1b);
    \end{tikzpicture}}};
\node (d) at (-1,0.75) {\scalebox{0.5}{\begin{tikzpicture}
    \node[gaugeBig, label=below:$1$] (1) at (0,0){};
    \node[gauger, label=below:$D_1$] (2) at (1,0){};
    \node[gaugeb, label=below:$C_1$] (3) at (2,0){};
    \node[gauger, label=below:$D_1$] (4) at (3,-0.5){};
    \node[gauger, label=below:$D_1$] (5) at (3,0.5){};
    \draw(1)--(2)--(3)--(4) (3)--(5);
\end{tikzpicture}}};

\end{tikzpicture}};
\node (3) at (9,0) {\begin{tikzpicture}
\node (0) at (0,5) {$\mathcal{H}$};
\node[hasse] (1a) at (0,0){};
\node[hasse] (2a) at (0,1.5){};
\node[hasse] (3a) at (0,3){};
\node[hasse] (4a) at (0,4.5){};
\draw[-] (1a)--(2a)node[midway, right]{$A_1$}--(3a) node[midway, right]{$A_3$}--(4a) node[midway, right]{$A_5$};
\end{tikzpicture}};
\end{tikzpicture}}
\label{quiv:D3_6S_Mag_FC_I}
\end{equation}
\begin{align}
\hscz\left(\ref{quiv:D3_6S_Mag_FC_I}\right)&=1+20t^{2}+340t^{4}+3926t^{6}+33526t^{8}+224534t^{10}+\cdots,\label{hs:D3_6S_Mag_FC_I_i}\\
\hsczh\left(\ref{quiv:D3_6S_Mag_FC_I}\right)&=16t^{2}+320t^{4}+3872t^{6}+33344t^{8}+224128t^{10}+\cdots,\label{hs:D3_6S_Mag_FC_I_ii}\\
\hsc\left(\ref{quiv:D3_6S_Mag_FC_I}\right)&=1+36t^{2}+660t^{4}+7798t^{6}+66870t^{8}+448662t^{10}+\cdots,\label{hs:D3_6S_Mag_FC_I_iii}\\
\hsh\left(\ref{quiv:D3_6S_Mag_FC_I}\right)&=\frac{\left(\begin{aligned}1+&3t^{2}+6t^{4}+2t^{5}+10t^{6}+4t^{7}+15t^{8}+6t^{9}\\+&21t^{10}+8t^{11}+22t^{12}+10t^{13}+25t^{14}+\cdots+t^{28}\end{aligned}\right)}{(1-t^{2})(1-t^{5})^{2}(1-t^{7})^{2}(1-t^{8})}.\;\label{hs:D3_6S_Mag_FC_I_iv}
\end{align}
The mesonic rightmost magnetic quiver of \eqref{quiv:spin6_6spinors} is considered in more detail below in \eqref{quiv:D3_6S_Mag_FC_II}. Direct computation shows that its Coulomb branch is the closure of the $\left[2^3\right]$ orbit in the $\mathfrak{sl}_6$ nilcone, and that its Higgs branch is the S\l odowy slice associated with the $\left[3^2\right]$ orbit. A conjectural Coulomb branch subtraction pattern is given in \eqref{quiv:spin6_6spinors} and \eqref{quiv:D3_6S_Mag_FC_II}, which appears to reproduce the stratification from the nilcone. Direct computation yields the Coulomb and Higgs branch Hilbert series in \eqref{hs:D3_6S_Mag_FC_II_i}, \eqref{hs:D3_6S_Mag_FC_II_ii}, \eqref{hs:D3_6S_Mag_FC_II_iii}, and \eqref{hs:D3_6S_Mag_FC_II_iv}.
\begin{equation}
\raisebox{-0.5\height}{\begin{tikzpicture}
\node (a) at (0,0) {\begin{tikzpicture}
    \node[gauger, label=below:$D_1$] (1) at (0,0){};
    \node[gaugeb, label=below:$C_1$] (2) at (1,0){};
    \node[gauger, label=below:$D_2$] (3) at (2,0){};
    \node[gaugeb, label=below:$C_2$] (4) at (3,0){};
    \node[gauger, label=below:$D_1$] (5) at (4,0){};
    \node[gaugeb, label=left:$C_1$] (6) at (3,1){};
    \node[gauger, label=left:$D_1$] (7) at (3,2){};
    \draw[-] (1)--(2)--(3)--(4)--(5) (4)--(6)--(7);
\end{tikzpicture}};
\node (b) at (5.5,0) {\begin{tikzpicture}
\node (0) at (0,5) {$\mathcal{C}$};
\node (aa1) at (0,-0.5) {$\bar{\mathcal{O}}^{A_5}_{\left[2^3\right]}$};
\node[hasse] (1a) at (0,0){};
\node[hasse] (2a) at (0,1.5){};
\node[hasse] (3a) at (0,3){};
\node[hasse] (4a) at (0,4.5){};
\node (ab1) at (0.5,4.5) {\tiny$\left[2^{3}\right]$};
\node (ab2) at (0.5,3) {\tiny$\left[2^2,1^{2}\right]$};
\node (ab3) at (0.5,1.5) {\tiny$\left[2,1^{4}\right]$};
\draw[-] (1a)--(2a)node[midway, right]{$a_5$}--(3a) node[midway, right]{$a_3$}--(4a) node[midway, right]{$a_1$};

\node (c) at (-1,3.75) {\scalebox{0.5}{\begin{tikzpicture}
    \node[gaugeBig, label=below:$1$] (1b) at (1,0){};
    \node[gaugeBig, label=below:$1$] (2b) at (2,0){};
    \draw[transform canvas={yshift=1.3pt}](1b)--(2b);
    \draw[transform canvas={yshift=-1.3pt}](1b)--(2b);
    \end{tikzpicture}}};
\node (d) at (-1,2.25) {\scalebox{0.5}{\begin{tikzpicture}
    \node[gauge, label=below:$1$] (1b) at (1,0){};
    \node[gauge, label=below:$1$] (2b) at (2,0){};
    \node[gauge, label=above:$1$] (3b) at (2,1){};
    \node[gauge, label=above:$1$] (4b) at (1,1){};
    \draw[-] (1b)--(2b)--(3b)--(4b)--(1b);
    \end{tikzpicture}}};
\node (d) at (-1,0.75) {\scalebox{0.5}{\begin{tikzpicture}
    \node[gaugeBig, label=below:$1$] (1) at (0,0){};
    \node[gauger, label=below:$D_1$] (2) at (1,0){};
    \node[gaugeb, label=below:$C_1$] (3) at (2,0){};
    \node[gauger, label=below:$D_1$] (4) at (3,-0.5){};
    \node[gauger, label=below:$D_1$] (5) at (3,0.5){};
    \draw(1)--(2)--(3)--(4) (3)--(5);
\end{tikzpicture}}};

\end{tikzpicture}};
\node (3) at (9,0) {\begin{tikzpicture}
\node (0) at (0,5) {$\mathcal{H}$};
\node (aa1) at (0,-0.5) {$\mathcal{S}^{A_5}_{\left[3^2\right]}$};
\node (ab1) at (0.5,1.5) {\tiny$\left[4,2\right]$};
\node (ab2) at (0.5,3) {\tiny$\left[5,1\right]$};
\node (ab3) at (0.4,4.5) {\tiny$\left[6\right]$};
\node[hasse] (1a) at (0,0){};
\node[hasse] (2a) at (0,1.5){};
\node[hasse] (3a) at (0,3){};
\node[hasse] (4a) at (0,4.5){};
\draw[-] (1a)--(2a)node[midway, right]{$A_1$}--(3a) node[midway, right]{$A_3$}--(4a) node[midway, right]{$A_5$};
\end{tikzpicture}};
\end{tikzpicture}}
\label{quiv:D3_6S_Mag_FC_II}
\end{equation}
\begin{align}
\hscz\left(\ref{quiv:D3_6S_Mag_FC_II}\right)&=\frac{\left(\begin{aligned} 1 +& 8 t^2 + 145 t^4 + 755 t^6 + 3364 t^8 + 9111 t^{10}\\ +& 19335 t^{12} + 28910 t^{14} + 33926 t^{16}+\cdots+t^{32}\end{aligned}\right)}{(1-t^{2})^{11}(1-t^{4})^{7}},\label{hs:D3_6S_Mag_FC_II_i}\\
\hsczh\left(\ref{quiv:D3_6S_Mag_FC_II}\right)&=\frac{16t^{2}\left(\begin{aligned}1+&7t^2+53 t^4 + 198 t^6 + 590 t^8 + 1179 t^{10}\\ +& 1844 t^{12} + 
 2080 t^{14}+\cdots+t^{28}\end{aligned}\right)}{(1-t^{2})^{11}(1-t^{4})^{7}},\label{hs:D3_6S_Mag_FC_II_ii}\\
\hsc\left(\ref{quiv:D3_6S_Mag_FC_II}\right)&=\frac{(1+t^{2})^{3}(1+14t^{2}+72t^{4}+133t^{6}+72t^{8}+14t^{10}+t^{12})}{(1-t^{2})^{18}},\label{hs:D3_6S_Mag_FC_II_iii}\\
\hsh\left(\ref{quiv:D3_6S_Mag_FC_II}\right)&=\frac{(1-t^{8})(1-t^{10})(1-t^{12})}{(1-t^{2})^{3}(1-t^{4})^{3}(1-t^{6})^{3}}.\;\label{hs:D3_6S_Mag_FC_II_iv}
\end{align}
The two magnetic quivers for $\spin(6)$ with six spinors lead to a further new orthosymplectic theory whose moduli spaces realise a nilpotent orbit closure and S\l odowy intersection inside the $\mathfrak{sl}_6$ nilpotent cone. The theory below in \eqref{quiv:D3_6S_Mag_FC_extra} is conjectured to be the subtraction result after an $A_1$ transition in either of the two magnetic quivers above. Explicit computation of its Coulomb branch's unrefined Hilbert series in \eqref{hs:D3_6S_Mag_FC_extra_I}, \eqref{hs:D3_6S_Mag_FC_extra_II}, and \eqref{hs:D3_6S_Mag_FC_extra_III} supports the notion that it constructs the closure next-to-minimal orbit in $\mathfrak{sl}_6$. As expected, explicit computation of its Higgs branch Hilbert series in \eqref{hs:D3_6S_Mag_FC_extra_IV} returns the S\l odowy intersection $\mathcal{S}^{A_5}_{\left[4,2\right]}$. Note that this theory also does not possess an outer automorphism symmetry.
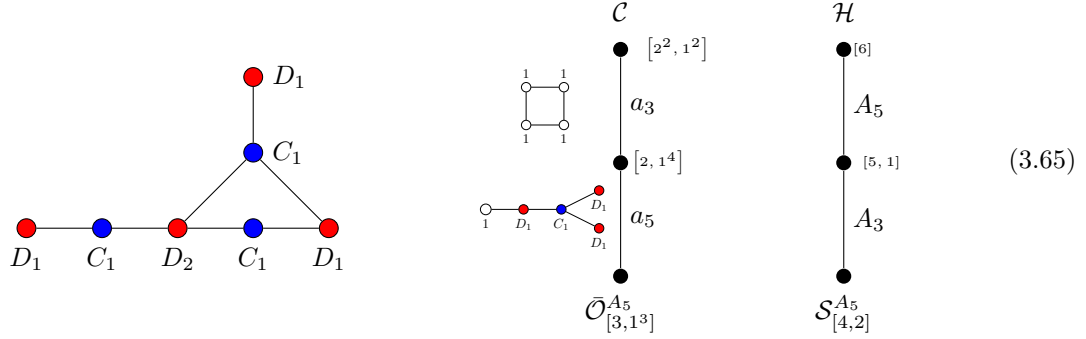
\begin{equation}
\raisebox{-0.5\height}{\begin{tikzpicture}
\node (a) at (0,0) {\begin{tikzpicture}
    \node[gauger, label=below:$D_1$] (1) at (0,0){};
    \node[gaugeb, label=below:$C_1$] (2) at (1,0){};
    \node[gauger, label=below:$D_2$] (3) at (2,0){};
    \node[gaugeb, label=below:$C_1$] (4) at (3,0){};
    \node[gauger, label=below:$D_1$] (5) at (4,0){};
    \node[gaugeb, label=right:$C_1$] (6) at (3,1){};
    \node[gauger, label=right:$D_1$] (7) at (3,2){};
    \draw[-] (1)--(2)--(3)--(4)--(5) (3)--(6)--(5) (6)--(7);
\end{tikzpicture}};
\node (b) at (5.5,0) {\begin{tikzpicture}
\node (0) at (0,3.5) {$\mathcal{C}$};
\node (aa1) at (0,-0.5) {$\bar{\mathcal{O}}^{A_5}_{\left[3,1^{3}\right]}$};
\node[hasse] (1a) at (0,0){};
\node[hasse] (2a) at (0,1.5){};
\node[hasse] (3a) at (0,3){};
\node (ab2) at (0.75,3) {\tiny$\left[2^2,1^{2}\right]$};
\node (ab3) at (0.5,1.5) {\tiny$\left[2,1^{4}\right]$};
\draw[-] (1a)--(2a)node[midway, right]{$a_5$}--(3a) node[midway, right]{$a_3$};

\node (c) at (-1,2.25) {\scalebox{0.5}{\begin{tikzpicture}
    \node[gauge, label=below:$1$] (1b) at (1,0){};
    \node[gauge, label=below:$1$] (2b) at (2,0){};
    \node[gauge, label=above:$1$] (3b) at (2,1){};
    \node[gauge, label=above:$1$] (4b) at (1,1){};
    \draw[-] (1b)--(2b)--(3b)--(4b)--(1b);
    \end{tikzpicture}}};
\node (d) at (-1,0.75) {\scalebox{0.5}{\begin{tikzpicture}
    \node[gaugeBig, label=below:$1$] (1) at (0,0){};
    \node[gauger, label=below:$D_1$] (2) at (1,0){};
    \node[gaugeb, label=below:$C_1$] (3) at (2,0){};
    \node[gauger, label=below:$D_1$] (4) at (3,-0.5){};
    \node[gauger, label=below:$D_1$] (5) at (3,0.5){};
    \draw(1)--(2)--(3)--(4) (3)--(5);
\end{tikzpicture}}};
\end{tikzpicture}};
\node (3) at (9,0) {\begin{tikzpicture}
\node (0) at (0,3.5) {$\mathcal{H}$};
\node (aa1) at (0,-0.5) {$\mathcal{S}^{A_5}_{\left[4,2\right]}$};
\node (ab1) at (0.5,1.5) {\tiny$\left[5,1\right]$};
\node (ab2) at (0.25,3) {\tiny$\left[6\right]$};
\node[hasse] (1a) at (0,0){};
\node[hasse] (2a) at (0,1.5){};
\node[hasse] (3a) at (0,3){};
\draw[-] (1a)--(2a)node[midway, right]{$A_3$}--(3a) node[midway, right]{$A_5$};
\end{tikzpicture}};
\end{tikzpicture}}
\label{quiv:D3_6S_Mag_FC_extra}
\end{equation}
\begin{align}
\hscz\left(\ref{quiv:D3_6S_Mag_FC_extra}\right)&=1 + 19 t^2 + 306 t^4 + 3202 t^6 + 24648 t^8 + 148086 t^{10}+\cdots,\label{hs:D3_6S_Mag_FC_extra_I}\\
\hsczh\left(\ref{quiv:D3_6S_Mag_FC_extra}\right)&=16 t^2 + 288 t^4 + 3168 t^6 + 24512 t^8 + 147888 t^{10}+\cdots,\label{hs:D3_6S_Mag_FC_extra_II}\\
\hsc\left(\ref{quiv:D3_6S_Mag_FC_extra}\right)&=1 + 35 t^2 + 594 t^4 + 6370 t^6 + 49160 t^8 + 295974 t^{10}+\cdots,\label{hs:D3_6S_Mag_FC_extra_III}\\
\hsh\left(\ref{quiv:D3_6S_Mag_FC_extra}\right)&=\frac{(1-t^{10})(1-t^{12})}{(1-t^2)(1-t^{4})^{3}(1-t^{6})^{2}}.\;\label{hs:D3_6S_Mag_FC_extra_IV}
\end{align}
Interestingly, there exists another unframed orthosymplectic quiver, closely resembling \eqref{quiv:D3_6S_Mag_FC_I} above whose moduli spaces are related to the \emph{other} nine quaternionic dimensional orbit in the $\mathfrak{sl}_{6}$ nilcone. This is given below in \eqref{quiv:D3_6S_Mag_FC_I_OTHER}. Direct computation in \eqref{hs:D3_6S_Mag_FC_I_OTHER_i}, \eqref{hs:D3_6S_Mag_FC_I_OTHER_ii} and \eqref{hs:D3_6S_Mag_FC_I_OTHER_iii} shows that its Coulomb branch is the closure of the $\left[3,1^{3}\right]$ nilpotent orbit, while \eqref{hs:D3_6S_Mag_FC_I_OTHER_iv} shows that its Higgs branch is the S\l odowy slice from the $\left[4,1^{2}\right]$ orbit. As with the other quivers above, it is possible to conjecture a Coulomb branch subtraction pattern for this theory that agrees with that of the nilcone. Interestingly, while it shares the same Hasse diagram with the magnetic quiver for the baryonic cone above, it does not have the $\urm(1)_{\mathrm{B}}$ symmetry factor on its Coulomb branch.
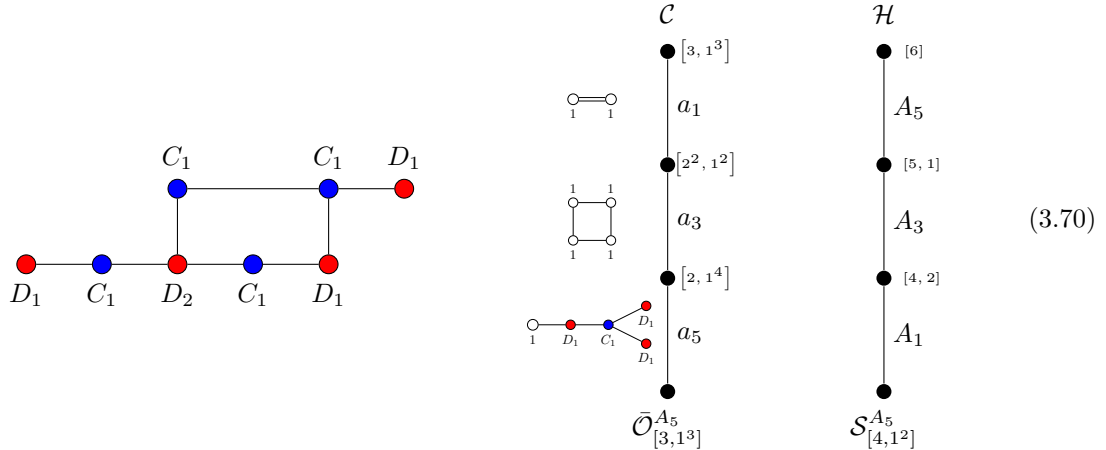
\begin{equation}
\raisebox{-0.5\height}{\begin{tikzpicture}
\node (a) at (0,0) {\begin{tikzpicture}
    \node[gauger, label=below:$D_1$] (1) at (0,0){};
    \node[gaugeb, label=below:$C_1$] (2) at (1,0){};
    \node[gauger, label=below:$D_2$] (3) at (2,0){};
    \node[gaugeb, label=below:$C_1$] (4) at (3,0){};
    \node[gauger, label=below:$D_1$] (5) at (4,0){};
    \node[gaugeb, label=above:$C_1$] (6) at (2,1){};
    \node[gaugeb, label=above:$C_1$] (7) at (4,1){};
    \node[gauger, label=above:$D_1$] (8) at (5,1){};
    \draw[-] (1)--(2)--(3)--(4)--(5) (3)--(6)--(7)--(8) (5)--(7);
\end{tikzpicture}};
\node (b) at (5.5,0) {\begin{tikzpicture}
\node (0) at (0,5) {$\mathcal{C}$};
\node (aa1) at (0,-0.5) {$\bar{\mathcal{O}}^{A_5}_{\left[3,1^{3}\right]}$};
\node[hasse] (1a) at (0,0){};
\node[hasse] (2a) at (0,1.5){};
\node[hasse] (3a) at (0,3){};
\node[hasse] (4a) at (0,4.5){};
\node (ab1) at (0.5,4.5) {\tiny$\left[3,1^{3}\right]$};
\node (ab2) at (0.5,3) {\tiny$\left[2^2,1^{2}\right]$};
\node (ab3) at (0.5,1.5) {\tiny$\left[2,1^{4}\right]$};
\draw[-] (1a)--(2a)node[midway, right]{$a_5$}--(3a) node[midway, right]{$a_3$}--(4a) node[midway, right]{$a_1$};

\node (c) at (-1,3.75) {\scalebox{0.5}{\begin{tikzpicture}
    \node[gaugeBig, label=below:$1$] (1b) at (1,0){};
    \node[gaugeBig, label=below:$1$] (2b) at (2,0){};
    \draw[transform canvas={yshift=1.3pt}](1b)--(2b);
    \draw[transform canvas={yshift=-1.3pt}](1b)--(2b);
    \end{tikzpicture}}};
\node (d) at (-1,2.25) {\scalebox{0.5}{\begin{tikzpicture}
    \node[gauge, label=below:$1$] (1b) at (1,0){};
    \node[gauge, label=below:$1$] (2b) at (2,0){};
    \node[gauge, label=above:$1$] (3b) at (2,1){};
    \node[gauge, label=above:$1$] (4b) at (1,1){};
    \draw[-] (1b)--(2b)--(3b)--(4b)--(1b);
    \end{tikzpicture}}};
\node (d) at (-1,0.75) {\scalebox{0.5}{\begin{tikzpicture}
    \node[gaugeBig, label=below:$1$] (1) at (0,0){};
    \node[gauger, label=below:$D_1$] (2) at (1,0){};
    \node[gaugeb, label=below:$C_1$] (3) at (2,0){};
    \node[gauger, label=below:$D_1$] (4) at (3,-0.5){};
    \node[gauger, label=below:$D_1$] (5) at (3,0.5){};
    \draw(1)--(2)--(3)--(4) (3)--(5);
\end{tikzpicture}}};

\end{tikzpicture}};
\node (3) at (9,0) {\begin{tikzpicture}
\node (0) at (0,5) {$\mathcal{H}$};
\node (aa1) at (0,-0.5) {$\mathcal{S}^{A_5}_{\left[4,1^{2}\right]}$};
\node (ab1) at (0.5,1.5) {\tiny$\left[4,2\right]$};
\node (ab2) at (0.5,3) {\tiny$\left[5,1\right]$};
\node (ab3) at (0.4,4.5) {\tiny$\left[6\right]$};
\node[hasse] (1a) at (0,0){};
\node[hasse] (2a) at (0,1.5){};
\node[hasse] (3a) at (0,3){};
\node[hasse] (4a) at (0,4.5){};
\draw[-] (1a)--(2a)node[midway, right]{$A_1$}--(3a) node[midway, right]{$A_3$}--(4a) node[midway, right]{$A_5$};
\end{tikzpicture}};
\end{tikzpicture}}
\label{quiv:D3_6S_Mag_FC_I_OTHER}
\end{equation}
\begin{align}
\hscz\left(\ref{quiv:D3_6S_Mag_FC_I_OTHER}\right)&=1+19t^{2}+325t^{4}+3687t^{6}+31263t^{8}+207917t^{10}+\cdots,\label{hs:D3_6S_Mag_FC_I_OTHER_i}\\
\hsczh\left(\ref{quiv:D3_6S_Mag_FC_I_OTHER}\right)&=16t^{2}+304t^{4}+3648t^{6}+31072t^{8}+207632t^{10}+\cdots,\label{hs:D3_6S_Mag_FC_I_OTHER_ii}\\
\hsc\left(\ref{quiv:D3_6S_Mag_FC_I_OTHER}\right)&=1+35t^{2}+629t^{4}+7335t^{6}+62335t^{8}+415549t^{10}+\cdots,\label{hs:D3_6S_Mag_FC_I_OTHER_iii}\\
\hsh\left(\ref{quiv:D3_6S_Mag_FC_I_OTHER}\right)&=\frac{(1-t^{10})(1-t^{12})}{(1-t^{2})^{4}(1-t^{5})^{4}}.\label{hs:D3_6S_Mag_FC_I_OTHER_iv}\;
\end{align}
\noindent\textbf{Eight Spinors  } This theory has a single cone, as shown in \cite{Bourget:2019rtl}. The integer and half-integer lattices of the Coulomb branch are calculated below in \eqref{HS:D3_8S_Mag_FC_I}, \eqref{HS:D3_8S_Mag_FC_II} and \eqref{HS:D3_8S_Mag_FC_III}, giving an $\surm(8)\times\urm(1)$ global symmetry. The Higgs branch is computed to be that given in \eqref{HS:D3_8S_Mag_FC_IV} -- a conjectural Higgs branch Hasse diagram, following the inversion rule, is given below. Note that this theory has an $S_2\times S_2$ outer automorphism. Gauging the $S_2$ that corresponds to a permutation of the two $\sprm(1)$ gauge nodes is expected to replace the $d_4$ at the top of the Coulomb branch Hasse diagram with a $\overline{n.min.B_3}$ structure, as is expected for such theories.
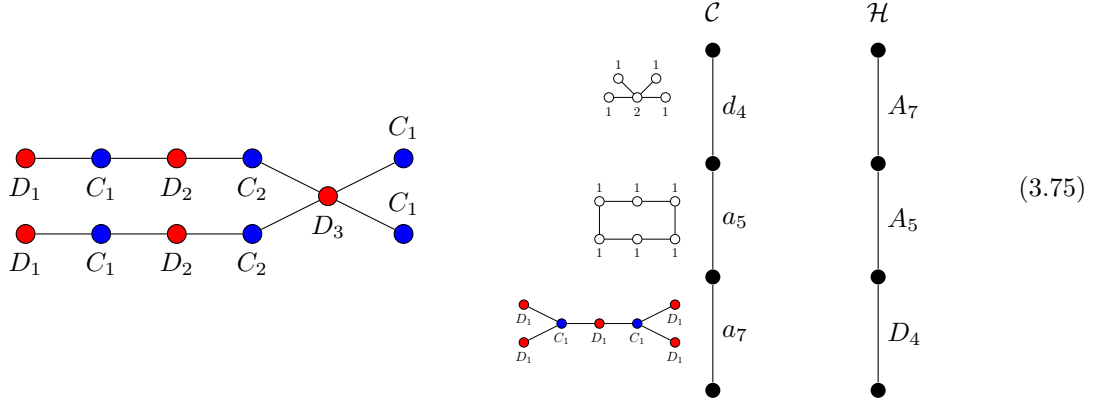
\begin{equation}
\raisebox{-0.5\height}{\begin{tikzpicture}
\node (a) at (0,0) {\begin{tikzpicture}
    \node[gauger, label=below:$D_1$] (1) at (0,0){};
    \node[gaugeb, label=below:$C_1$] (2) at (1,0){};
    \node[gauger, label=below:$D_2$] (3) at (2,0){};
    \node[gaugeb, label=below:$C_2$] (4) at (3,0){};
    \node[gauger, label=below:$D_3$] (5) at (4,-0.5){};
    \node[gaugeb, label=below:$C_2$] (6) at (3,-1){};
    \node[gauger, label=below:$D_2$] (7) at (2,-1){};
    \node[gaugeb, label=below:$C_1$] (8) at (1,-1){};
    \node[gauger, label=below:$D_1$] (9) at (0,-1){};
    \node[gaugeb, label=above:$C_1$] (10) at (5,0){};
    \node[gaugeb, label=above:$C_1$] (11) at (5,-1){};
    \draw[-] (1)--(2)--(3)--(4)--(5)--(6)--(7)--(8)--(9) (10)--(5)--(11);
\end{tikzpicture}};
\node (b) at (5.5,0) {\begin{tikzpicture}
    \node (0) at (0,5) {$\mathcal{C}$};
    \node[hasse] (1) at (0,0){};
    \node[hasse] (2) at (0,1.5){};
    \node[hasse] (3) at (0,3){};
    \node[hasse] (4) at (0,4.5){};
    \draw[-] (1)--(2)node[midway, right]{$a_{7}$}--(3)node[midway, right]{$a_5$}--(4)node[midway, right]{$d_4$};
\node (c) at (-1,4) {\scalebox{0.5}{\begin{tikzpicture}
    \node[gauge, label=below:$2$] (1b) at (0,0){};
    \node[gauge, label=below:$1$] (2b) at (0.75,0){};
    \node[gauge, label=below:$1$] (3b) at (-0.75,0){};
    \node[gauge, label=above:$1$] (4b) at (0.5,0.5){};
    \node[gauge, label=above:$1$] (5b) at (-0.5,0.5){};
    \draw[-] (1b)--(2b) (1b)--(3b) (1b)--(4b) (5b)--(1b);
    \end{tikzpicture}}};
\node (d) at (-1,2.25) {\scalebox{0.5}{\begin{tikzpicture}
    \node[gauge, label=below:$1$] (1b) at (1,0){};
    \node[gauge, label=below:$1$] (2b) at (2,0){};
    \node[gauge, label=below:$1$] (3b) at (3,0){};
    \node[gauge, label=above:$1$] (4b) at (1,1){};
    \node[gauge, label=above:$1$] (5b) at (2,1){};
    \node[gauge, label=above:$1$] (6b) at (3,1){};
    \draw[-] (1b)--(2b)--(3b)--(6b)--(5b)--(4b)--(1b);
    \end{tikzpicture}}};
\node (e) at (-1.5,0.75) {\scalebox{0.5}{\begin{tikzpicture}
    \node[gauger, label=below:$D_1$] (1) at (0,-0.5){};
    \node[gaugeb, label=below:$C_1$] (2) at (1,0){};
    \node[gauger, label=below:$D_1$] (3) at (2,0){};
    \node[gaugeb, label=below:$C_1$] (4) at (3,0){};
    \node[gauger, label=below:$D_1$] (5) at (4,-0.5){};
    \node[gauger, label=below:$D_1$] (6) at (0,0.5){};
    \node[gauger, label=below:$D_1$] (7) at (4,0.5){};
    \draw(1)--(2)--(3)--(4)--(5) (2)--(6) (4)--(7);
    \end{tikzpicture}}};
\end{tikzpicture}};
\node (3) at (9,0) {\begin{tikzpicture}
\node (0) at (0,5) {$\mathcal{H}$};
\node[hasse] (1a) at (0,0){};
\node[hasse] (2a) at (0,1.5){};
\node[hasse] (3a) at (0,3){};
\node[hasse] (4a) at (0,4.5){};
\draw[-] (1a)--(2a)node[midway, right]{$D_4$}--(3a) node[midway, right]{$A_5$}--(4a) node[midway, right]{$A_7$};
\end{tikzpicture}};
\end{tikzpicture}}
\label{quiv:D3_8S_Mag_FC}
\end{equation}
\begin{align}
\hscz\left(\ref{quiv:D3_8S_Mag_FC}\right)&=1+32t^{2}+1100t^{4}+24523t^{6}+416632t^8+5585451t^{10}+\cdots \label{HS:D3_8S_Mag_FC_I}\\
\hsczh\left(\ref{quiv:D3_8S_Mag_FC}\right)&=32t^2+1056t^4+24512t^6+415744t^8+5585088t^{10}+\cdots \label{HS:D3_8S_Mag_FC_II}\\
\hsc\left(\ref{quiv:D3_8S_Mag_FC}\right)&=1+64t^2+2156t^4+49035t^6+832376t^8+11170539t^{10}+\cdots \label{HS:D3_8S_Mag_FC_III}\\
\hsh\left(\ref{quiv:D3_8S_Mag_FC}\right)&=\frac{1-t^{2}+t^{6}+2t^{8}-t^{12}+2t^{16}+t^{18}-t^{22}+t^{24}}{(1-t^{2})(1-t^{4})^{2}(1-t^{6})^{2}(1-t^{8})}.\label{HS:D3_8S_Mag_FC_IV}\;
\end{align}
A basic observation regarding the quiver in \eqref{quiv:D3_8S_Mag_FC} is that the lower two slices in its Coulomb branch Hasse diagram occur in the nilpotent cone of $\mathfrak{sl}_{8}$. Hence by performing a $d_4$ subtraction a quiver can be derived whose Coulomb branch realises this space (and whose Higgs branch is the associated S\l odowy slice). This is given below in \eqref{quiv:D3_6S_Mag_FC_related_theory}.
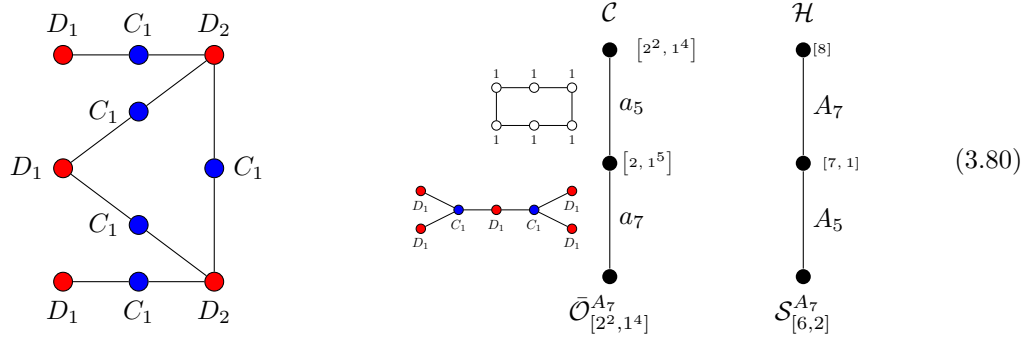
\begin{equation}
\raisebox{-0.5\height}{\begin{tikzpicture}
\node (a) at (0,0) {\begin{tikzpicture}
    \node[gauger, label=above:$D_1$] (1) at (0,1.5){};
    \node[gaugeb, label=above:$C_1$] (2) at (1,1.5){};
    \node[gauger, label=above:$D_2$] (3) at (2,1.5){};
    \node[gauger, label=below:$D_1$] (4) at (0,-1.5){};
    \node[gaugeb, label=below:$C_1$] (5) at (1,-1.5){};
    \node[gauger, label=below:$D_2$] (6) at (2,-1.5){};
    \node[gaugeb, label=left:$C_1$] (7) at (1,0.75){};
    \node[gauger, label=left:$D_1$] (8) at (0,0){};
    \node[gaugeb, label=left:$C_1$] (9) at (1,-0.75){};
    \node[gaugeb, label=right:$C_1$] (10) at (2,0){};
    \draw[-] (1)--(2)--(3)--(7)--(8)--(9)--(6)--(5)--(4) (3)--(10)--(6);

\end{tikzpicture}};
\node (b) at (5.5,0) {\begin{tikzpicture}
\node (0) at (0,3.5) {$\mathcal{C}$};
\node (aa1) at (0,-0.5) {$\bar{\mathcal{O}}^{A_7}_{\left[2^2,1^{4}\right]}$};
\node[hasse] (1a) at (0,0){};
\node[hasse] (2a) at (0,1.5){};
\node[hasse] (3a) at (0,3){};
\node (ab2) at (0.75,3) {\tiny$\left[2^2,1^{4}\right]$};
\node (ab3) at (0.5,1.5) {\tiny$\left[2,1^{5}\right]$};
\draw[-] (1a)--(2a)node[midway, right]{$a_7$}--(3a) node[midway, right]{$a_5$};

\node (d) at (-1,2.25) {\scalebox{0.5}{\begin{tikzpicture}
    \node[gauge, label=below:$1$] (1b) at (1,0){};
    \node[gauge, label=below:$1$] (2b) at (2,0){};
    \node[gauge, label=below:$1$] (3b) at (3,0){};
    \node[gauge, label=above:$1$] (4b) at (1,1){};
    \node[gauge, label=above:$1$] (5b) at (2,1){};
    \node[gauge, label=above:$1$] (6b) at (3,1){};
    \draw[-] (1b)--(2b)--(3b)--(6b)--(5b)--(4b)--(1b);
    \end{tikzpicture}}};
\node (e) at (-1.5,0.75) {\scalebox{0.5}{\begin{tikzpicture}
    \node[gauger, label=below:$D_1$] (1) at (0,-0.5){};
    \node[gaugeb, label=below:$C_1$] (2) at (1,0){};
    \node[gauger, label=below:$D_1$] (3) at (2,0){};
    \node[gaugeb, label=below:$C_1$] (4) at (3,0){};
    \node[gauger, label=below:$D_1$] (5) at (4,-0.5){};
    \node[gauger, label=below:$D_1$] (6) at (0,0.5){};
    \node[gauger, label=below:$D_1$] (7) at (4,0.5){};
    \draw(1)--(2)--(3)--(4)--(5) (2)--(6) (4)--(7);
    \end{tikzpicture}}};
\end{tikzpicture}};
\node (3) at (9,0) {\begin{tikzpicture}
\node (0) at (0,3.5) {$\mathcal{H}$};
\node (aa1) at (0,-0.5) {$\mathcal{S}^{A_7}_{\left[6,2\right]}$};
\node (ab1) at (0.5,1.5) {\tiny$\left[7,1\right]$};
\node (ab2) at (0.25,3) {\tiny$\left[8\right]$};
\node[hasse] (1a) at (0,0){};
\node[hasse] (2a) at (0,1.5){};
\node[hasse] (3a) at (0,3){};
\draw[-] (1a)--(2a)node[midway, right]{$A_5$}--(3a) node[midway, right]{$A_7$};
\end{tikzpicture}};
\end{tikzpicture}}
\label{quiv:D3_6S_Mag_FC_related_theory}
\end{equation}
\begin{align}
\hscz\left(\ref{quiv:D3_6S_Mag_FC_related_theory}\right)&=1+31t^{2}+992t^4+18671t^6+250025 t^8 + 2530760 t^{10}+\cdots\\
\hsczh\left(\ref{quiv:D3_6S_Mag_FC_related_theory}\right)&=32 t^2 + 960 t^4 + 18688 t^6 + 249600 t^8 + 2530912 t^{10}+\cdots\\
\hsc\left(\ref{quiv:D3_6S_Mag_FC_related_theory}\right)&=1 + 63 t^2 + 1952 t^4 + 37359 t^6 + 499625 t^8 + 5061672 t^{10}+\cdots\\
\hsh\left(\ref{quiv:D3_6S_Mag_FC_related_theory}\right)&=\frac{(1-t^{14})(1-t^{16})}{(1-t^{2})(1-t^{4})(1-t^{6})^{2}(1-t^{8})^{2}}\;
\end{align}
\subsubsection{$\spin(7)$}
This section considers $\spin(7)$ gauge theory with three and four spinors. As for the $\spin(6)$ examples, reducing the number of flavours leads to incomplete Higgsing (and familiar magnetic quivers), while adding more matter leads to the UV fixed point entering a decompactification limit. To the best of the authors' knowledge, none of the magnetic quivers in this section (finite or infinite coupling) are previously featured in the literature.

\noindent\textbf{Three Spinors  } The brane system for $\spin(7)$ with three spinors at finite coupling is given below in \eqref{brane:spin7_3s_FC} -- its associated magnetic quiver is presented alongside its Coulomb and Higgs branch Hasse diagrams in \eqref{quiv:spin73S_magquiv_FC}. 
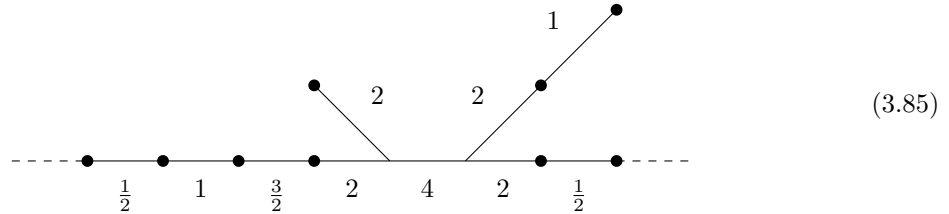
\begin{equation}
\raisebox{-0.5\height}{\begin{tikzpicture}
    \draw(0,0)--(7,0);
    \draw(4,0)--(3,1);
    \draw(5,0)--(7,2);
    \draw[dashed](-1,0)--(0,0);
    \draw[dashed](7,0)--(8,0);
    \node[7brane]at(0,0){};
    \node[7brane]at(1,0){};
    \node[7brane]at(2,0){};
    \node[7brane]at(3,0){};
    \node[7brane]at(6,0){};
    \node[7brane]at(7,0){};
    \node[7brane]at(3,1){};
    \node[7brane]at(6,1){};
    \node[7brane]at(7,2){};
    \node[label=below:{$\frac{1}{2}$}]at(0.5,0){};
    \node[label=below:{$1$}]at(1.5,0){};
    \node[label=below:{$\frac{3}{2}$}]at(2.5,0){};
    \node[label=below:{$2$}]at(3.5,0){};
    \node[label=below:{$4$}]at(4.5,0){};
    \node[label=below:{$2$}] at (5.5,0){};
    \node[label=below:{$\frac{1}{2}$}] at (6.5,0){};
    \node[label=above right:{$2$}] at (3.5,0.5){};
    \node[label=above left:{$2$}] at (5.5,0.5){};
    \node[label=above left:{$1$}] at (6.5,1.5){};
\end{tikzpicture}}
\label{brane:spin7_3s_FC}
\end{equation}
The magnetic quiver's Coulomb branch is $\bar{\mathcal{O}}^{C_3}_{\left[2^3\right]}$ and Higgs branch is straightforwardly computed as $\mathcal{S}^{B_3}_{\left[3^{2},1\right]}$, as follows from Barbasch-Vogan duality \cite{Barbasch1980TheLS,barbasch_vogan2,BARBASCH_vogan3,barbasch_vogan_4}. As is often the case with orthosymplectic quivers (and theories that realise nilpotent orbit closures of $BCD$-type more broadly), the naive inversion algorithm \cite{Grimminger:2020dmg} does not apply here. In contrast to many of the examples considered in this note, the Coulomb branch quiver subtraction pattern for this example is unclear. Nevertheless, this is a new example of an orthosymplectic quiver with Coulomb and Higgs branches realising spaces in classical nilpotent cones. Note that all nodes in the theory above have balance greater than or equal to zero and that the quiver is simply laced. It's conceivable that these quivers sit in a family of related theories.
\begin{equation}
\raisebox{-0.5\height}{\begin{tikzpicture}
\node (a) at (0,0) {\begin{tikzpicture}
    \node[gauger, label=below:$D_1$] (1) at (0,0){};
    \node[gaugeb, label=below:$C_1$] (2) at (1,0){};
    \node[gauger, label=below:$D_2$] (3) at (2,0){};
    \node[gaugeb, label=below:$C_1$] (4) at (3,0){};
    \node[gauger, label=below:$D_1$] (5) at (4,0){};
    \draw[-] (1)--(2)--(3) (4)--(5);
    \draw[transform canvas={yshift=1.3pt}](3)--(4);
    \draw[transform canvas={yshift=-1.3pt}](3)--(4);
\end{tikzpicture}}; 
\node (b) at (6,0) {\begin{tikzpicture}
\node (0) at (0,5) {$\mathcal{C}$};
    \node[hasse] (1a) at (0,0){};
    \node[hasse] (2a) at (0,1.5){};
    \node[hasse] (3a) at (0,3){};
    \node[hasse] (4a) at (0,4.5){};
    \node (ab1) at (0.6,1) {\tiny$\left[2,1^4\right]$};
    \node (ab2) at (0.6,3) {\tiny$\left[2^2,1^2\right]$};
    \node (ab3) at (0.6,4.5) {\tiny$\left[2^3\right]$};
    \draw[-] (1a)--(2a)node[midway, right]{$c_3$}--(3a) node[midway, right]{$c_2$}--(4a) node[midway, right]{$c_1$};
    \node (4) at (0,-0.5) {$\bar{\mathcal{O}}^{C_3}_{\left[2^{3}\right]}$};
\end{tikzpicture}};
\node (3) at (10.5,0) {\begin{tikzpicture}
    \node (0) at (0,3.5) {$\mathcal{H}$};
    \node (1) at (0,-0.5) {$\mathcal{S}^{B_3}_{\left[3^{2},1\right]}$};
    \node[hasse] (1a) at (0,0){};
    \node[hasse] (2a) at (0,1.5){};
    \node[hasse] (3a) at (0,3){};
    \node (ab1) at (0.4,3) {\tiny$\left[7\right]$};
    \node (ab2) at (0.6,1.5) {\tiny$\left[5,1^2\right]$};
    \draw[-] (1a)--(2a)node[midway, right]{$D_3$}--(3a) node[midway, right]{$A_5$};
    \end{tikzpicture}};
\end{tikzpicture}}
\label{quiv:spin73S_magquiv_FC}
\end{equation}
\begin{align}
    \hscz\left[\ref{quiv:spin73S_magquiv_FC}\right]=&\frac{1 + 5 t^2 + 40 t^4 + 102 t^6 + 223 t^8 + 250 t^{10}+\cdots+t^{20}}{(1-t^{2})^8(1-t^4)^4}\\
    \hsczh\left[\ref{quiv:spin73S_magquiv_FC}\right]=&\frac{8t^2(1 + 4 t^2 + 15 t^4 + 25 t^6 + 34 t^8+\cdots+t^{16})}{(1-t^{2})^8(1-t^4)^4}\\\hsc\left[\ref{quiv:spin73S_magquiv_FC}\right]=&\frac{(1+t^2)^2(1+7t^2+15t^4+7t^6+t^8)}{(1-t^2)^{12}}\rightarrow\bar{\mathcal{O}}^{C_3}_{\left[2^3\right]}\\
    \hsh\left[\ref{quiv:spin73S_magquiv_FC}\right]=&\frac{(1-t^{8})(1-t^{12})}{(1-t^{2})(1-t^{4})^{4}(1-t^{6})}
\end{align}
An example of a related quiver is given by \eqref{quiv:nminB2_random}. Although this theory is not itself a magnetic quiver for any 5d $\mathcal{N}=1$ $\spin(N)$ gauge theory, it merits a short treatment; to the authors' knowledge, it does not appear in the quiver literature. Its Coulomb branch is $\mathbb{H}^{2}\times\overline{n.min.B_2}$, as shown by the Hilbert series calculations below, and its Higgs branch is $\mathcal{S}^{B_2}_{\left[3,1^{2}\right]}$. This is again of course an example of Barbasch-Vogan duality. Note that like for \eqref{quiv:spin73S_magquiv_FC} the full stratification of the Coulomb branch of \eqref{quiv:nminB2_random} under quiver subtraction is also unclear. However, this theory is notable for offering a construction of $\overline{n.min.C_{2}}$ that does not contain any non-simple lacing -- this is a Lagrangian theory.
\begin{equation}
\raisebox{-0.5\height}{\begin{tikzpicture}
\node (a) at (0,0) {\begin{tikzpicture}
    \node[gauger, label=below:$D_1$] (1) at (0,0){};
    \node[gaugeb, label=below:$C_1$] (2) at (1,0){};
    \node[gauger, label=below:$D_2$] (3) at (2,0){};
    \node[gaugeb, label=below:$C_1$] (4) at (3,0){};
    \draw[-] (1)--(2)--(3);
    \draw[transform canvas={yshift=1.3pt}](3)--(4);
    \draw[transform canvas={yshift=-1.3pt}](3)--(4);
\end{tikzpicture}}; 
\node (b) at (6,0) {\begin{tikzpicture}
\node (0) at (0,3.5) {$\mathcal{C}$};
    \node[hasse] (1a) at (0,0){};
    \node[hasse] (2a) at (0,1.5){};
    \node[hasse] (3a) at (0,3){};
    \node (ab1) at (0.4,3) {\tiny$\left[2^2\right]$};
    \node (ab2) at (0.6,1.5) {\tiny$\left[2,1^2\right]$};
    \draw[-] (1a)--(2a)node[midway, right]{$c_2$}--(3a) node[midway, right]{$a_1$};
    \node (4) at (0,-0.5) {$\bar{\mathcal{O}}^{C_2}_{\left[2^{2}\right]}$};
\end{tikzpicture}};
\node (3) at (10.5,0) {\begin{tikzpicture}
    \node (0) at (0,2) {$\mathcal{H}$};
    \node[hasse] (1a) at (0,0){};
    \node[hasse] (2a) at (0,1.5){};
    \node (ab1) at (0.6,1.5) {\tiny$\left[3,1^2\right]$};
    \draw[-] (1a)--(2a)node[midway, right]{$A_3$};
    \node (1) at (0,-0.5) {$\mathcal{S}^{B_2}_{\left[3,1^{2}\right]}$};
    \end{tikzpicture}};
\end{tikzpicture}}
\label{quiv:nminB2_random}
\end{equation}
\begin{align}
    \hscz\left[\ref{quiv:nminB2_random}\right]=&\frac{(1+t^2)(1+9t^{2}+20t^4+9t^6+t^8)}{(1-t^{2})^{10}}\\
    \hsczh\left[\ref{quiv:nminB2_random}\right]=&\frac{4t(1+t^2)^2(1+3t^2+t^4)}{(1-t^{2})^{10}}\\
    \hsc\left[\ref{quiv:nminB2_random}\right]=&\frac{1}{(1-t)^4}\times\frac{1+4t^2+4t^4+t^6}{(1-t^2)^{6}}\rightarrow\mathbb{H}^{2}\times\bar{\mathcal{O}}^{B_2}_{\left[3,1^{2}\right]}\\
    \hsh\left[\ref{quiv:nminB2_random}\right]=&\frac{1-t^{8}}{(1-t^{2})(1-t^{4})^{2}}
\end{align}
Returning to the $\spin(7)$ theory with three spinors, the theory at infinite coupling is realized by the brane system \eqref{brane:spin7_3s_IC} below.
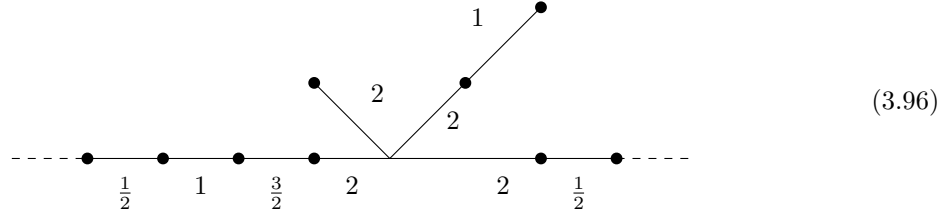
\begin{equation}
    \raisebox{-0.5\height}{\begin{tikzpicture}
        \draw(0,0)--(7,0);
        \draw(4,0)--(3,1);
        \draw(4,0)--(6,2);
        \draw[dashed](-1,0)--(0,0);
        \draw[dashed](7,0)--(8,0);
        \node[7brane]at(0,0){};
        \node[7brane]at(1,0){};
        \node[7brane]at(2,0){};
        \node[7brane]at(3,0){};
        \node[7brane]at(6,0){};
        \node[7brane]at(7,0){};
        \node[7brane]at(3,1){};
        \node[7brane]at(5,1){};
        \node[7brane]at(6,2){};
        \node[label=below:{$\frac{1}{2}$}]at(0.5,0){};
        \node[label=below:{$1$}]at(1.5,0){};
        \node[label=below:{$\frac{3}{2}$}]at(2.5,0){};
        \node[label=below:{$2$}]at(3.5,0){};
        \node[label=below:{$2$}]at(5.5,0){};
        \node[label=below:{$\frac{1}{2}$}]at(6.5,0){};
        \node[label=above right:{$2$}]at(3.5,0.5){};
        \node[label=right:{$2$}]at(4.5,0.5){};
        \node[label=above left:{$1$}]at(5.5,1.5){};
    \end{tikzpicture}}
\label{brane:spin7_3s_IC}
\end{equation}
The associated magnetic quiver is given in \eqref{quiv_spin7_3s_inf}. This theory is now unitary-orthosymplectic, and contains two hypermultiplets in the second-rank antisymmetric representation of $\urm(2)$. Note that its rank has increased by one compared to the finite-coupling quiver \eqref{quiv:spin73S_magquiv_FC} and the global symmetry is now $\sprm(3)\times\urm(1)$.
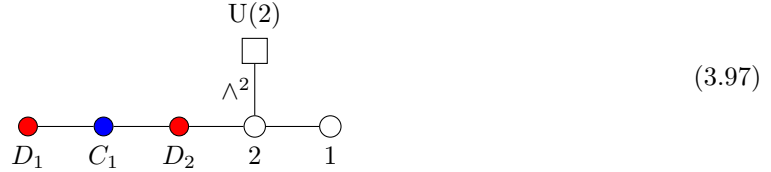
\begin{equation}
\raisebox{-0.5\height}{\begin{tikzpicture}
    \node[gauger, label=below:$D_1$] (1) at (0,0){};
    \node[gaugeb, label=below:$C_1$] (2) at (1,0){};
    \node[gauger, label=below:$D_2$] (3) at (2,0){};
    \node[gaugeBig, label=below:$2$] (4) at (3,0){};
    \node[gaugeBig, label=below:$1$] (5) at (4,0){};
    \node[flavour, label=above:$\urm(2)$] (6) at (3,1){};
    \node at (2.75,0.5) {$\wedge^2$};
    \draw[-] (1)--(2)--(3)--(4)--(5) (4)--(6);
\end{tikzpicture}}
\label{quiv_spin7_3s_inf}
\end{equation}
\begin{align}
    \hscz\left[\ref{quiv_spin7_3s_inf}\right]=&1+14t^{2}+134t^{4}+20t^{5}+908t^{6}+260t^{7}+4791t^{8}+2076t^{9}+20861t^{10}+\cdots, \label{hs:quiv_spin7_3s_inf_i}\\
    \hsczh\left[\ref{quiv_spin7_3s_inf}\right]=&8t^{2}+104t^{4}+8t^{5}+800t^{6}+200t^{7}+4456t^{8}+1816t^{9}+19928t^{10}+\cdots, \label{hs:quiv_spin7_3s_inf_ii}\\
    \hsc\left[\ref{quiv_spin7_3s_inf}\right]=&1+22t^{2}+238t^{4}+28t^{5}+1708t^{6}+460t^{7}+9247t^{8}+3892t^{9}+40789t^{10}+\cdots, \label{hs:quiv_spin7_3s_inf_iii}\\
    &\hwg\left[\hsc\left(\mathcal{Q_{\eqref{quiv_spin7_3s_inf}}}\right)\right]=\pe\left[\sum_{j=1}^{3}\mu_j^2t^{2j}+t^2+\mu_3(q+q^{-1})-\mu_3^2t^{10}\right]
    , \label{hs:quiv_spin7_3s_inf_iv}\\
    \hsh\left[\ref{quiv_spin7_3s_inf}\right]=&\frac{1+t^2-t^3+2t^6+t^7+2t^8-t^{11}+t^{12}+t^{14}}{(1-t^{3})(1-t^{4})(1-t^{5})(1-t^{6})}.\label{hs:quiv_spin7_3s_inf_v}
\end{align}
Note that $\mu_j$ are $\mathfrak{sp}(3)$ highest weight fugacities, while q is a $\mathfrak{u}(1)$ fuagcity. The Coulomb and Higgs branches of this theory are the same as those of $\sorm(7)$ with three vectors \cite{Akhond:2021ffo}. Direct comparison with $\overline{n.min.B_{4}}$ and $\overline{n.min.C_{4}}$, along with the extra $\urm(1)$ global symmetry factor, shows that the Coulomb branch above is not the Hilbert series of the closure of a nilpotent orbit of classical type \cite{Hanany:2016gbz,Cabrera:2017njm}. Note that while \eqref{hs:quiv_spin7_3s_inf_i}, \eqref{hs:quiv_spin7_3s_inf_ii}, \eqref{hs:quiv_spin7_3s_inf_iii} and \eqref{hs:quiv_spin7_3s_inf_v} are direct computations, \eqref{hs:quiv_spin7_3s_inf_iv} is conjectural.

\noindent\textbf{Four Spinors  } The brane system for $\spin(7)+4\mathrm{S}$ at finite coupling is given below in \eqref{brane:spin7_4s_FC}, alongside the two magnetic quivers and the Coulomb branch they realise in \eqref{quiv:spin7_4spinors}. 
\begin{equation}
\begin{array}{c}
    \begin{tikzpicture}
    \draw[dashed](0,0)--(-1,0);
    \draw(0,0)--(11,0);
    \draw[dashed](12,0)--(11,0);
    \node[7brane]at(0,0){};
    \node[7brane]at(1,0){};
    \node[7brane]at(2,0){};
    \node[7brane]at(3,0){};
    \node[7brane]at(6,0){};
    \node[7brane]at(7,0){};
    \node[7brane]at(8,0){};
    \node[7brane]at(9,0){};
    \node[7brane]at(10,0){};
    \node[7brane]at(11,0){};
    \node[label=below:{$\frac{1}{2}$}]at(0.5,0){};
    \node[label=below:{$1$}]at(1.5,0){};
    \node[label=below:{$\frac{3}{2}$}]at(2.5,0){};
    \node[label=below:{$2$}]at(3.5,0){};
    \node[label=below:{$4$}]at(4.5,0){};
    \node[label=below:{$4$}]at(5.5,0){};
    \node[label=below:{$\frac{5}{2}$}]at(6.5,0){};
    \node[label=below:{$2$}]at(7.5,0){};
    \node[label=below:{$\frac{3}{2}$}]at(8.5,0){};
    \node[label=below:{$1$}]at(9.5,0){};
    \node[label=below:{$\frac{1}{2}$}]at(10.5,0){};
    \draw(4,0)--(3,1);
    \draw(5,0)--(5,1);
    \node[7brane]at(3,1){};
    \node[7brane]at(5,1){};
    \node[label=right:{$2$}]at(5,.5){};
    \node[label=right:{$2$}]at(2.75,.5){};
\end{tikzpicture}
\end{array}
\label{brane:spin7_4s_FC}
\end{equation}
Importantly, this space is a union of two cones, owing to the fact that the theory Higgses to $\sprm(1)$ with two fundamental flavours, well-known to have a finite-coupling Higgs branch which is the union of two $A_1$ hyper-K\"ahler cones \cite{ferlito2016taleconeshiggsbranch,bourget2023talencones}. As such, the two magnetic quivers have identical Coulomb branch  Hasse diagrams, although direct calculation of their Coulomb branch Hilbert series shows that the spaces differ as varieties. The Coulomb branch Hilbert series of the magnetic quiver on the left of \eqref{quiv:spin7_4spinors} is given in \eqref{hs:spin74s_fc_L_i}, \eqref{hs:spin74s_fc_L_ii} and \eqref{hs:spin74s_fc_L_iii}, while that for the magnetic quiver on the right of \eqref{quiv:spin7_4spinors} is given in \eqref{hs:spin74s_fc_R_i}, \eqref{hs:spin74s_fc_R_ii} and \eqref{hs:spin74s_fc_R_iii}. As for the magnetic quiver \eqref{quiv:spin73S_magquiv_FC} for $\spin(7)$ with three spinors, the Hasse diagram for the Coulomb branches of the magnetic quivers for the four-spinor case on the right of \eqref{quiv:spin7_4spinors} is obtained through Higgsing in the electric theory. At present, quiver subtraction techniques are unable to reliably reproduce this Hasse diagram. However, several of its features can be realistically motivated by considering basic components of the quivers themselves. For each cone, the top-most $A_1$ transition appears to correspond to a $C_1-C_1$ subtraction -- in each case, the resulting rebalancing (presumably with a $C_1$ gauge node) creates an affine $A_3$ Dynkin diagram-shaped sub-quiver. Further steps realising the $\overline{n.min.\sprm(4)}$ structure of the Hasse diagram in \eqref{quiv:spin7_4spinors} are unclear.
% \begin{equation}
% \raisebox{-0.5\height}{\begin{tikzpicture}
%     \node[gauger, label=below:$D_1$] (1) at (0,0){};
%     \node[gaugeb, label=below:$C_1$] (2) at (1,0){};
%     \node[gauger, label=below:$D_2$] (3) at (2,0){};
%     \node[gaugeb, label=below:$C_1$] (4) at (3,0){};
%     \node[gauger, label=below:$D_2$] (5) at (4,0){};
%     \node[gaugeb, label=below:$C_1$] (6) at (5,0){};
%     \node[gauger, label=below:$D_1$] (7) at (6,0){};
%     \node[gaugeb, label=above:$C_1$] (8) at (2,1){};
%     \node[gaugeb, label=above:$C_1$] (9) at (4,1){};
%     \draw[-] (1)--(2)--(3)--(4)--(5)--(6)--(7) (3)--(8) (5)--(9)--(8);
% \end{tikzpicture}}
% \end{equation}
\begin{equation}
\raisebox{-0.5\height}{\begin{tikzpicture}
\node (a) at (0,0) {\scalebox{0.75}{\begin{tikzpicture}
    \node[gauger, label=below:$D_1$] (1) at (0,0){};
    \node[gaugeb, label=below:$C_1$] (2) at (1,0){};
    \node[gauger, label=below:$D_2$] (3) at (2,0){};
    \node[gaugeb, label=below:$C_2$] (4) at (3,0){};
    \node[gauger, label=below:$D_2$] (5) at (4,0){};
    \node[gaugeb, label=below:$C_1$] (6) at (5,0){};
    \node[gauger, label=below:$D_1$] (7) at (6,0){};
    \node[gaugeb, label=above:$C_1$] (8) at (3,1){};
    \draw[-] (1)--(2)--(3)--(4)--(5)--(6)--(7) (4)--(8);
\end{tikzpicture}}};
\node (b) at (3,0) {\scalebox{0.75}{\begin{tikzpicture}
\node at (0,0){\Large $\cup$};
\end{tikzpicture}}};
\node (c) at (6,0) {\scalebox{0.75}{\begin{tikzpicture}
    \node[gauger, label=below:$D_1$] (1a) at (0,0){};
    \node[gaugeb, label=below:$C_1$] (2a) at (1,0){};
    \node[gauger, label=below:$D_2$] (3a) at (2,0){};
    \node[gaugeb, label=below:$C_1$] (4a) at (3,0){};
    \node[gauger, label=below:$D_2$] (5a) at (4,0){};
    \node[gaugeb, label=below:$C_1$] (6a) at (5,0){};
    \node[gauger, label=below:$D_1$] (7a) at (6,0){};
    \node[gaugeb, label=above:$C_1$] (8a) at (2,1){};
    \node[gaugeb, label=above:$C_1$] (9a) at (4,1){};
    \draw[-] (1a)--(2a)--(3a)--(4a)--(5a)--(6a)--(7a) (3a)--(8a)--(9a)--(5a);
\end{tikzpicture}}};
\node (c) at (10,0) {\begin{tikzpicture}
\node (0) at (0,4.5) {$\mathcal{C}$};
    \node[hasse] (1a) at (0,0){};
    \node[hasse] (2a) at (0,1){};
    \node[hasse] (3a) at (0,2){};
    \node[hasse] (4a) at (0,3){};
    \node[hasse] (5a) at (1,4){};
    \node[hasse] (6a) at (-1,4){};
\draw[-] (1a)--(2a)node[midway, right]{$c_4$}--(3a) node[midway, right]{$c_3$}--(4a) node[midway, right]{$a_3$}--(5a) node[midway, right]{$A_1$} (4a)--(6a) node[midway, left]{$A_1$};
\end{tikzpicture}};
\end{tikzpicture}}
\label{quiv:spin7_4spinors}
\end{equation}
Each magnetic quiver merits closer study. Consider first the magnetic quiver on the left of \eqref{quiv:spin7_4spinors}, reproduced alongside its Coulomb and (conjectural) Higgs branch Hasse diagrams in \eqref{quiv:spin7_4spinors_i}. Although the Hilbert series of both these spaces are immediately calculable, and are given below in \eqref{hs:spin74s_fc_L_i}, \eqref{hs:spin74s_fc_L_ii}, \eqref{hs:spin74s_fc_L_iii} and \eqref{hs:spin74s_fc_L_iv}, the Higgs branch Hasse diagram is an educated guess from various pieces of evidence. Firstly, that the lowest slice is $A_1$ is supported by the fact that \eqref{hs:spin74s_fc_L_iv} shows that the Higgs branch itself has an $\mathfrak{sl}_2$ global symmetry algebra. Secondly, jumping ahead slightly, the intersection of the two cones, given in \eqref{quiv:spin7_4spinors_intersection}, has a Higgs branch which is immediately identifiable to be (modulo possible monodromies) $D_4\times D_5$ -- a product of Klein $D$-type singularities. Hence from the small pool of possible Hasse diagram structures available, this expectation seems reasonable. Nevertheless, it is important to stress that this is purely conjectural -- it would be helpful to find other information that would pin down this Higgsing pattern explicitly.
\begin{equation}
\raisebox{-0.5\height}{\begin{tikzpicture}
\node (a) at (0,0) {\begin{tikzpicture}
    \node[gauger, label=below:$D_1$] (1) at (0,0){};
    \node[gaugeb, label=below:$C_1$] (2) at (1,0){};
    \node[gauger, label=below:$D_2$] (3) at (2,0){};
    \node[gaugeb, label=below:$C_2$] (4) at (3,0){};
    \node[gauger, label=below:$D_2$] (5) at (4,0){};
    \node[gaugeb, label=below:$C_1$] (6) at (5,0){};
    \node[gauger, label=below:$D_1$] (7) at (6,0){};
    \node[gaugeb, label=above:$C_1$] (8) at (3,1){};
    \draw[-] (1)--(2)--(3)--(4)--(5)--(6)--(7) (4)--(8);
\end{tikzpicture}}; 
\node (b) at (5,0) {\begin{tikzpicture}
\node (0) at (0,4.5) {$\mathcal{C}$};
    \node[hasse] (1a) at (0,0){};
    \node[hasse] (2a) at (0,1){};
    \node[hasse] (3a) at (0,2){};
    \node[hasse] (4a) at (0,3){};
    \node[hasse] (5a) at (0,4){};
    \draw[-] (1a)--(2a)node[midway, right]{$c_4$}--(3a) node[midway, right]{$c_3$}--(4a)node[midway, right]{$a_3$}--(5a)node[midway, right]{$A_1$};
\end{tikzpicture}};
\node (3) at (8.5,0) {\begin{tikzpicture}
    \node (0) at (0,2.5) {$\mathcal{H}?$};
    \node[hasse] (0a) at (0,-1){};
    \node[hasse] (1a) at (0,0){};
    \node[hasse] (2a) at (1,1){};
    \node[hasse] (3a) at (-1,1){};
    \node[hasse] (4a) at (0,2){};
    \draw[-] (0a)--(1a) node[midway, right]{$A_1$} --(2a) node[midway, right]{$D_4$}--(4a) node[midway, right]{$D_5$} (1a)--(3a) node[midway, left]{$D_5$}--(4a) node[midway, left]{$D_4$};
    \end{tikzpicture}};
\end{tikzpicture}}
\label{quiv:spin7_4spinors_i}
\end{equation}
% \begin{align}
%     \hscz\left[\mathcal{Q}_{\ref{quiv:spin7_4spinors_i}}\right]=&\frac{\left(\begin{aligned}
%     1&+9t^{2}+185t^{4}+1281t^{6}+8272t^{8}+33543t^{10}+114973t^{12}\\&+287939t^{14}+606568t^{16}+987732t^{18}+1359345t^{20}\\&+1477040t^{22}+\cdots+t^{44}\end{aligned}\right)}{(1-t^2)^{11}(1-t^{4})^{11}} \label{hs:spin74s_fc_L_i}\\
%     \hsczh\left[\mathcal{Q}_{\ref{quiv:spin7_4spinors_i}}\right]=&\frac{16t^{2}\left(\begin{aligned}1&+9t^{2}+91t^{4}+481t^{6}+2197t^{8}+6949t^{10}+18464t^{12}\\&+37119t^{14}++62887t^{16}+83518t^{18}+93864t^{20}\\&+\cdots+t^{40}\end{aligned}\right)}{(1-t^{2})^{11}(1-t^{4})^{11}}\label{hs:spin74s_fc_L_ii}\\
%     \hsc\left[\mathcal{Q}_{\ref{quiv:spin7_4spinors_i}}\right]=&\frac{\left(\begin{aligned}1&+25t^{2}+329t^{4}+2737t^{6}+15968t^{8}+68695t^{10}+226157t^{12}\\&+583363t^{14}+1200472t^{16}++1993924t^{18}+2695633t^{20}\\&+2978864t^{22}+\cdots+t^{44}\end{aligned}\right)}{(1-t^{2})^{11}(1-t^{4})^{11}}\label{hs:spin74s_fc_L_iii}\\
%     \hsh\left(\mathcal{Q}_{\ref{quiv:spin7_4spinors_i}}\right)=&\frac{\left(\begin{aligned}1&-t^4+3t^8-3t^{10}-6t^{14}+6t^{16}+2t^{18}\\-&2t^{20}-6t^{22}+6t^{24}+3t^{28} -3t^{30}+t^{34}-t^{38}\end{aligned}\right)}{(1-t^2)^3(1-t^4)(1-t^6)^3(1-t^8)^2}
% \end{align}
\begin{align}
    \hscz\left[\mathcal{Q}_{\ref{quiv:spin7_4spinors_i}}\right]
    &=
    \frac{P_{0}(t)}
    {(1-t^2)^{11}(1-t^{4})^{11}}
    \label{hs:spin74s_fc_L_i}
    \\
    \hsczh\left[\mathcal{Q}_{\ref{quiv:spin7_4spinors_i}}\right]
    &=
    \frac{16t^{2}P_{1}(t)}
    {(1-t^{2})^{11}(1-t^{4})^{11}}
    \label{hs:spin74s_fc_L_ii}
    \\
    \hsc\left[\mathcal{Q}_{\ref{quiv:spin7_4spinors_i}}\right]
    &=
    \frac{P_{2}(t)}
    {(1-t^{2})^{11}(1-t^{4})^{11}}
    \label{hs:spin74s_fc_L_iii}
    \\
    \hsh\left(\mathcal{Q}_{\ref{quiv:spin7_4spinors_i}}\right)
    &=
    \frac{P_{H}(t)}
    {(1-t^2)^3(1-t^4)(1-t^6)^3(1-t^8)^2}.
    \label{hs:spin74s_fc_L_iv}
\end{align}
where,
\[
\begin{aligned}
P_{0}(t)
={}&
1+9t^{2}+185t^{4}+1281t^{6}+8272t^{8}
+33543t^{10}+114973t^{12}
\\
&\quad
+287939t^{14}+606568t^{16}+987732t^{18}
+1359345t^{20}+1477040t^{22}
+\cdots+t^{44},
\\[1ex]
P_{1}(t)
={}&
1+9t^{2}+91t^{4}+481t^{6}+2197t^{8}
+6949t^{10}+18464t^{12}
\\
&\quad
+37119t^{14}+62887t^{16}+83518t^{18}
+93864t^{20}
+\cdots+t^{40},
\\[1ex]
P_{2}(t)
={}&
1+25t^{2}+329t^{4}+2737t^{6}+15968t^{8}
+68695t^{10}+226157t^{12}
\\
&\quad
+583363t^{14}+1200472t^{16}+1993924t^{18}
+2695633t^{20}+2978864t^{22}
+\cdots+t^{44},
\\[1ex]
P_{H}(t)
={}&
1-t^4+3t^8-3t^{10}-6t^{14}+6t^{16}
+2t^{18}-2t^{20}
\\
&\quad
-6t^{22}+6t^{24}+3t^{28}
-3t^{30}+t^{34}-t^{38}.
\end{aligned}
\]
Turning to the second cone, drawn on the right-hand side of \eqref{quiv:spin7_4spinors} and reproduced below in \eqref{quiv:spin7_4spinors_ii}, much of the same analysis applies. It's important to stress that although the Coulomb and Higgs branch Hasse diagrams of the two magnetic quivers are expected to be the same, direct calculation of their Hilbert series shows that these are distinct spaces. This of course accords with expectation regarding the structure of $\sprm(1)$ with two fundamental flavours. Note that the $C_1-C_1$ bifundamental is again expected to contribute the $A_1$ transition at the top of the Coulomb branch Hasse diagram and at the bottom of the Higgs branch Hasse diagram.
\begin{equation}
\raisebox{-0.5\height}{\begin{tikzpicture}
\node (a) at (0,0) {\begin{tikzpicture}
    \node[gauger, label=below:$D_1$] (1a) at (0,0){};
    \node[gaugeb, label=below:$C_1$] (2a) at (1,0){};
    \node[gauger, label=below:$D_2$] (3a) at (2,0){};
    \node[gaugeb, label=below:$C_1$] (4a) at (3,0){};
    \node[gauger, label=below:$D_2$] (5a) at (4,0){};
    \node[gaugeb, label=below:$C_1$] (6a) at (5,0){};
    \node[gauger, label=below:$D_1$] (7a) at (6,0){};
    \node[gaugeb, label=above:$C_1$] (8a) at (2,1){};
    \node[gaugeb, label=above:$C_1$] (9a) at (4,1){};
    \draw[-] (1a)--(2a)--(3a)--(4a)--(5a)--(6a)--(7a) (3a)--(8a)--(9a)--(5a);
\end{tikzpicture}}; 
\node (b) at (5,0) {\begin{tikzpicture}
\node (0) at (0,4.5) {$\mathcal{C}$};
    \node[hasse] (1a) at (0,0){};
    \node[hasse] (2a) at (0,1){};
    \node[hasse] (3a) at (0,2){};
    \node[hasse] (4a) at (0,3){};
    \node[hasse] (5a) at (0,4){};
    \draw[-] (1a)--(2a)node[midway, right]{$c_4$}--(3a) node[midway, right]{$c_3$}--(4a)node[midway, right]{$a_3$}--(5a)node[midway, right]{$A_1$};
\end{tikzpicture}};
\node (3) at (8.5,0) {\begin{tikzpicture}
    \node (0) at (0,2.5) {$\mathcal{H}?$};
    \node[hasse] (0a) at (0,-1){};
    \node[hasse] (1a) at (0,0){};
    \node[hasse] (2a) at (1,1){};
    \node[hasse] (3a) at (-1,1){};
    \node[hasse] (4a) at (0,2){};
    \draw[-] (0a)--(1a) node[midway, right]{$A_1$} --(2a) node[midway, right]{$D_4$}--(4a) node[midway, right]{$D_5$} (1a)--(3a) node[midway, left]{$D_5$}--(4a) node[midway, left]{$D_4$};
    \end{tikzpicture}};
\end{tikzpicture}}
\label{quiv:spin7_4spinors_ii}
\end{equation}
\begin{align} \hscz\left[\mathcal{Q}_{\ref{quiv:spin7_4spinors_ii}}\right] &= \frac{P_{0}(t)} {(1-t^{2})^{11}(1-t^{4})^{11}} \label{hs:spin74s_fc_R_i} \\ \hsczh\left[\mathcal{Q}_{\ref{quiv:spin7_4spinors_ii}}\right] &= \frac{16t^{2}P_{1}(t)} {(1-t^{2})^{11}(1-t^{4})^{11}} \label{hs:spin74s_fc_R_ii} \\ \hsc\left[\mathcal{Q}_{\ref{quiv:spin7_4spinors_ii}}\right] &= \frac{ (1+t^{2}) \left( 1+13t^{2}+133t^{4}+608t^{6}+1478t^{8} +2002t^{10}+\cdots+t^{20} \right) } {(1-t^{2})^{22}} \label{hs:spin74s_fc_R_iii} \\ \hsh\left(\mathcal{Q}_{\ref{quiv:spin7_4spinors_ii}}\right) &= \frac{(1-t^{12})(1-t^{16})} {(1-t^{2})^{3}(1-t^{4})(1-t^{5})^{2}(1-t^{7})^{2}} \label{hs:spin7_4s_fc_R_iv} \end{align} where, \[ \begin{aligned} P_{0}(t) ={}& 1+9t^{2}+195t^{4}+1570t^{6}+10667t^{8} +47665t^{10}+168009t^{12} \\ &\quad +445360t^{14}+949366t^{16}+1596550t^{18} +2193618t^{20}+2420444t^{22} +\cdots+t^{44}, \\[1ex] P_{1}(t) ={}& 1+10t^{2}+107t^{4}+640t^{6}+3043t^{8} +10366t^{10}+28086t^{12} \\ &\quad +58930t^{14}+100356t^{16}+136390t^{18} +152046t^{20} +\cdots+t^{44}. \end{aligned} \]
% \begin{align}
%     \hscz\left[\mathcal{Q}_{\ref{quiv:spin7_4spinors_ii}}\right]=&\frac{\left(\begin{aligned}1+&9t^{2}+195t^{4}+1570t^{6}+10667t^{8}+47665t^{10}+168009t^{12}+445360t^{14}\\+&949366t^{16}+1596550t^{18}+2193618t^{20}+2420444t^{22}+\cdots+t^{44}\end{aligned}\right)}{(1-t^{2})^{11}(1-t^{4})^{11}}\label{hs:spin74s_fc_R_i}\\\hsczh\left[\mathcal{Q}_{\ref{quiv:spin7_4spinors_ii}}\right]=&\frac{16t^{2}\left(\begin{aligned}1+&10t^{2}+107t^{4}+640t^{6}+3043t^{8}+10366t^{10}+28086t^{12}\\+&58930t^{14}+100356t^{16}+136390t^{18}+152046t^{20}+\cdots+t^{44}\end{aligned}\right)}{(1-t^{2})^{11}(1-t^{4})^{11}}\label{hs:spin74s_fc_R_ii}\\
%     \hsc\left[\mathcal{Q}_{\ref{quiv:spin7_4spinors_ii}}\right]=&\frac{(1+t^{2})(1+13t^{2}+133t^{4}+608t^{6}+1478t^{8}+2002t^{10}+\cdots+t^{20})}{(1-t^{2})^{22}}
%     \label{hs:spin74s_fc_R_iii}\\
%     \hsh\left(\mathcal{Q}_{\ref{quiv:spin7_4spinors_ii}}\right)=&\frac{(1-t^{12})(1-t^{16})}{(1-t^{2})^{3}(1-t^{4})(1-t^{5})^{2}(1-t^{7})^{2}}
% \label{hs:spin7_4s_ii_HB}
% \end{align}
The intersection of the two Coulomb branches in \eqref{quiv:spin7_4spinors} is given by the magnetic quiver below in \eqref{quiv:spin7_4spinors_intersection}. This is derived by finding the shared nodes in the two quivers (equivalently the shared brane lockings in the web) and checked via Hilbert series using the algorithm in \cite{bourget2023talencones}. Explicit computation of the Hilbert series supports the identification of its Higgs branch as a product of $D$-type Klein singularities -- $D_4 \times D_5$. However, this may be complicated by global properties such as monodromy \cite{Bennett:2024loi}, which the Hilbert series cannot currently detect. The shape of the resulting quiver also motivates the comment above regarding the $C_1-C_1$ bifundamental giving rise to the $A_1$ transitions at the top and bottom of the Coulomb and Higgs branch Hasse diagrams respectively in \eqref{quiv:spin7_4spinors_i} and \eqref{quiv:spin7_4spinors_ii}.
\begin{equation}
\raisebox{-0.5\height}{\begin{tikzpicture}
\node (a) at (0,0) {\begin{tikzpicture}
    \node[gauger, label=below:$D_1$] (1) at (0,0){};
    \node[gaugeb, label=below:$C_1$] (2) at (1,0){};
    \node[gauger, label=below:$D_2$] (3) at (2,0){};
    \node[gaugeb, label=below:$C_1$] (4) at (3,0){};
    \node[gauger, label=below:$D_2$] (5) at (4,0){};
    \node[gaugeb, label=below:$C_1$] (6) at (5,0){};
    \node[gauger, label=below:$D_1$] (7) at (6,0){};
    \node[gaugeb, label=above:$C_1$] (8) at (3,1){};
    \draw[-] (1)--(2)--(3)--(4)--(5)--(6)--(7) (3)--(8)--(5);
\end{tikzpicture}}; 
\node (b) at (5.5,0) {\begin{tikzpicture}
\node (0) at (0,3.5) {$\mathcal{C}$};
    \node[hasse] (1a) at (0,0){};
    \node[hasse] (2a) at (0,1){};
    \node[hasse] (3a) at (0,2){};
    \node[hasse] (4a) at (0,3){};
    \draw[-] (1a)--(2a)node[midway, right]{$c_4$}--(3a) node[midway, right]{$c_3$}--(4a) node[midway, right]{$a_3$};
\end{tikzpicture}};
\node (3) at (8.5,0) {\begin{tikzpicture}
    \node (0) at (0,3) {$\mathcal{H}?$};
    \node[hasse] (1a) at (0,0){};
    \node[hasse] (2a) at (1.25,1.25){};
    \node[hasse] (3a) at (-1.25,1.25){};
    \node[hasse] (4a) at (0,2.5){};
    \draw[-] (1a)--(2a) node[midway, right]{$D_4$}--(4a) node[midway, right]{$D_5$} (1a)--(3a) node[midway, left]{$D_5$}--(4a) node[midway, left]{$D_4$};
    \end{tikzpicture}};
\end{tikzpicture}}
\label{quiv:spin7_4spinors_intersection}
\end{equation}
\begin{align}
    \hscz\left[\mathcal{Q}_{\ref{quiv:spin7_4spinors_intersection}}\right]
    &=
    \frac{P^{\cap}_{0}(t)}
    {(1-t^{2})^{10}(1-t^{4})^{10}}
    \\
    \hsczh\left[\mathcal{Q}_{\ref{quiv:spin7_4spinors_intersection}}\right]
    &=
    \frac{16t^{2}P^{\cap}_{1}(t)}
    {(1-t^{2})^{10}(1-t^{4})^{10}}
    \\
    \hsc\left[\mathcal{Q}_{\ref{quiv:spin7_4spinors_intersection}}\right]
    &=
    \frac{
        (1+t^{2})^{2}
        \left(
        1+14t^{2}+121t^{4}+454t^{6}+722t^{8}
        +454t^{10}+121t^{12}+14t^{14}+t^{16}
        \right)
    }
    {(1-t^{2})^{20}}
    \\
    \hsh\left[\mathcal{Q}_{\ref{quiv:spin7_4spinors_intersection}}\right]
    &=
    \frac{(1-t^{12})(1-t^{16})}
    {(1-t^{4})^{3}(1-t^{6})^{2}(1-t^{8})} \rightarrow D_4 \times D_5.
\end{align}
where
\[
\begin{aligned}
P^{\cap}_{0}(t)
={}&
1+10t^{2}+195t^{4}+1450t^{6}+9091t^{8}
+36160t^{10}+113648t^{12}
\\
&\quad
+260320t^{14}+477020t^{16}+669740t^{18}
+760026t^{20}
+\cdots+t^{40},
\\[1ex]
P^{\cap}_{1}(t)
={}&
1+10t^{2}+100t^{4}+540t^{6}+2329t^{8}
+6960t^{10}
\\
&\quad
+29460t^{14}+42303t^{16}+47020t^{18}
+\cdots+t^{36}.
\end{aligned}
\]
% \begin{align}
%     \hscz\left[\mathcal{Q}_{\ref{quiv:spin7_4spinors_intersection}}\right]=&\frac{\left(\begin{aligned}1+&10t^{2}+195t^{4}+1450t^{6}+9091t^{8}+36160t^{10}+113648t^{12}\\+&260320t^{14}+477020t^{16}+669740t^{18}+760026t^{20}+\cdots+t^{40}\end{aligned}\right)}{(1-t^{2})^{10}(1-t^{4})^{10}}\\
%     \hsczh\left[\mathcal{Q}_{\ref{quiv:spin7_4spinors_intersection}}\right]=&\frac{16t^{2}\left(\begin{aligned}1+&10t^{2}+100t^{4}+540t^{6}+2329t^{8}+6960t^{10}\\+&29460t^{14}+42303t^{16}+47020t^{18}+\cdots+t^{36}\end{aligned}\right)}{(1-t^{2})^{10}(1-t^{4})^{10}}\\
%     \hsc\left[\mathcal{Q}_{\ref{quiv:spin7_4spinors_intersection}}\right]=&\frac{(1+t^{2})^{2}(1+14t^{2}+121t^{4}+454t^{6}+722t^{8}+454t^{10}+121t^{12}+14t^{14}+t^{16})}{(1-t^{2})^{20}}\\
%     \hsh\left[\mathcal{Q}_{}\right]=&\frac{(1-t^{12})(1-t^{16})}{(1-t^{4})^{3}(1-t^{6})^{2}(1-t^{8})}
% \end{align}
The gauge theory at infinite coupling is realized by the modified brane system given below in \eqref{brane:spin7_4s_IC}. This augments the right-hand side cone of \eqref{quiv:spin7_4spinors} to that given in \eqref{quiv:spin7_4s_infcoup}, while keeping the left-hand cone intact. Note that the new magnetic quiver is unitary-orthosymplectic and contains a hypermultiplet in the $\Lambda^{2}$ representation of $\urm(2)$. The Coulomb branch Hilbert series of the augmented cone is given in \eqref{hs:spin74s_IC_i}, \eqref{hs:spin74s_IC_ii} and \eqref{hs:spin74s_IC_iii}. At present, neither the Hasse diagram nor the precise identity of the moduli space is known.
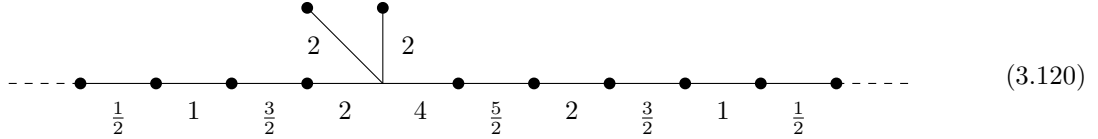
\begin{equation}
\begin{array}{c}
    \begin{tikzpicture}
    \draw[dashed](0,0)--(-1,0);
    \draw(0,0)--(10,0);
    \draw[dashed](10,0)--(11,0);
    \node[7brane]at(0,0){};
    \node[7brane]at(1,0){};
    \node[7brane]at(2,0){};
    \node[7brane]at(3,0){};
    \node[7brane]at(5,0){};
    \node[7brane]at(6,0){};
    \node[7brane]at(7,0){};
    \node[7brane]at(8,0){};
    \node[7brane]at(9,0){};
    \node[7brane]at(10,0){};
    \node[label=below:{$\frac{1}{2}$}]at(0.5,0){};
    \node[label=below:{$1$}]at(1.5,0){};
    \node[label=below:{$\frac{3}{2}$}]at(2.5,0){};
    \node[label=below:{$2$}]at(3.5,0){};
    \node[label=below:{$4$}]at(4.5,0){};
    \node[label=below:{$\frac{5}{2}$}]at(5.5,0){};
    \node[label=below:{$2$}]at(6.5,0){};
    \node[label=below:{$\frac{3}{2}$}]at(7.5,0){};
    \node[label=below:{$1$}]at(8.5,0){};
    \node[label=below:{$\frac{1}{2}$}]at(9.5,0){};
    \draw(4,0)--(3,1);
    \draw(4,0)--(4,1);
    \node[7brane]at(3,1){};
    \node[7brane]at(4,1){};
    \node[label=right:{$2$}]at(4,.5){};
    \node[label=right:{$2$}]at(2.75,.5){};
\end{tikzpicture}
\end{array}
\label{brane:spin7_4s_IC}
\end{equation}
\begin{equation}
\raisebox{-0.5\height}{
\scalebox{0.9}{\begin{tikzpicture}
\node (a) at (0,0) {\begin{tikzpicture}
    \node[gauger, label=below:$D_1$] (1) at (0,0){};
    \node[gaugeb, label=below:$C_1$] (2) at (1,0){};
    \node[gauger, label=below:$D_2$] (3) at (2,0){};
    \node[gaugeb, label=below:$C_2$] (4) at (3,0){};
    \node[gauger, label=below:$D_2$] (5) at (4,0){};
    \node[gaugeb, label=below:$C_1$] (6) at (5,0){};
    \node[gauger, label=below:$D_1$] (7) at (6,0){};
    \node[gaugeb, label=above:$C_1$] (8) at (3,1){};
    \draw[-] (1)--(2)--(3)--(4)--(5)--(6)--(7) (4)--(8);
\end{tikzpicture}};
\node (b) at (4,0) {\begin{tikzpicture}
\node at (0,0){\Large $\cup$};
\end{tikzpicture}};
\node (c) at (8,0.5) {\begin{tikzpicture}
    \node[gauger, label=below:$D_1$] (1) at (0,0){};
    \node[gaugeb, label=below:$C_1$] (2) at (1,0){};
    \node[gauger, label=below:$D_2$] (3) at (2,0){};
    \node[gaugeb, label=below:$C_2$] (4) at (3,0){};
    \node[gauger, label=below:$D_2$] (5) at (4,0){};
    \node[gaugeb, label=below:$C_1$] (6) at (5,0){};
    \node[gauger, label=below:$D_1$] (7) at (6,0){};
    \node[gaugeBig, label=left:$2$] (8) at (3,1){};
    \node[flavour, label=above:$1$] (9) at (3,2){};
    \node at (2.75,1.5){$\wedge^{2}$};
    \draw[-] (1)--(2)--(3)--(4)--(5)--(6)--(7) (4)--(8)--(9);
\end{tikzpicture}};
\end{tikzpicture}}}
\label{quiv:spin7_4s_infcoup}
\end{equation}
\begin{align}
    \hscz\left[\ref{quiv:spin7_4s_infcoup}\right]=&1+21t^{2}+382t^{4}+4800t^{6}+46956t^{8}+370347t^{10}+\cdots \label{hs:spin74s_IC_i}\\
    \hsczh\left[\ref{quiv:spin7_4s_infcoup}\right]=&16t^{2}+336t^{4}+4608t^{6}+45984t^{8}+366960t^{10}+\cdots \label{hs:spin74s_IC_ii}\\
    \hsc\left[\ref{quiv:spin7_4s_infcoup}\right]=&1+37t^{2}+718t^{4}+9408t^{6}+92940t^{8}+737307t^{10}+\cdots \label{hs:spin74s_IC_iii}\\
    \hsh\left[\ref{quiv:spin7_4s_infcoup}\right]=&\frac{(1+t)\left(\begin{aligned}1-&t+2t^2+t^4+t^5+t^6+2t^7+2t^8+3t^9+3t^{10}+4t^{11}\\+&6t^{12}+2t^{13}+7t^{14}+3t^{15}+5t^{16}+3t^{17}+\cdots+t^{34}\end{aligned}\right)}{(1-t^{4})(1-t^{6})^{2}(1-t^{7})(1-t^{8})(1-t^{10})}
\end{align}
% \paragraph{More than Four Spinors}
% Although matter constraints in five dimensions preclude models with more than four \todo{find correct number} spinors, it's possible to `bootstrap' candidate orthosymplectic quivers for the three-dimensional theory using the finite-coupling Hasse diagram for the five-dimensional theory.
%%%%%%%%%%%%%%%%%%%%%%%%%%%%%%%
% \subsection{Theories above Rank-3}
% \subsubsection{$\spin(8)$}
% \left\{2s,2s+2c,\right\}
% \paragraph{Finite coupling analysis}
% The Hasse diagram for $\spin(8)$ with $n$ spinors (of the same type) is: 
% \begin{equation}
% \raisebox{-0.5\height}{\begin{tikzpicture}
% \node (1) [hasse] at (0,0) {};
% \node (1a) at (1.8,0) {$\text{SO}(8)+n$ spinors};
% \node (2a) at (1.8,1) {$\text{G}_2$};
% \node (2) [hasse] at (0,1) {};
% \node (3) [hasse] at (0,2) {};
% \node (3a) at (1.8,2) {$\text{SU}(3)$};
% \node (4) [hasse] at (0,3) {};
% \node (3a) at (1.8,3) {$\text{SU}(2)\; \text{Not sure if this HD is correct}$};
% \node (5) [hasse] at (0,4) {};
% \draw[-] (1)--(2) node[pos=0.5,midway, right]{$c_n$} --(3) node[pos=0.5,midway, right]{$c_{n-2}$}--(4) node[pos=0.5,midway, right]{$a_{2n-7}$} --(5) node[pos=0.5,midway, right]{$d_{2n-8}$};
% \end{tikzpicture}}
% \end{equation}
% The dimension of the Higgs branch is $\text{dim}_{\mathbb{H}}(\mathcal{H})=8n-28$.
\section{New Quivers for $\mathfrak{so}_{2n+1}$ and $\mathfrak{sp}_{n}$ Nilpotent Orbits}
\label{sec:BC_wreathing}
The emergence of new quivers providing Coulomb-branch constructions of $\mathfrak{sl}_{k}$ and $\mathfrak{so}_{2k}$ minimal nilpotent orbit closures leads to the question of whether closely-related theories exist for which the Coulomb branch is an orbifold. It turns out that such theories do exist, obtained from the magnetic quivers in Tables \ref{tab:spin2_magquivs_i}, \ref{tab:spin2_magquivs_ii} and \ref{tab:spin3_magquivs} via `wreathing' \cite{Bourget_2021,grimminger2025wreathingdiscretegaugingnoninvertible}.

Wreathing is a discrete gauging of an outer automorphism symmetry under which the Coulomb branch is quotiented by a finite group factor. Several common examples coincide with well-known finite hyper-K\"ahler quotients, such as $\overline{min.D_{n}} \xrightarrow{\mathbb{Z}_{2}}\overline{n.min.B_{n-1}}$ and $\overline{min.A_{2n+1}} \xrightarrow{\mathbb{Z}_{2}}\overline{n.min.C_{n}}$ \cite{kostant-brylinski}. Interestingly, these hyper-K\"ahler quotients are precisely realised by wreathings on various of the magnetic quivers derived in this work. This gives non-trivial new examples of 3d $\mathcal{N}=4$ gauge theories whose Coulomb branch is a $B$- or $C$-type nilpotent orbit closure, given below in Tables \ref{tab:wreathing} and \ref{tab:folding}.

Note that there is no standard notation for wreathed quivers in the literature. The following examples use the shorthand $G\sim$ to denote the $G$-wreathing of the sub-quiver to the right of the $\sim$. Using the decoration notation, this is equivalently expressed as
\begin{equation}\raisebox{-0.5\height}{\begin{tikzpicture}
\node (a) at (0,0.5) {\begin{tikzpicture}
    \node[gauge] (Q) at (0,0){$Q$};
    \node at (1,0) (wreath) {$S_n \sim$};
    \node[gauge] (q) at (2,0){$q$};
    \draw(Q)--(wreath)--(q);
\end{tikzpicture}};
\node (a) at (3,0.5) {\begin{tikzpicture}
\node (1d) at (0,0) {$=$};
\end{tikzpicture}};
\node (c) at (6,0.5) {\begin{tikzpicture}
    \node[gauge] (Q) at (0,0){$Q$};
    \node[gauge] (q1) at (1.5,1){$q$};
    \node[gauge] (q2) at (1.5,-1){$q$};
    \node at (1.5,0.1) {$\vdots$};
    \draw[-] (Q)--(q1) (Q)--(q2);
    \draw[red] (q1) circle (0.3cm);
    \draw[red] (q2) circle (0.3cm);
    \draw [decorate, decoration = {brace, raise=5pt, amplitude=5pt}] (2,1) --  (2,-1) node[pos=0.5,right=10pt,black]{$n$};
\end{tikzpicture}};
\end{tikzpicture}}
\end{equation}
The $C$-type wreathings and foldings in Tables \ref{tab:wreathing} and \ref{tab:folding} arise from polymerisation examples of the form 
\begin{equation}
    \frac{\mathbb{H}^{n}\times\mathbb{H}^{n}}{\spin(2)}
\end{equation}
In these cases, the wreathed outer automorphism arises from the $S_2$ factor permuting the two copies of $\mathbb{H}^{n}$. Unlike other outer automorphisms present in these magnetic quivers, this symmetry emerges from the polymerisation itself -- it is not inherited from one of the quivers in Table \ref{tab:free_magnetic_theories}. This differs from the $B$-type wreathings and foldings in Tables \ref{tab:wreathing} and \ref{tab:folding}. For these examples, the magnetic quivers arise from polymerisations of the form
\begin{equation}
    \frac{\mathbb{H}^{4}\times\mathbb{H}^{n}}{\spin(3)}
\end{equation}For these theories, the outer automorphism whose wreathing effects $\overline{min.D_{n}} \xrightarrow{\mathbb{Z}_{2}}\overline{n.min.B_{n-1}}$ is inherited from teh $\mathbb{H}^{4}$ factor -- the $S_3$ outer automorphism of the $\mathbb{H}^{4}$ theory in Table \ref{tab:free_magnetic_theories} is broken to $S_2$ under the polymerisation.

Note that the Higgs branch calculations given in Table \ref{tab:wreathing} are strictly \emph{local}; that is, they do not account for global effects arising from monodromy actions on resolution fibres. Under current techniques it is unclear whether such effects might promote the Klein $D$-type singularities identified here to those of $C$-type or $B$-type, as studied in \cite{Bennett:2024loi}.
The following examples focus on cases in which the wreathed structure of the Coulomb branch is particularly simple to identify, using the results of Kostant and Brylinski \cite{kostant-brylinski}. Clearly, other magnetic quivers introduced in this work admit analogous wreathings along their outer automorphism symmetries. Such cases, whose resulting geometries are considerably more complicated to describe than those included here, will be studied further in future work.
\begin{table}[h!]
\centering
\begin{tabular}{|c|c|c|c|}
\hline
$\mathcal{T}$ & Wreathed Quiver & CB & HB \\ \hline
$\begin{aligned}&\spin(2)\\&N_{\mathrm{S}}=4\end{aligned}$ & $\raisebox{-0.5\height}{\begin{tikzpicture}
    \node[gauger, label=below:$D_1$] (1) at (0,0){};
    \node at (1,0) (wreath) {$\mathbb Z_2 \sim$};
    \node[gauge, label=below:$1$] (2) at (2,0){};
    \draw(1)--(wreath)--(2);
\end{tikzpicture}}$ & $\overline{n.min \;C_2}$ & $D_4$ \\ \hline
$\begin{aligned}&\spin(2)\\&N_{\mathrm{S}}=8\end{aligned}$ & $\raisebox{-0.5\height}{\begin{tikzpicture}
    \node[gauger, label=below:$D_1$] (1) at (0,0){};
    \node at (1,0) (wreath) {$\mathbb Z_2 \sim$};
    \node[gaugeb, label=below:$C_1$] (2) at (2,0){};
    \node[gauger, label=below:$D_1$] (3) at (3,0){};
    \node[gauger, label=above:$D_1$] (4) at (2,1){};
    \draw(1)--(wreath)--(2)--(3) (2)--(4);
\end{tikzpicture}}$ & $\overline{n.min \;C_4}$ & $D_6$ \\ \hline
$\begin{aligned}&\spin(2)\\&N_{\mathrm{S}}=16\end{aligned}$ & $\raisebox{-0.5\height}{\begin{tikzpicture}
    \node[gauger, label=below:$D_1$] (1) at (0,0){};
    \node at (1,0) (wreath) {$\mathbb Z_2 \sim$};
    \node[gaugeb, label=below:$C_1$] (2) at (2,0){};
    \node[gauger, label=below:$D_2$] (3) at (3,0){};
    \node[gaugeb, label=below:$C_1$] (4) at (4,0){};
    \node[gauger, label=below:$D_1$] (5) at (5,0){};
    \node[gaugeb, label=left:$C_1$] (6) at (3,1){};
    \node[gauger, label=above:$D_1$] (7) at (3,2){};
    \draw(1)--(wreath)--(2)--(3)--(4)--(5) (3)--(6)--(7);
\end{tikzpicture}}$ & $\overline{n.min \;C_8}$ & $D_{10}$  \\ \hline
$\begin{aligned}&\spin(2)\\&N_{\mathrm{S}}=32\end{aligned}$ & $\raisebox{-0.5\height}{\begin{tikzpicture}
    \node[gauger, label=below:$D_1$] (1) at (0,0){};
    \node at (1,0) (wreath) {$\mathbb Z_2 \sim$};
    \node[gaugeb, label=below:$C_1$] (2) at (2,0){};
    \node[gauger, label=below:$D_2$] (3) at (3,0){};
    \node[gaugeb, label=below:$C_2$] (4) at (4,0){};
    \node[gauger, label=below:$D_3$] (5) at (5,0){};
    \node[gaugeb, label=below:$C_2$] (6) at (6,0){};
    \node[gauger, label=below:$D_2$] (7) at (7,0){};
    \node[gaugeb, label=below:$C_1$] (8) at (8,0){};
    \node[gauger, label=below:$D_1$] (9) at (9,0){};
    \node[gaugeb, label=above:$C_1$] (10) at (5,1){};
    \draw[-] (1)--(wreath)--(2)--(3)--(4)--(5)--(6)--(7)--(8)--(9) (5)--(10);
\end{tikzpicture}}$ & $\overline{n.min \;C_{16}}$ & $D_{18}$  \\ \hline
$\begin{aligned}&\spin(3)\\&N_{\mathrm{S}}=6\end{aligned}$ & $\raisebox{-0.5\height}{\begin{tikzpicture}
    \node[gauger, label=below:$D_1$] (1) at (0,0){};
    \node at (1,0) (wreath) {$\sim \mathbb Z_2$};
    \node[gaugeb, label=below:$C_1$] (2) at (2,0){};
    \node[gauger, label=below:$D_2$] (3) at (3,0){};
    \node[gaugeb, label=below:$C_1$] (4) at (4,0){};
    \node[gauger, label=below:$D_1$] (5) at (5,0){};
    \node[gaugeb, label=left:$C_1$] (6) at (3,1){};
    \node[gauger, label=left:$D_1$] (7) at (3,2){};
    \draw[-] (1)--(wreath)--(2)--(3)--(4)--(5) (3)--(6)--(7);
\end{tikzpicture}}$ & $\overline{n.min \;B_{5}}$ & $D_{10}$ \\ \hline
 $\begin{aligned}&\spin(3)\\&N_{\mathrm{S}}=10\end{aligned}$& $\raisebox{-0.5\height}{\begin{tikzpicture}
    \node[gauger, label=below:$D_1$] (1a) at (0,0){};
    \node at (1,0) (wreath) {$\sim \mathbb Z_2$};
    \node[gaugeb, label=below:$C_1$] (2) at (2,0){};
    \node[gauger, label=below:$D_2$] (3) at (3,0){};
    \node[gaugeb, label=below:$C_2$] (4) at (4,0){};
    \node[gauger, label=below:$D_3$] (5) at (5,0){};
    \node[gaugeb, label=below:$C_2$] (6) at (6,0){};
    \node[gauger, label=below:$D_2$] (7) at (7,0){};
    \node[gaugeb, label=below:$C_1$] (8) at (8,0){};
    \node[gauger, label=below:$D_1$] (9) at (9,0){};
    \node[gaugeb, label=above:$C_1$] (10) at (5,1){};
    \draw (1a)--(wreath)--(2)--(3)--(4)--(5)--(6)--(7)--(8)--(9) (5)--(10);
\end{tikzpicture}}$ & $\overline{n.min \;B_{9}}$ & $D_{18}$ \\ \hline
\end{tabular}
\caption{Wreathings of various quivers studied in Tables \ref{tab:spin2_magquivs_i}, \ref{tab:spin2_magquivs_ii} and \ref{tab:spin3_magquivs}. These give new theories whose Coulomb branch is the closure of a classical nilpotent orbit of types $B$ and $C$. The Higgs branches of these theories are also straightforward to compute -- they evaluate as Klein singularities of type-$D$. A subtlety remains regarding the global structure of these Higgs branches -- namely, whether there exists a monodromy permuting their resolution fibres. This extra data is not captured in the Hilbert series. Note that although the Coulomb branches of these theories are of height 2, the Higgs branches are of height 1.}
\label{tab:wreathing}
\end{table}
\begin{table}[h!]
\centering
\begin{tabular}{|c|c|c|}
\hline
$\mathcal{T}$ & Folded Quiver & CB  \\ \hline
$\begin{aligned}&\spin(2)\\&N_{\mathrm{S}}=4\end{aligned}$& $\raisebox{-0.5\height}{\begin{tikzpicture}
    \node[gauger, label=below:$D_1$] (1) at (0,0){};
    \node[gaugeBig, label=below:$1$] (2) at (1,0){};
    \draw[transform canvas={yshift=1.3pt}] (1)--(2);
    \draw[transform canvas={yshift=-1.3pt}] (1)--(2);
    \draw[-] (0.6,0.2)--(0.4,0)--(0.6,-0.2);
\end{tikzpicture}}$ & $c_2$  \\ \hline
$\begin{aligned}&\spin(2)\\&N_{\mathrm{S}}=8\end{aligned}$ & $\raisebox{-0.5\height}{\begin{tikzpicture}
    \node[gauger, label=below:$D_1$] (1) at (0,0){};
    \node[gaugeb, label=below:$C_1$] (2) at (1,0){};
    \node[gauger, label=left:$D_1$] (3) at (1,1){};
    \node[gauger, label=below:$D_1$] (4) at (2,0){};
    \draw[transform canvas={yshift=1.3pt}] (4)--(2);
    \draw[transform canvas={yshift=-1.3pt}] (4)--(2);
    \draw[-] (1.6,0.2)--(1.4,0)--(1.6,-0.2);
    \draw(1)--(2) (2)--(3);
\end{tikzpicture}}$ & $c_4$  \\ \hline
$\begin{aligned}&\spin(2)\\&N_{\mathrm{S}}=16\end{aligned}$ & $\raisebox{-0.5\height}{\begin{tikzpicture}
    \node[gauger, label=below:$D_1$] (1) at (0,0){};
    \node[gaugeb, label=below:$C_1$] (2) at (1,0){};
    \node[gauger, label=below:$D_2$] (3) at (2,0){};
    \node[gaugeb, label=below:$C_1$] (4) at (3,0){};
    \node[gauger, label=below:$D_1$] (5) at (4,0){};
    \node[gaugeb, label=left:$C_1$] (10) at (2,1){};
    \node[gauger, label=above:$D_1$] (11) at (2,2){};
    \draw(1)--(2)--(3)--(4) (3)--(10)--(11);
    \draw[transform canvas={yshift=1.3pt}] (4)--(5);
    \draw[transform canvas={yshift=-1.3pt}] (4)--(5);
    \draw[-] (3.6,0.2)--(3.4,0)--(3.6,-0.2);
\end{tikzpicture}}$ & $c_8$  \\ \hline
$\begin{aligned}&\spin(2)\\&N_{\mathrm{S}}=32\end{aligned}$ & $\raisebox{-0.5\height}{\begin{tikzpicture}
    \node[gauger, label=below:$D_1$] (1) at (0,0){};
    \node[gaugeb, label=below:$C_1$] (2) at (1,0){};
    \node[gauger, label=below:$D_2$] (3) at (2,0){};
    \node[gaugeb, label=below:$C_2$] (4) at (3,0){};
    \node[gauger, label=below:$D_3$] (5) at (4,0){};
    \node[gaugeb, label=below:$C_2$] (6) at (5,0){};
    \node[gauger, label=below:$D_2$] (7) at (6,0){};
    \node[gaugeb, label=below:$C_1$] (8) at (7,0){};
    \node[gauger, label=below:$D_1$] (9) at (8,0){};
    \node[gaugeb, label=above:$C_1$] (10) at (4,1){};
    \draw[-] (1)--(2)--(3)--(4)--(5)--(6)--(7)--(8) (5)--(10);
    \draw[transform canvas={yshift=1.3pt}] (8)--(9);
    \draw[transform canvas={yshift=-1.3pt}] (8)--(9);
    \draw[-] (7.6,0.2)--(7.4,0)--(7.6,-0.2);
\end{tikzpicture}}$ & $c_{16}$  \\ \hline
$\begin{aligned}&\spin(3)\\&N_{\mathrm{S}}=6\end{aligned}$ & $\raisebox{-0.5\height}{\begin{tikzpicture}
    \node[gauger, label=below:$D_1$] (1) at (0,0){};
    \node[gaugeb, label=below:$C_1$] (2) at (1,0){};
    \node[gauger, label=below:$D_2$] (3) at (2,0){};
    \node[gaugeb, label=below:$C_1$] (4) at (3,0){};
    \node[gauger, label=below:$D_1$] (5) at (4,0){};
    \node[gaugeb, label=left:$C_1$] (10) at (2,1){};
    \node[gauger, label=above:$D_1$] (11) at (2,2){};
    \draw(1)--(2)--(3)--(4) (3)--(10)--(11);
    \draw[transform canvas={yshift=1.3pt}] (4)--(5);
    \draw[transform canvas={yshift=-1.3pt}] (4)--(5);
    \draw[-] (3.4,0.2)--(3.6,0)--(3.4,-0.2);
\end{tikzpicture}}$ & $b_5$  \\ \hline
$\begin{aligned}&\spin(3)\\&N_{\mathrm{S}}=10\end{aligned}$ & $\raisebox{-0.5\height}{\begin{tikzpicture}
    \node[gauger, label=below:$D_1$] (1) at (0,0){};
    \node[gaugeb, label=below:$C_1$] (2) at (1,0){};
    \node[gauger, label=below:$D_2$] (3) at (2,0){};
    \node[gaugeb, label=below:$C_2$] (4) at (3,0){};
    \node[gauger, label=below:$D_3$] (5) at (4,0){};
    \node[gaugeb, label=below:$C_2$] (6) at (5,0){};
    \node[gauger, label=below:$D_2$] (7) at (6,0){};
    \node[gaugeb, label=below:$C_1$] (8) at (7,0){};
    \node[gauger, label=below:$D_1$] (9) at (8,0){};
    \node[gaugeb, label=above:$C_1$] (10) at (4,1){};
    \draw[-] (1)--(2)--(3)--(4)--(5)--(6)--(7)--(8) (5)--(10);
    \draw[transform canvas={yshift=1.3pt}] (8)--(9);
    \draw[transform canvas={yshift=-1.3pt}] (8)--(9);
    \draw[-] (7.4,0.2)--(7.6,0)--(7.4,-0.2);
\end{tikzpicture}}$ & $b_9$  \\ \hline
\end{tabular}
\caption{Various foldings of several of the theories in Tables \ref{tab:spin2_magquivs_i}, \ref{tab:spin2_magquivs_ii} and \ref{tab:spin3_magquivs}. Note that this can equivalently be thought of as introducing a single non-simple lacing between two rank-1 nodes in the theories in Table \ref{tab:free_magnetic_theories}. Interestingly, either $B$- or $C$-type minimal nilpotent orbit closures arise depending on the set of nodes made short.}
\label{tab:folding}
\end{table}
\FloatBarrier
\section{Outlook}
This work investigates the magnetic quivers for various 5d $\mathcal{N}=1$ $\spin(N)$ SQCDs with spinor matter at finite and infinite coupling, paying particular attention to those for $\spin(2)$ and $\spin(3)$ gauge theory. Interestingly, the magnetic quivers for these theories give a large number of new unframed orthosymplectic 3d $\mathcal{N}=4$ quivers whose Coulomb branch is the closure of a minimal nilpotent orbit of $\surm(k)$ or $\sorm(2k)$. These new quivers are particularly useful in quiver subtraction, as shown for higher-rank theories in Sections \ref{sec:rank2theories} and \ref{sec:rank3theories}, in which the Higgs branch Hasse diagram of the electric theory at finite coupling is reproduced using an orthosymplectic quiver subtraction algorithm.

It is worth noting that at least some of the magnetic quivers presented in this work, are full-fledged 3d mirrors of certain $\spin(N)$ gauge theories with spinor matter. Typically this fails to be the case when multiple cones are present, or when the electric theory is not completely Higgsable. Providing further tests of mirror symmetry, for instance by matching the sphere partition function or the superconformal index in the former cases is a natural extension of our study.

Similarly, it is tempting to investigate the existence of brane systems for $\spin(N)$ gauge theory with spinor matter in three, four, or six dimensions. It would be interesting to consider whether the same magnetic quivers -- this time as full 3d mirrors, emerge from D3-D5-NS5 brane systems under some modifications.

Some of the magnetic quivers given in this work do not explicitly display the outer automorphism symmetries present for their unitary cousins. It would be interesting to systematically understand the effect of wreathing on these theories and provide an interpretation in terms of the electric theories.

One of the interesting features of the construction explored in this work is the fact that UV physics places an upper bound on the amount of matter in the 5d electric theory. Usually, the brane systems which violate this bound are deemed uninteresting, and rarely studied. On the other hand, this work has demondstrated that even when the theory is not UV-complete in 5d, the brane system can still be used to extract interesting physics. A perfect example is the magnetic quivers presented here for Spin(2) gauge theories. While the theory has a Landau pole and UV incomplete in 5d. The brane system is useful for obtaining interesting magnetic quivers at finite coupling.
%%%%%%%%%%%%%%%%%%%%%%%%%%%%%%%%%%%%%%%%%%%%%%%%%%%%%%%%%%%%%%%%%&
%%%%%%%%%%%%%%%%%%%%%%%%%%%%%%%&&
\FloatBarrier
\section*{Acknowledgements}
We thank Rudolph Kalveks and Guhesh Kumaran for helpful comments and discussions. In particualar, we thank Elias Van den Driessche for invaluable discussions, collaboration and comments. The work of SB and AH is partially supported by STFC Consolidated Grant ST/X000575/1. The work of SB is supported by the STFC DTP research studentship grant ST/Y509231/1. The research of MA is supported in part by an INFN fellowship under the “Iniziativa Specifica” 24 ST\&FI. The authors thank the Simons Center for Geometry and Physics for their hospitality during various stages of this project
\appendix
\section{Higgs Branch Hilbert Series for $\spin(7)$ and $\spin(8)$ Theories}
\label{app:spin7spin8}
\begin{landscape}
\begin{table}[h!]
    \centering
    \begin{tabular}{|c|c|c|c|}
    \hline 
    $\left(N_f,N_s\right)$ & Hilbert Series & PLog & Symmetry\\\hline
    $\left(0,2\right)$& Incomplete Higgsing & \textemdash & \textemdash \\\hline
    $\left(0,3\right)$& Incomplete Higgsing & \textemdash & \textemdash \\\hline
    $\left(0,4\right)$& \raisebox{-0.5\height}{$\frac{1+16t^{2}+100t^{4}+247t^{6}+177t^{8}+2t^{10}-14t^{12}-t^{14}}{(1-t^{2})^{12}}$}& $t^{2}\left(\begin{aligned}28&-36t^{2}+7t^{4}\\+&455t^{6}+\cdots\end{aligned}\right)$ & $\mathfrak{so}_{8}$\\\hline
    $\left(0,5\right)$& \raisebox{-0.5\height}{$\frac{(1+t^{2})^{2}(1+23t^{2}+223t^{4}+925t^{6}+1532t^{8}+925t^{10}+223t^{12}+23t^{14}+t^{16})}{(1-t^{2})^{20}}$}& $t^{2}\left(\begin{aligned}45&-55t^{2}-156t^{4}\\+&3420t^{6}+\cdots\end{aligned}\right)$ & $\mathfrak{so}_{10}$\\\hline
    $\left(1,2\right)$& \raisebox{-0.5\height}{$\frac{1+t+6t^{2}+15t^{3}+29t^{4}+40t^{5}+61t^{6}+51t^{7}+22t^{8}-9t^{9}-15t^{10}-8t^{11}+2t^{13}}{(1-t)^{2}(1-t^{2})^{4}(1+t+t^{2})^{3}}$} & $t^{2}\left(\begin{aligned}9&+12t-t^{2}\\-&34t^{3}+\cdots\end{aligned}\right)$ & $\mathfrak{su}_2 \oplus \mathfrak{su}_2 \oplus \mathfrak{su}_2$\\\hline
    $\left(2,1\right)$& \raisebox{-0.5\height}{$\frac{1+7t^{2}+4t^{3}+33t^{4}+20t^{5}+84t^{6}+40t^{7}+127t^{8}+40t^{9}+104t^{10}+20t^{11}+43t^{12}+4t^{13}+2t^{14}-4t^{16}-t^{18}}{(1-t^{2})^{4}(1-t^{4})^{4}}$} & $t^{2}\left(\begin{aligned}11&+4t+9t^{2}\\-&28t^{3}+\cdots\end{aligned}\right)$ & $\mathfrak{sp}_2\oplus\mathfrak{u}_{1}$\\\hline
    $\left(3,1\right)$& \raisebox{-0.5\height}{$\frac{\begin{aligned} 1-& 2t + 15 t^2 - 22 t^3 + 126 t^4 - 136 t^5 + 668 t^6 - 564 t^7 + 2458 t^8 - 1676 t^9 + 6514 t^{10}\\ -& 3696 t^{11} + 12848 t^{12} - 6186 t^{13} + 19165 t^{14} - 7974 t^{15} + 21886 t^{16} - 7974 t^{17}\\ +& 19165 t^{18} - 6186 t^{19} + 12848 t^{20} - 3696 t^{21} + 6514 t^{22} - 1676 t^{23} + 2458 t^{24}\\ -& 564 t^{25} + 668 t^{26} - 136 t^{27} + 126 t^{28} - 22 t^{29} + 15 t^{30} - 2 t^{31} + t^{32}\end{aligned}}{(1-t)^2(1-t^2)^{8}(1-t^4)^8}$} & $t^{2}\left(\begin{aligned}22&+6t+27t^{2}\\+&22t^{3}+\cdots\end{aligned}\right)$ & $\mathfrak{sp}_{3}\oplus\mathfrak{u}_{1}$\\\hline
    $\left(2,2\right)$& \raisebox{-0.5\height}{$\frac{\begin{aligned} 1 +& t + 11 t^2 + 30 t^3 + 123 t^4 + 281 t^5 + 796 t^6 + 1651 t^7 + 3627 t^8 + 6428 t^9 + 11546 t^{10}\\ +& 17738 t^{11} + 26993 t^{12} + 35963 t^{13} + 46896 t^{14} + 54445 t^{15} + 61687 t^{16} + 62542 t^{17}\\ +& 61687 t^{18} + 54445 t^{19} + 46896 t^{20} + 35963 t^{21} + 26993 t^{22} + 17738 t^{23} + 11546 t^{24}\\ +& 6428 t^{25} + 3627 t^{26} + 1651 t^{27} + 796 t^{28} + 281 t^{29} + 123 t^{30} + 30 t^{31} + 11 t^{32} + t^{33} + t^{34}
    \end{aligned}}{(1-t)^4(1-t^2)^6(1-t^4)^6(1+t+t^2)^5}$} & $t^{2}\left(\begin{aligned}16&+24t+44t^{2}\\-&32t^{3}+\cdots\end{aligned}\right)$ & $\mathfrak{sp}_2\oplus\mathfrak{su}_2 \oplus \mathfrak{su}_2$ \\\hline
    \end{tabular}
    \caption{Exact unrefined results for the Higgs branch Hilbert series of $\spin(7)$ with various spinor and vectors. Above $N_s=N_f=2$ the computational complexity of the Weyl integral becomes prohibitive. Exact results for three and two spinors are calculable using direct algebraic techniques, although the singular structure of these spaces is expected to be a nilpotent orbit closure.}
    \label{tab:spin7_bits}
\end{table}
\end{landscape}
\begin{table}[h!]
    \centering
    \begin{tabular}{|c|c|c|}
    \hline 
    $\left(N_f,N_s,N_c\right)$ & Hilbert Series & Symmetry\\\hline
    $\left(5\right)$& $1+55t^{2}+1495t^{4}+26898t^{6}+358160t^{8}+\cdots$& $\mathfrak{sp}_{5}$\\\hline
    $\left(6\right)$& $1 + 78 t^2 + 3015 t^4 + 77077 t^6 + 1467102 t^8+\cdots$& $\mathfrak{sp}_{6}$\\\hline
    $\left(7\right)$& $1 + 105 t^2 + 5474 t^4 + 189020 t^6+\cdots$& $\mathfrak{sp}_{7}$\\\hline
    $\left(8\right)$& $1 + 136 t^2 + 9196 t^4 + 412335 t^6+\cdots$& $\mathfrak{sp}_{8}$\\\hline
    $\left(2,3\right)$& $1 + 31 t^2 + 565 t^4 + 7420 t^6+\cdots$& $\mathfrak{sp}_{2}\oplus\mathfrak{sp}_{3}$\\\hline
    $\left(3,3\right)$& $1 + 42 t^2 + 1098 t^4 + 21238 t^6+\cdots$& $\mathfrak{sp}_{3}\oplus\mathfrak{sp}_{3}$\\\hline
    $\left(4,2\right)$& $1 + 46 t^2 + 1215 t^4 + 23565 t^6+\cdots$& $\mathfrak{sp}_{2}\oplus\mathfrak{sp}_{4}$\\\hline
    $\left(5,1\right)$& $1 + 58 t^2 + 1710 t^4 + 34602 t^6+\cdots$& $\mathfrak{sp}_{5}\oplus\mathfrak{sp}_{1}$\\\hline
    $\left(2,2,2\right)$& $1 + 30 t^2 + 64 t^3 + 555 t^4 + 1984 t^5 + 9565 t^6$& $\mathfrak{sp}_{2}\oplus\mathfrak{sp}_{2}\oplus\mathfrak{sp}_{2}$\\\hline
    \end{tabular}
    \caption{Perturbative calculations of the finite-coupling Higgs branch Hilbert series of $\spin(8)$ with $N_v$ vectors, $N_s$ spinors and $N_c$ cospinors.}
    \label{tab:spin8_bits}
\end{table}
\section{Coulomb Branch Hilbert Series for Magnetic Quivers}
\label{app:CB_hilbert_series}
This appendix contains the results of various Hilbert series calculations for the magnetic quivers studied in this note. For unframed orthosymplectic theories, monopoles valued in both the integer and half-integer lattices contribute to the Coulomb branch Hilbert series. Table \ref{tab:spin2_hs} contains Hilbert series for the $\spin(2)$ magnetic quivers in Tables \ref{tab:spin2_magquivs_i} and \ref{tab:spin2_magquivs_ii}, Table \ref{tab:spin3_hs} contains Hilbert series for the $\spin(3)$ magnetic quivers in Table \ref{tab:spin3_magquivs}, Table \ref{tab:spin4_hs} contains Hilbert series for the $\spin(4)$ magnetic quivers in Table \ref{tab:spin4_magquivs} and Table \ref{tab:wreath_fold_hs} contains Hilbert series for the wreathed and folded quivers in Tables \ref{tab:wreathing} and \ref{tab:folding}.
\begin{landscape}
\begin{table}[h!]
\centering
\begin{tabular}{|c|c|c|c|}
\hline
$N_\textrm{S}$ & $\Lambda_{\mathbb{Z}}$ & $\Lambda_{\mathbb{Z}+\frac{1}{2}}$ & CB \\\hline 
4 & \raisebox{-0.5\height}{$\frac{1+4t^{2}+23t^{4}+24t^{6}+23t^{8}+4t^{10}+t^{12}}{(1-t^{2})^{3}(1-t^{4})^{3}}$} & \raisebox{-0.5\height}{$\frac{8t^{2}\left(1+2t^{2}+4t^{4}+2t^{6}+t^{8}\right)}{(1-t^{2})^{3}(1-t^{4})^{3}}$} & $a_3$\\\hline
5 & \raisebox{-0.5\height}{$\frac{1+12t^{2}+58t^{4}+124t^{6}+170t^{8}+\cdots+t^{16}}{(1-t^{2})^{4}(1-t^{4})^{4}}$} & \raisebox{-0.5\height}{$\frac{8t^{2}\left(1+6t^{2}+17t^{4}+22t^{6}+17t^{8}+6t^{10}+t^{12}\right)}{(1-t^{2})^{4}(1-t^{4})^{4}}$} & $a_4$\\\hline
6 & \raisebox{-0.5\height}{$\frac{(1+14t^{2}+123t^{4}+412t^{6}+916t^{8}+1100t^{10}+\cdots+t^{20})}{(1-t^{2})^{5}(1-t^{4})^{5}}$} & \raisebox{-0.5\height}{$\frac{16t^{2}(1+7t^{2}+28t^{4}+54t^{6}+72t^{8}+\cdots+t^{16})}{(1-t^{2})^{5}(1-t^{4})^{5}}$} & $a_5$\\\hline
8 & \raisebox{-0.5\height}{$\frac{1+24t^{2}+421t^{4}+2624t^{6}+10571t^{8}+25640t^{10}+44559t^{12}+51968t^{14}+\cdots+t^{28}}{(1-t^{2})^{7}(1-t^{4})^{7}}$} & \raisebox{-0.5\height}{$\frac{32t^{2}\left(1+12t^{2}+86t^{4}+320t^{6}+821t^{8}+1364t^{10}+1656t^{12}+\cdots+t^{24}\right)}{(1-t^{2})^{7}(1-t^{4})^{7}}$} & $a_7$\\\hline
9 & \raisebox{-0.5\height}{$\frac{\left(\begin{aligned}1&+56t^{2}+876t^{4}+6664t^{6}+30506t^{8}+92904t^{10}\\&+199612t^{12}+312152t^{14}+361818t^{16}+\cdots+t^{32}\end{aligned}\right)}{(1-t^{2})^{8}(1-t^{4})^{8}}$} & \raisebox{-0.5\height}{$\frac{16t^{2}\left(\begin{aligned}1&+28t^{2}+287t^{4}+1568t^{6}+5355t^{8}+12348t^{10}\\&+20101t^{12}+23584t^{14}+\cdots+t^{28}\end{aligned}\right)}{(1-t^{2})^{8}(1-t^{4})^{8}}$} & $a_8$\\\hline
10 & \raisebox{-0.5\height}{$\frac{\left(\begin{aligned}1&+58t^{2}+1133t^{4}+11160t^{6}+66886t^{8}\\&+265412t^{10}+742518t^{12}+1512760t^{14}\\&+2302166t^{16}+2642532t^{18}+\cdots+t^{36}\end{aligned}\right)}{(1-t^{2})^{9}(1-t^{4})^{9}}$} & \raisebox{-0.5\height}{$\frac{32t^{2}\left(\begin{aligned}1&+29t^{2}+330t^{4}+2065t^{6}+8330t^{8}\\&+23289t^{10}+47386t^{12}+71825t^{14}\\&+82450t^{16}+\cdots+t^{32}\end{aligned}\right)}{(1-t^{2})^{9}(1-t^{4})^{9}}$} & $a_9$\\\hline
12 & \raisebox{-0.5\height}{$\frac{\left(\begin{aligned}1&+68t^{2}+2235t^{4}+33528t^{6}+297957t^{8}\\&+1722348t^{10}+7053727t^{12}+21291248t^{14}\\&+49086312t^{16}+87897808t^{18}+124211800t^{20}\\&+139168304t^{22}+\cdots+t^{44}\end{aligned}\right)}{(1-t^{2})^{11}(1-t^{4})^{11}}$} & \raisebox{-0.5\height}{$\frac{64t^{2}\left(\begin{aligned}1&+34t^{2}+528t^{4}+4642t^{6}+26967t^{8}\\&+110060t^{10}+332992t^{12}+766392t^{14}\\&+1374328t^{16}+1939640t^{18}\\&+2175744t^{20}+\cdots+t^{40}\end{aligned}\right)}{(1-t^{2})^{11}(1-t^{4})^{11}}$} & $a_{11}$\\\hline
16 & $1+127t^{2}+9152t^{4}+323648t^{6}+\cdots$ &$128t^{2}+9088t^{4}+323712t^{6}+\cdots$ & $a_{15}$\\\hline
17 & $1+256t^{2}+18768t^{4}+691968t^{6}+\cdots$& $32t^{2}+4352t^{4}+223584t^{6}+\cdots$ & $a_{16}$\\\hline
18 & $1+259t^{2}+20085t^{4}+802327t^{6}+\cdots$& $64t^{2}+8832t^{4}+468032t^{6}+\cdots$ & $a_{17}$\\\hline
20 & $1+271t^{2}+25140t^{4}+1245260t^{6}+\cdots$& $128t^{2}+18560t^{4}+1082240t^{6}+\cdots$ & $a_{19}$\\\hline
24 & $1+319t^{2}+45648t^{4}+3350960t^{6}+\cdots$& $256t^{2}+43776t^{4}+3319040t^{6}+\cdots$ & $a_{23}$\\\hline
32 & $1+511t^{2}+139008t^{4}+\cdots$& $512t^{2}+138752t^{4}+\cdots$ & $a_{31}$\\\hline
\end{tabular}
\caption{Coulomb branch Hilbert series calculations for the $\spin(2)$ magnetic quivers in Tables \ref{tab:spin2_magquivs_i} and \ref{tab:spin2_magquivs_ii}. Each example contains a significant enhancement from the half-integer lattice at order $t^{2}$, which augments the integer-lattice contribution (typically a product of several factors) to a single Lie group of type $\surm(N)$.}
\label{tab:spin2_hs}
\end{table}
\end{landscape}

\begin{landscape}
\begin{table}[h!]
\centering
\begin{tabular}{|c|c|c|c|}
\hline
$N_\textrm{S}$ & $\Lambda_{\mathbb{Z}}$ & $\Lambda_{\mathbb{Z}+\frac{1}{2}}$ & CB \\\hline 
6 & \raisebox{-0.5\height}{$\frac{\left(\begin{aligned}1+&25t^{2}+559t^4+4953t^6+29450t^8\\+&110446t^{10}+305214t^{12}+602903t^{14}\\+&916616t^{16}+1038354t^{18}+\cdots+t^{36}\end{aligned}\right)}{(1-t^{2})^{9}(1-t^{4})^{9}}$} & \raisebox{-0.5\height}{$\frac{32t^{2}\left(\begin{aligned}1&+16t^{2}+162t^{4}+896t^{6}+3514t^{8}\\+&9408t^{10}+19058t^{12}+28352t^{14}\\+&32770t^{16}+\cdots+t^{32}\end{aligned}\right)}{(1-t^{2})^{9}(1-t^{4})^{9}}$} & $d_6$ \\\hline
8 & \raisebox{-0.5\height}{$\frac{\left(\begin{aligned}1&+43t^{2}+2017t^{4}+33398t^{6}+377672t^{8}\\+&2734880t^{10}+14688568t^{12}+58436302t^{14}\\+&182721703t^{16}+449894397t^{18}+899279783t^{20}\\+&1457613116t^{22}+1949633360t^{24}\\+&2140843120t^{26}+\cdots+t^{52}\end{aligned}\right)}{(1-t^{2})^{13}(1-t^{4})^{13}}$} & \raisebox{-0.5\height}{$\frac{64t^{2}\left(\begin{aligned}1&+28t^{2}+547t^{4}+5768t^{6}+43282t^{8}\\+&227680t^{10}+918088t^{12}+2843472t^{14}\\+&7052153t^{16}+14013596t^{18}+22829285t^{20}\\+&30395920t^{22}+33522760t^{24}+\cdots+t^{48}\end{aligned}\right)}{(1-t^{2})^{13}(1-t^{4})^{13}}$} & $d_8$ \\\hline
9 & $1+121t^{2}+5830t^{4}+162280t^{6}+3065576t^{8}+\cdots$ & $1+153t^{2}+8550t^{4}+261800t^{6}+5264424t^{8}\cdots$ & $d_9$ \\\hline
10 & $1+126t^{2}+7394t^{4}+259274t^{6}+\cdots$ & $64t^{2}+5696t^{4}+231040t^{6}+\cdots$ & $d_{10}$\\\hline
12 & $1+148t^{2}+13860t^{4}+\cdots$ & $128t^{2}+13440t^{4}+\cdots$ & $d_{12}$\\\hline
16 & $1+240t^{2}+43504t^{4}+\cdots$ & $256t^{2}+43264t^{4}+\cdots$ & $d_{16}$\\\hline
\end{tabular}
\caption{Coulomb branch Hilbert series calculations for the $\spin(3)$ magnetic quivers in Table \ref{tab:spin3_magquivs}. Each example contains a significant enhancement from the half-integer lattice at order $t^{2}$, which augments the integer-lattice contribution (typically a product of several factors) to a single Lie group of type $\sorm(2N)$.}
\label{tab:spin3_hs}
\end{table}
\end{landscape}
\begin{landscape}
\begin{table}[h!]
\centering
\begin{tabular}{|c|c|c|c|}
\hline
$(N_\textrm{S},N_\textrm{C})$ & $\Lambda_{\mathbb{Z}}$ & $\Lambda_{\mathbb{Z}+\frac{1}{2}}$ & CB \\\hline 
$(2,3)$ & \raisebox{-0.5\height}{$\frac{1+4t^{2}+23t^{4}+24t^{6}+23t^{8}+4t^{10}+t^{12}}{(1-t^{2})^{6}(1-t^{4})^{2}}$} & \raisebox{-0.5\height}{$\frac{8t^{2}(1+2t^{2}+4t^{4}+2t^{6}+t^{8})}{(1-t^{2})^{6}(1-t^{4})^{2}}$} & $a_1\times a_3$\\\hline
$(3,3)$ & \raisebox{-0.5\height}{$\frac{(1+8t^{2}+126t^{4}+488t^{6}+1535t^{8}+2576t^{10}+3332t^{12}+t^{24})}{(1-t^{2})^{6}(1-t^{4})^{6}}$} & \raisebox{-0.5\height}{$\frac{16t^{2}(1+2t^{2}+4t^{4}+2t^{6}+t^{8})(1+4t^{2}+23t^{4}+24t^{6}+\cdots+t^{12})}{(1-t^{2})^{6}(1-t^{4})^{6}}$} & $a_3\times a_3$ \\\hline
$(4,4)$ & \raisebox{-0.5\height}{$\frac{\left(\begin{aligned}1&+14t^{2}+507t^{4}+3950t^{6}+29797t^{8}\\&+119304t^{10}+409556t^{12}+933064t^{14}\\&+1796442t^{16}+2482612t^{18}+2900194t^{20}\\&+\cdots+t^{40}\end{aligned}\right)}{(1-t^{2})^{10}(1-t^{4})^{10}}$} & \raisebox{-0.5\height}{$\frac{32t^{2}\left(\begin{aligned}1&+11t^{2}+150t^{4}+827t^{6}+4039t^{8}\\&+12066t^{10}+30557t^{12}+53941t^{14}\\&+80453t^{16}+\cdots+t^{36}\end{aligned}\right)}{(1-t^{2})^{10}(1-t^{4})^{10}}$} & $d_4\times d_4$ \\\hline
$(5,4)$ & $1+57t^{2}+1546t^{4}+26837t^{6}+336007t^{8}+\cdots$ & $16t^{2}+784t^{4}+17792t^{6}+257488t^{8}+\cdots$ & $d_4\times d_5$ \\\hline
$(5,5)$ & $1+58t^{2}+1965t^{4}+44044t^{6}+\cdots$ & $32t^{2}+1600t^{4}+40544t^{6}+\cdots$ & $d_5\times d_5$\\\hline
$(6,6)$ & $1+68t^{2}+3856t^{4}+131344t^{6}+\cdots$ & $64t^{2}+3776t^{4}+131072t^{6}+\cdots$ & $d_6\times d_6$\\\hline
$(8,8)$ & $1+112t^{2}+12592t^{4}+765462t^{6}+\cdots$ & $128t^{2}+12416t^{4}+766848t^{6}+\cdots$ & $d_8\times d_8$\\\hline
\end{tabular}
\caption{Coulomb branch Hilbert series calculations for the $\spin(4)$ magnetic quivers in Table \ref{tab:spin4_magquivs}. Each example contains a significant enhancement from the half-integer lattice at order $t^{2}$, which augments the integer-lattice contribution (typically a product of several factors) to a product of two Lie groups of the form $\sorm(2N)\times\sorm(2N')$.}
\label{tab:spin4_hs}
\end{table}
\end{landscape}
\begin{landscape}
\begin{table}[h!]
\centering
\begin{tabular}{|c|c|c|c|}
\multicolumn{4}{c}{Wreathed Theories} \\\hline
$\mathcal{T}$ & $\Lambda_{\mathbb{Z}}$ & $\Lambda_{\mathbb{Z}+\frac{1}{2}}$ & CB \\\hline 
$\begin{aligned}&\spin(2)\\&N_{\mathrm{S}}=4\end{aligned}$ & $1 + 4 t^2 + 27 t^4 + 79 t^6 + 226 t^8 + 493 t^{10}+\cdots$& $6 t^2 + 22 t^4 + 86 t^6 + 214 t^8 + 508 t^{10}$& \\\hline
$\begin{aligned}&\spin(2)\\&N_{\mathrm{S}}=8\end{aligned}$ & $1+16t^{2}+330t^{4}+3290t^{6}+23815t^{8}+129731t^{10}+\cdots$ & $20t^{2}+308t^{4}+3340t^{6}+23660t^{8}+130012t^{10}+\cdots$ & $\overline{n.min.C_4}$ \\\hline
$\begin{aligned}&\spin(2)\\&N_{\mathrm{S}}=16\end{aligned}$ & $1+64t^{2}+4644t^{4}+161892t^{6}+3591642t^{8}+56339274t^{10}+\cdots$ & $72t^{2}+4552t^{4}+162264t^{6}+3589464t^{8}+56345736t^{10}+\cdots$ & $\overline{n.min.C_8}$ \\\hline
$\begin{aligned}&\spin(2)\\&N_{\mathrm{S}}=32\end{aligned}$ & $1+256t^{2}+69768t^{4}+\cdots$ & $272t^{2}+69392t^{4}+\cdots$ & $\overline{n.min.C_{16}}$\\\hline
$\begin{aligned}&\spin(3)\\&N_{\mathrm{S}}=6\end{aligned}$ & $1+31t^{2}+641t^{4}+7995t^{6}+71725t^{8}+495575t^{10}+\cdots$ & $24t^{2}+568t^{4}+76870t^{6}+70208t^{8}+490696t^{10}+\cdots$ & $\overline{n.min.B_{5}}$\\\hline
$\begin{aligned}&\spin(3)\\&N_{\mathrm{S}}=10\end{aligned}$ & $1+123t^{2}+6621t^{4}+212767t^{6}+\cdots$ & $48t^{2}+4208t^{4}+166112t^{6}+\cdots$ & $\overline{n.min.B_{9}}$\\\hline
\multicolumn{4}{c}{Folded Theories} \\\hline
$\begin{aligned}&\spin(2)\\&N_{\mathrm{S}}=4\end{aligned}$ & $\frac{1 + 2 t^2 + 10 t^4 + 2 t^6 + t^8}{(1-t^{2})^2(1-t^{4})^{2}}$ & $\frac{6 t^2 + 4 t^4 + 6 t^6}{(1-t^2)^2(1-t^4)^2}$ & \\\hline
$\begin{aligned}&\spin(2)\\&N_{\mathrm{S}}=8\end{aligned}$ & \raisebox{-0.5\height}{$\frac{\left(\begin{aligned}1+&12t^{2}+108t^{4}+212t^{6}+358t^{8}\\+&212t^{10}+108t^{12}+12t^{14}+t^{16}\end{aligned}\right)}{(1-t^{2})^{4}(1-t^{4})^{4}}$} & \raisebox{-0.5\height}{$\frac{4t^{2}\left(5+20t^{2}+67t^{4}+72t^{6}+67t^{8}+20t^{10}+5t^{12}\right)}{(1-t^{2})^{4}(1-t^{4})^{4}}$} & $c_4$\\\hline
$\begin{aligned}&\spin(2)\\&N_{\mathrm{S}}=16\end{aligned}$ & \raisebox{-0.5\height}{$\frac{\left(\begin{aligned}1+&56t^{2}+1464t^{4}+12712t^{6}+68252t^{8}+220600t^{10}\\+&508424t^{12}+807016t^{14}+957254t^{16}+\cdots+t^{32}\end{aligned}\right)}{(1-t^{2})^{8}(1-t^{4})^{8}}$} & \raisebox{-0.5\height}{$\frac{8t^{2}\left(\begin{aligned}9&+168t^{2}+1659t^{4}+8304t^{6}+28121t^{8}\\+&62552t^{10}+102307t^{12}+118048t^{14}+\cdots+t^{28}\end{aligned}\right)}{(1-t^{2})^{8}(1-t^{4})^{8}}$} & $c_8$\\\hline
$\begin{aligned}&\spin(2)\\&N_{\mathrm{S}}=32\end{aligned}$ & $\begin{aligned}1&+256t^{2}+26248t^{4}+1161984t^{6}\\+&30763812t^{8}+560541952t^{10}+\cdots \end{aligned}$ & $\begin{aligned} 272t^{2}&+26112t^{4}+1162800t^{6}\\ &+30759936t^{8}+560557456t^{10}+\cdots\end{aligned}$ & $c_{16}$\\\hline
$\begin{aligned}&\spin(3)\\&N_{\mathrm{S}}=6\end{aligned}$ & $\begin{aligned}1+&31t^{2}+600t^{4}+6930t^{6}\\+&56520t^{8}+352902t^{10}+\cdots\end{aligned}$ & $\begin{aligned}24t^{2}&+544t^{4}+6720t^{6}\\+&55680t^{8}+350592t^{10}+\cdots\end{aligned}$ & $b_5$\\\hline
$\begin{aligned}&\spin(3)\\&N_{\mathrm{S}}=10\end{aligned}$ & $1+123t^{2}+6480t^{4}+201892t^{6}+4282776t^{8}+67260228t^{10}+\cdots$& $48t^{2}+4160t^{4}+159392t^{6}+3740544t^{8}+61984992t^{10}+\cdots$ & $b_{10}$\\\hline
\end{tabular}
\caption{Coulomb branch Hilbert series calculations for the wreathed and folded magnetic quivers in Table \ref{tab:wreathing}. Each example contains a significant enhancement from the half-integer lattice at order $t^{2}$, which augments the integer-lattice contribution (typically a product of several factors) to a symmetry of the form $\sprm(N)$ or $\sorm(2N+1)$.}
\label{tab:wreath_fold_hs}
\end{table}
\end{landscape}
\bibliographystyle{JHEP}
\bibliography{ref}
\end{document}